\documentclass[lettersize,journal]{IEEEtran}
\usepackage{amsmath,amsfonts}
\usepackage{algorithmic}
\usepackage{algorithm}
\usepackage{array}
\usepackage[caption=false,font=normalsize,labelfont=sf,textfont=sf]{subfig}
\usepackage{textcomp}
\usepackage{stfloats}
\usepackage{url}
\usepackage{verbatim}
\usepackage{cite}

\usepackage[table,xcdraw]{xcolor}
\usepackage[nameinlink, capitalize]{cleveref}
\usepackage{booktabs}
\usepackage{multirow}
\usepackage{subfig}
\usepackage{graphicx}
\usepackage{tcolorbox}
\usepackage{balance}
\usepackage{subfiles}    
\usepackage{amssymb}
\newcommand{\INDSTATE}[1][1]{\STATE\hspace{#1\algorithmicindent}}
\newcommand{\insight}[2]
  {\vspace{-1ex}
    \begin{tcolorbox}[boxrule=1pt,left=1pt, right=1pt, top=1pt,bottom=1pt,]
      \textbf{Discovery #1:} #2
    \end{tcolorbox}\vspace{-1ex}
  }

\newcommand{\answer}[2]
  {\vspace{-1ex}
    \begin{tcolorbox}[boxrule=1pt,left=1pt, right=1pt, top=1pt,bottom=1pt,]
      \textbf{Answer to RQ#1:} #2
    \end{tcolorbox}\vspace{-1ex}
  }

\begin{document}

\title{Lorica: A Synergistic Fine-Tuning Framework for Advancing Personalized Adversarial Robustness}

\author{
Tianyu~Qi,~Lei~Xue,~Yufeng~Zhan,~and~Xiaobo~Ma
\IEEEcompsocitemizethanks{
\IEEEcompsocthanksitem T.~Qi, and L.~Xue are with the School of Cyber Science and Technology, Sun Yat-sen University, Shenzhen, China. 
E-mail: qity9@mail2.sysu.edu.cn,~xuelei3@mail.sysu.edu.cn.
\IEEEcompsocthanksitem Y.~Zhan is with the School of Automation, Beijing Institute of Technology, Beijing, China. 
E-mail: yu-feng.zhan@bit.edu.cn.
\IEEEcompsocthanksitem X.~Ma is with the School of Cyber Science and Engineering, Xi'an Jiaotong University, Xi'an, China. 
E-mail: xma.cs@xjtu.edu.cn.
}
\thanks{This work is supported by the National Natural Science Foundation of China (No. 62372490, W2412110, 62473047, U23A20332, 62272381, 623B2081), Natural Science Basic Research Program of Shaanxi Province (2023-JC-JQ-50). We also thank the Guangdong Key Laboratory of Information Security Technology for their support. Lei Xue is the corresponding author. The conference version of this paper is ``Sylva: Tailoring Personalized Adversarial Defense in Pre-trained Models via Collaborative Fine-tuning," in Proc. of ACM CCS, 2025: 1679-1693.}}

\maketitle

\begin{abstract}
The growing use of large pre-trained models in edge computing has made model inference on mobile clients both feasible and popular. Yet these devices remain vulnerable to adversarial attacks, threatening model robustness and security. Federated adversarial training (FAT) offers a promising solution by enhancing robustness while preserving client privacy. However, FAT often yields a generalized global model that struggles with heterogeneous client data, leading to limited personalization and significant communication overhead, while largely overlooking robustness against Byzantine adversarial clients.
%
%
In this paper, we propose \textit{Lorica}, a personalized synergistic adversarial training framework that delivers customized defense models through a two-phase process. In Phase 1, \textit{Lorica} applies LoRA-FA for local adversarial fine-tuning, enabling personalized robustness while reducing communication by uploading only LoRA-FA parameters. In Phase 2, a forward-gating selection strategy improves benign accuracy, further refining the personalized model. This yields tailored defense models that effectively balance robustness and accuracy.
Extensive experiments on benchmark datasets demonstrate that \textit{Lorica} can achieve up to 68$\times$ improvements in communication efficiency compared to state-of-the-art algorithms, while achieving up to 35.3\% and 55.0\% enhancements in adversarial robustness and benign accuracy, respectively.
\end{abstract}


\section{Introduction}

With the rapid advancement of large language models (LLM), large-scale pre-trained models have garnered widespread attention across various fields, including computer vision~\cite{dosovitskiy2020image} and autonomous driving~\cite{yang2024unipad}, etc. 
Fine-tuning pre-trained models for downstream tasks has gradually established itself as a novel learning paradigm~\cite{hu2021lora}. 
Meanwhile, the increasing computational power of edge devices has facilitated the localized deployment of the pre-trained models, unlocking their potential for various applications on devices~\cite{liu2024mobilellm}.
%

However, recent studies have revealed substantial security risks associated with deploying pre-trained models on edge devices.
%
The weak protection mechanisms of these devices make permissions vulnerable to unauthorized access, exposing sensitive resources, while locally stored model parameters are often inadequately secured, heightening the risk of theft or manipulation~\cite{deng2022understanding}.
On the other hand, many pre-trained model weights and architectures are open-source, allowing attackers to easily obtain original parameters and designs from public repositories, such as embedding backdoors or executing poisoning attacks on fine-tuned models for downstream tasks~\cite{liu2022poisonedencoder}.
%
These vulnerabilities threaten the security of edge-based learning, highlighting the need for robust defenses.

\begin{figure*}[!tb]
    \centering
    \includegraphics[width=6.8  in]{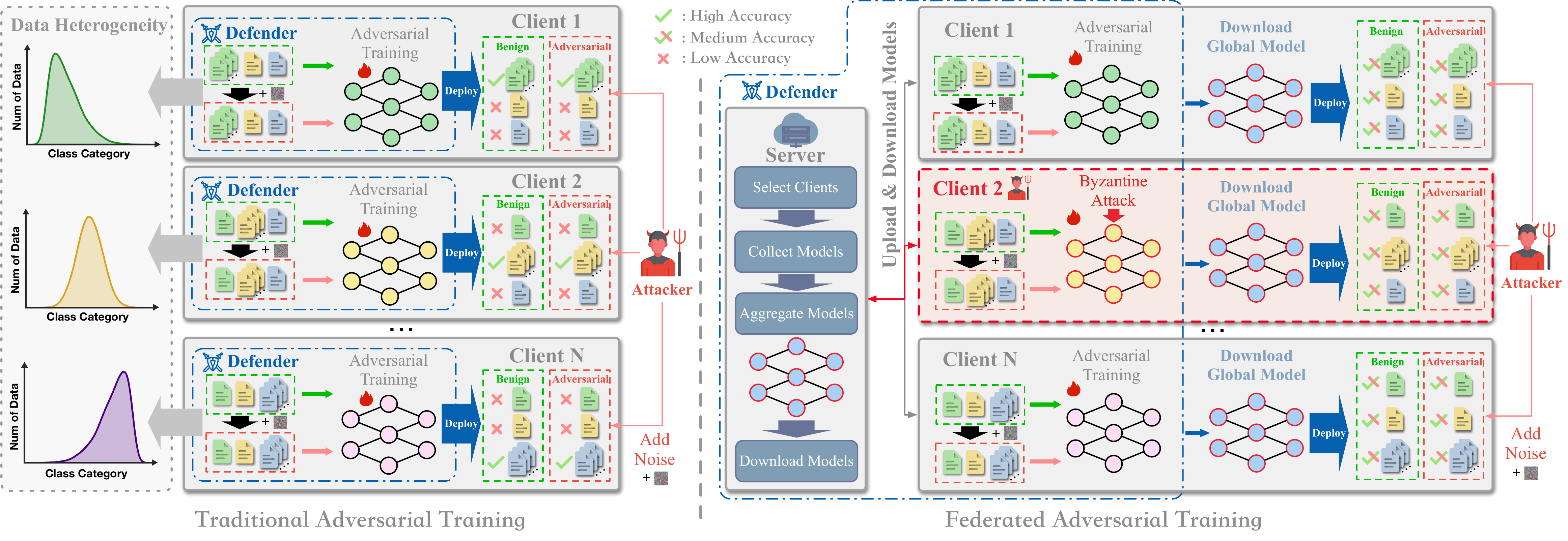}
    \vspace{-3mm} 
    \caption{Challenges of adversarial training in multi-client scenario. (1) Left: Local adversarial training performs well on abundant classes but fails on rare or unseen ones. (2) Right: Federated adversarial training improves generalization across clients but lacks personalization for unique client data distributions.}
    \label{fig:Challenge}
    \vspace{-5mm}
\end{figure*}

To mitigate these challenges, adversarial training (AT) has become a widely utilized defense strategy~\cite{madry2017towards, zhang2019theoretically}. By incorporating benign and adversarial data, this approach enhances model robustness while preserving accuracy, yet traditional adversarial training shows notable limitations in real-world scenarios.
%
%
As shown on the left side of \cref{fig:Challenge}, the challenges of adversarial training stem from several key factors. 
%
First, the limited data of local clients often hinders adversarial training, causing suboptimal performance or overfitting during downstream fine-tuning.
%
Second, client data is typically non-IID; for example, autonomous driving data under extreme weather is scarce, while healthcare datasets vary significantly across hospitals.
%
Such variability exposes weaknesses exploitable by attackers, and privacy concerns further prevent centralized training, exacerbating these challenges.

In multi-client environments, federated adversarial training (FAT) has proven effective for enhancing robustness while preserving privacy~\cite{zizzo2020fat}. With large pre-trained models gaining prominence, FAT-based fine-tuning is expected to dominate, allowing devices to contribute limited data through periodic synchronization and knowledge sharing.
%
Despite its potential, most studies focus on generalized global models for non-IID data~\cite{zhou2022adversarial, chen2022calfat}. Yet under significant heterogeneity, such models often fail to provide effective client-specific defense, as their strong generalization limits adaptation to unique data distributions, leading to diminished performance, as illustrated on the right side of \cref{fig:Challenge}.
Moreover, FAT typically assumes benign participants. Under Byzantine clients uploading arbitrary updates, plug-in defenses can help, but integrating them into the FAT pipeline often adds extra computation overhead in large model training~\cite{ma2022shieldfl}.
%
Integrating FAT with large pre-trained models introduces communication overhead that can hinder training speed, especially in resource-constrained settings. These limitations highlight a critical gap: can we design a personalized adversarial training framework that delivers robust, tailored defense for each client while maintaining efficiency on edge devices and simultaneously being resilient to Byzantine attacks?

This study aims to address three critical challenges in adversarial fine-tuning pre-trained models in multi-client environments:
(1) \textbf{Designing a personalized defense framework}: Given the heterogeneous data distributions across clients, a collaborative adversarial training framework is required to provide each client with a tailored defense model while ensuring data privacy.
%
(2) \textbf{Enhancing robustness under adversarial and Byzantine attacks}: Ensuring high benign accuracy and adversarial robustness in the presence of Byzantine clients uploading malicious updates is critical for reliable deployment.
%
%
(3) \textbf{Balancing efficiency for clients}: Achieving improved performance on resource-constrained edge devices requires addressing communication and training efficiency challenges.
By achieving these objectives, each client obtains a robust and accurate model tailored to its unique data, excelling in downstream tasks while offering strong personalized defenses against adversarial attacks.

To address the challenges above, we propose \textit{Lorica}, a synergistic framework that tailors pre-trained models with robust defense capabilities through personalized adversarial fine-tuning for edge devices.
\textit{Lorica} is designed to fine-tune models in a way that simultaneously enhances robustness and accuracy for each client, even in environments with heterogeneous data distributions.
%
To validate feasibility, we conduct preliminary experiments using the memory-efficient low-rank adaptation (LoRA-FA) algorithm for personalized adversarial fine-tuning. These experiments demonstrate its potential in adversarial training while exposing critical limitations of existing defenses.
Then we present \textit{Lorica}'s two-phase training framework. 
In the first phase, personalized adversarial training divides the model into two modules: the LoRA-FA module and the classifier module. The parameters of the LoRA-FA module are shared and aggregated across clients to enhance the generalization ability of the model backbone for adversarial feature extraction. 
Meanwhile, the classifier module is fine-tuned locally on each client, enabling personalized improvements in classifying adversarial samples based on each client's unique data distribution.
%
We propose a ball-tree-based aggregation algorithm to accelerate adversarial training convergence and a MoE-inspired adaptive class-balanced weighted loss to address data heterogeneity. Meanwhile, the aggregation also provides Byzantine awareness by mitigating the influence of malicious client updates.
In the second phase, we employ a forward-gating network to adaptively select and freeze layers, enhancing each client’s benign accuracy while preserving robustness.
%
%
This selective adjustment enables a higher degree of personalization, effectively balancing benign performance with adversarial defense.
%
%
The main contributions of this paper can be summarized as follows:
\begin{itemize}
    \item We propose a personalized synergistic adversarial training framework for pre-trained models, enabling client-specific defenses while preserving privacy and minimizing overhead on edge devices.
    \item We introduce LoRA-FA adversarial fine-tuning, integrating MoE-inspired adaptive optimization with a ball-tree-based model aggregation with Byzantine-aware and a newly designed loss to enhance robust generalization under heterogeneous data, along with theoretical convergence guarantees.
    \item We propose a cooperative a forward-gating layer selection method for pre-trained models, where the gating network adaptively selects layers to optimize the trade-off between benign accuracy and adversarial robustness.
    \item We conduct experiments on widely used datasets and pre-trained models, comparing \textit{Lorica} with popular defenses under various attack scenarios, demonstrating its effectiveness in personalized defense.
\end{itemize}

\section{Preliminaries}\label{sec2}

\subsection{Federated Adversarial Training}

Federated learning is an effective algorithm for protecting client privacy~\cite{mcmahan2017communication, qi2025robin}. In a distributed system with $N$ clients and a server, each client $i$ holds a private dataset $(x,y)\in \mathcal{D}_i$, downloads the initialized model $w_i$, and trains it on local data.
The training objective for each client can be expressed as
\begin{equation}
    f_i(w_i)=\frac{1}{|\mathcal{D}_i|}\sum_{(x,y)\in \mathcal{D}_i}f_i(w_i,x,y),
\end{equation}
where $|\mathcal{D}_i|$ denotes the size of the dataset for client $i$.

In federated adversarial training, the local objective is to train models to recognize both benign and adversarial samples, thereby improving robustness. For a benign sample $x$, an adversarial sample is generated as $x^{adv}=x+\sigma$, where the noise $\sigma$ maximizes the loss $\mathcal{L}(w_i, x+\sigma, y)$.
The local training objective for clients in federated adversarial training can be expressed as
\begin{equation}
    f_{i}(w_{i},x,x^{adv},y)=\min \mathbb{E}_{x\sim \mathcal{D}_{i}}\left [\max \mathcal{L}(w_{i},x,x^{adv},y)\right ].
\end{equation}

After local training, clients upload their models to the server, which aggregates parameters weighted by the data size of each device, represented as
\begin{equation}\label{equ:aggre}
    w_{g}=\sum_{i=1}^{N}\frac{|\mathcal{D}_{i}|w_{i}}{\sum_{i=1}^{N}|\mathcal{D}_{i}|}.
\end{equation}
Subsequently, clients can download the global model $w_g$ for further local training, repeating until final convergence.

Most FAT methods assume benign clients during adversarial training, yet the training process itself is vulnerable to Byzantine behaviors. We consider a threat model where a fraction $\rho$ of clients are malicious and can upload arbitrary LoRA-FA updates (or parameters) to disrupt global or expert aggregation, degrading robustness and benign accuracy. The set of honest clients is denoted as $\mathbf{H} \subseteq \{1,2,\dots,N\}$.

\subsection{Memory-efficient Low-Rank Adaptation}

Fine-tuning pre-trained models in mobile computing is challenging due to limited GPU memory, which makes full-parameter training impractical. LoRA-FA~\cite{zhang2023lora}, built on low-rank adaptation (LoRA)~\cite{hu2021lora}, addresses this issue by freezing pre-trained parameters and introducing lightweight trainable layers into transformer modules, thereby reducing computational overhead.

During fine-tuning, the updated parameters exhibit low intrinsic rank. For a LoRA-FA model $w \in \mathbb{R}^{r_{in}\times C}$, where $r_{in}$ is the input dimension and $C$ the number of classes, the model consists of a backbone $w^B \in \mathbb{R}^{r_{in}\times r_{out}}$ and a classifier $w^C \in \mathbb{R}^{r_{out}\times C}$. The backbone is composed of frozen pre-trained parameters $w^P$ and LoRA-FA parameters $w^L$, with $w^L = w^{LA}w^{LB}$, where $w^{LA}$ is fixed and only $w^{LB}$ is updated.
We define the overall mapping as $\mathcal{F}(\cdot)$, where $\mathcal{F}^P(\cdot)$, $\mathcal{F}^L(\cdot)$, and $\mathcal{F}^C(\cdot)$ denote the pre-trained backbone, LoRA-FA module, and classifier, respectively. For an input $x$, the model output is
\begin{equation}
\hat{y} = \mathcal{F}(x) = \mathcal{F}^C(\mathcal{F}^P(x) + \mathcal{F}^L(x)).
\end{equation}
During training, $w^P$ and $w^{LA}$ remain frozen, while only $w^{LB}$ is updated, achieving efficiency with performance close to full-parameter fine-tuning.

\subsection{Threat Model}

We analyze the security risks associated with deploying large pre-trained models in distributed multi-client scenarios, including perspectives from both attackers and defenders.

\subsubsection{Attacker}


We posit that deploying models on client devices introduces system-level vulnerabilities, allowing attackers to exploit client permissions and compromise model inference through adversarial attacks. Depending on the attacker’s access level, such attacks can be classified into two categories:
(1) \textbf{White-box attacks}. The attacker has full access to the deployed model parameters and can generate strong adversarial examples.
(2) \textbf{Grey-box attacks}. The attacker has access to the pre-trained model (e.g., from public platforms such as Hugging Face) but not the client-specific fine-tuned parameters.
The adversarial samples generated must meet two critical criteria:
(1) \textbf{Inconspicuousness}. The perturbations in the adversarial samples should be subtle enough to go unnoticed by humans.
(2) \textbf{Impactfulness}. The adversarial attacks must effectively compromise the model’s downstream tasks, even when standard defense mechanisms are in place.

In addition to inference-time adversaries, we also consider Byzantine attackers during federated adversarial training, a setting rarely explored in existing FAT studies. We assume that a fraction $\rho$ of clients can behave maliciously and upload arbitrary LoRA-FA updates or parameters to the server, aiming to disrupt model aggregation and degrade robustness and benign accuracy.

\subsubsection{Defender}

We propose involving defenders in the fine-tuning process of client models for downstream tasks to enhance robustness. Collaborative fine-tuning is assumed to take place within a distributed framework based on a parameter server (PS) architecture~\cite{li2014scaling}. In this setup, the defender operates at the server level, providing macro-level oversight and regulation of client activities to ensure robust and secure model training.
%

Defenders should satisfy three key properties: (1) \textbf{Data Privacy}, ensuring that each client’s private data remains protected throughout training and deployment; (2) \textbf{Heterogeneity}, supporting personalized defenses that adapt to diverse client data distributions, adversarial capabilities, and the presence of Byzantine or malicious participants; and (3) \textbf{Efficiency}, minimizing computation, memory usage, and communication overhead to enable deployment on resource-constrained devices and networks.

In contrast to conventional FAT settings that focus on inference-time adversarial attacks, we consider a defender operating during federated adversarial training under heterogeneous client data distributions. The primary goal remains effective coordination of adversarial training for personalized robustness, while also accounting for possible Byzantine behaviors arising from a small fraction of clients during training.

\begin{table}[!tb]
\centering
\caption{Adversarial training with/without LoRA and LoRA-FA}\label{moti0}
\setlength{\tabcolsep}{8pt}
\scalebox{0.9}{
\renewcommand{\arraystretch}{0.85}  
\begin{tabular}{@{}c|c|cccc@{}}
\toprule
   Metrics                            & Methods         & ResNet18 & ViT-T & ViT-B & ViT-L \\ \midrule
\multirow{3}{*}{AR$\uparrow$ (\%)}        & w/o-LoRA & \textbf{54.36}        & 53.28     & 59.02      & 62.18      \\
                                        & w-LoRA   & 48.22        & \textbf{58.72}     & 61.14      & \textbf{64.69}      \\
                                            & w-LoRA-FA & 46.73        & 57.28     & \textbf{62.10}      & 62.98      \\ \midrule
\multirow{3}{*}{BA$\uparrow$ (\%)}    & w/o-LoRA & \textbf{62.05}        & 60.14     & 64.37      & 70.66   \\  
                                & w-LoRA   & 59.35        & 63.27    & \textbf{69.73}      & 75.91     \\
                               & w-LoRA-FA & 60.04        & \textbf{63.38}     & 69.43      & \textbf{75.94}      \\ \midrule
\multirow{3}{*}{\begin{tabular}[c]{@{}c@{}}Time$\downarrow$ \\ ($10^3$s/epoch)\end{tabular}} & w/o-LoRA & 0.38         & 1.13      & 1.90       & 6.38       \\
                                    & w-LoRA & 0.31 & 0.84 & 1.60 & 4.90 \\
                               & w-LoRA-FA & \textbf{0.30}         & \textbf{0.75}     & \textbf{1.46}       & \textbf{4.39}       \\ \midrule
\multirow{3}{*}{Paras$\downarrow$ (M)}      & w/o-LoRA & 11.16        & 7.37      & 84.84      & 292.14     \\
                                & w-LoRA   & 0.51         & 1.95      & 2.06       & 2.29       \\
                               & w-LoRA-FA & \textbf{0.26}        & \textbf{0.98}      & \textbf{1.04}      & \textbf{1.15}     \\ \midrule
\multirow{3}{*}{Mem$\downarrow$ (G)}   & w/o-LoRA & 1.25         & 1.04      & 4.86       & 11.43       \\
                        & w-LoRA   & 0.98         & 0.89      & 3.14       & 6.63       \\
                       & w-LoRA-FA & \textbf{0.49}         & \textbf{0.45}      & \textbf{1.57}       &\textbf{3.32}       \\ \bottomrule
\end{tabular}
}
\vspace{-5mm}
\end{table}

\begin{figure}[!tb]
\centering
\subfloat[CIFAR-10]{\includegraphics[width=1.1in]{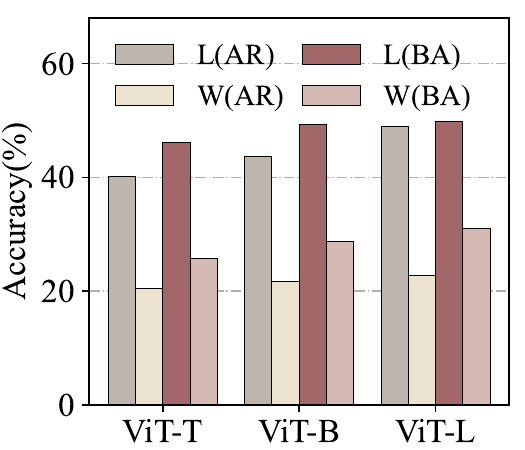}%
\label{fig:moti1_cifar10}}
\subfloat[STL-10]{\includegraphics[width=1.1in]{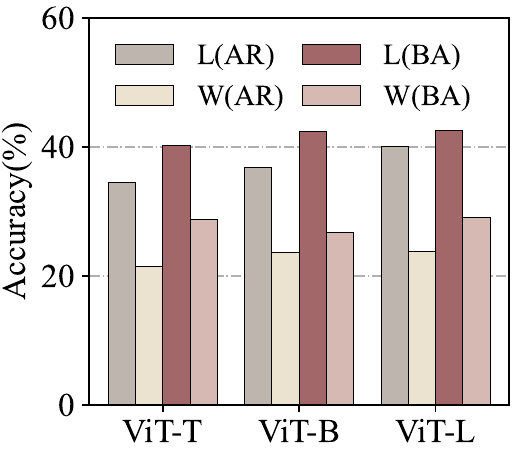}%
\label{fig:moti1_stl10}}
\subfloat[GTSRB]{\includegraphics[width=1.1in]{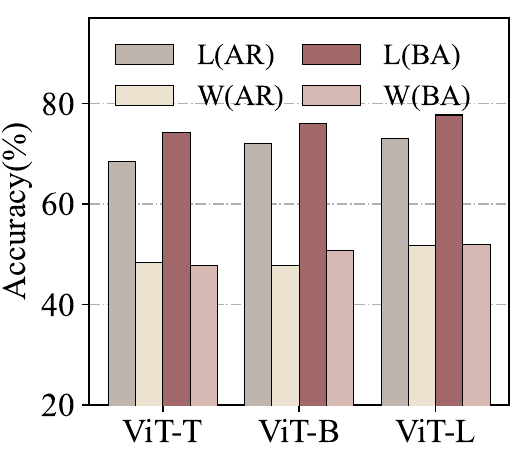}%
\label{fig:moti1_gtsrb}}
\caption{Performance of personalized LoRA-FA in adversarial training under heterogeneous environments}
\label{fig:moti1}
\vspace{-6mm}
\end{figure}

\begin{figure}[!tb]
\centering
\subfloat[CIFAR-10]{\includegraphics[width=1.1in]{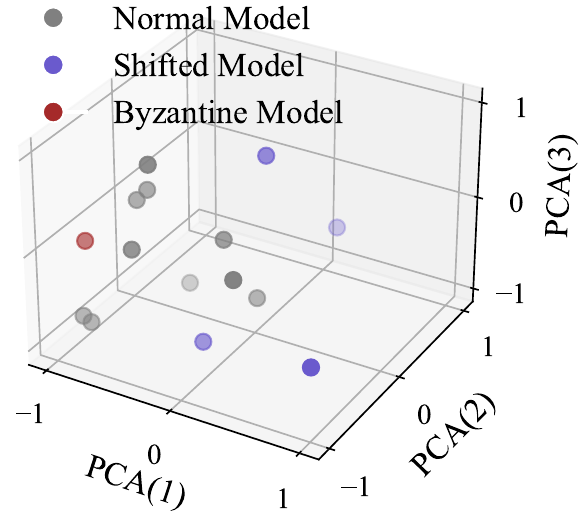}%
\label{fig:moti2_cifar10_scatter}}
\subfloat[STL-10]{\includegraphics[width=1.1in]{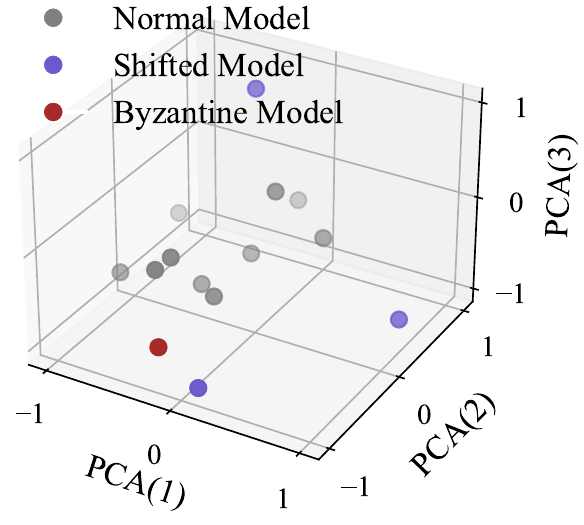}%
\label{fig:moti2_stl10_scatter}}
\subfloat[GTSRB]{\includegraphics[width=1.1in]{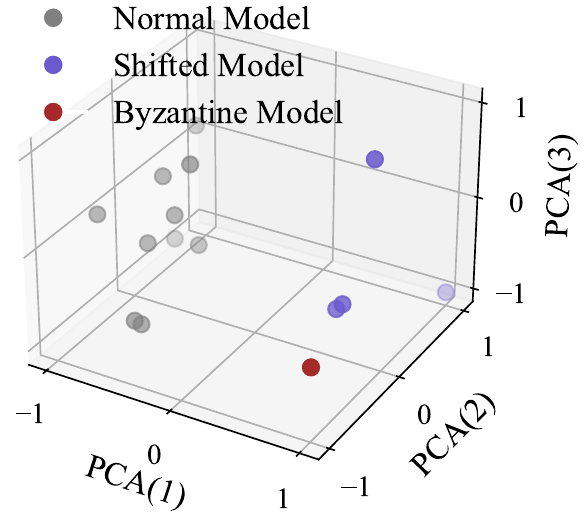}%
\label{fig:moti2_gtsrb_scatter}}
\caption{PCA embedding of the LoRA-FA after federated adversarial training}
\label{fig:moti2_scatter}
\vspace{-6mm}
\end{figure}

\begin{figure}[!tb]
\centering
\subfloat[CIFAR-10]{\includegraphics[width=1.1in]{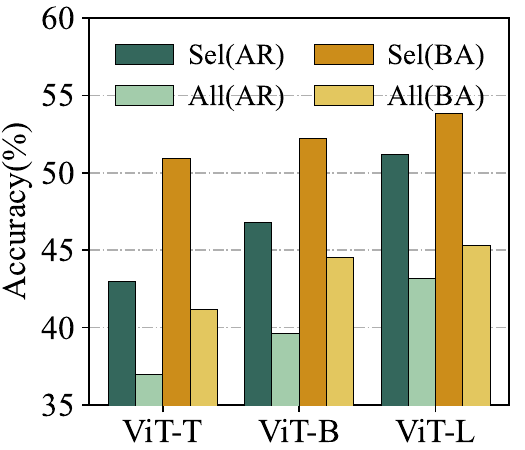}%
\label{fig:moti2_cifar10}}
\subfloat[STL-10]{\includegraphics[width=1.1in]{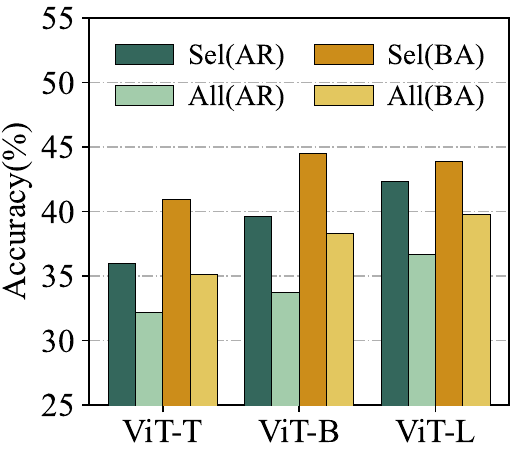}%
\label{fig:moti2_stl10}}
\subfloat[GTSRB]{\includegraphics[width=1.1in]{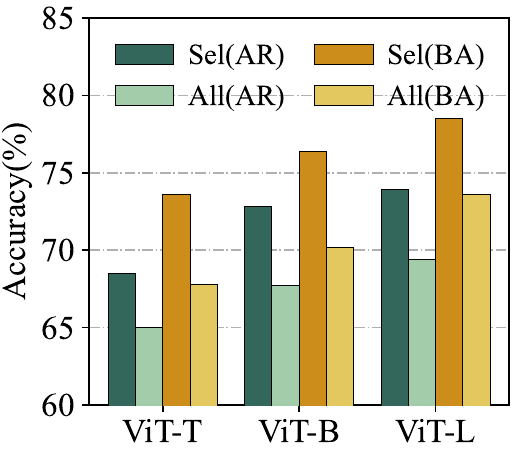}%
\label{fig:moti2_gtsrb}}
\caption{Comparison of aggregation methods for adversarial training in heterogeneous environments}
\label{fig:moti2}
\vspace{-6mm}
\end{figure}

\subsection{Preliminary Analysis of Existing Issues}

In this section, we investigate the critical challenges of fine-tuning pre-trained models for multi-client attack-defense systems, including the efficient use of edge resources for adversarial training, the need for personalized frameworks, the extraction of knowledge from heterogeneous data, and the trade-off between robustness and benign accuracy. We design preliminary experiments to examine these factors, highlighting key findings and open challenges that motivate the development of \textit{Lorica}.

\subsubsection{Impact of Adversarial Training with LoRA-FA}

Adversarial training via full model training has demonstrated effectiveness, but it incurs high training costs, rendering it inefficient for edge devices with limited computational resources. To address this, we conduct experiments to evaluate the effectiveness of adversarial fine-tuning using LoRA-FA.

We utilize the widely-used ResNet~\cite{he2016deep} and ViT~\cite{dosovitskiy2020image} pre-trained models, applying the TRADES algorithm~\cite{zhang2019theoretically} for adversarial fine-tuning via LoRA-FA on the CIFAR-10 dataset~\cite{krizhevsky2009learning}. ResNet and ViT-T are trained for 50 epochs, while ViT-B and ViT-L are trained for 100 epochs.
The experimental results are evaluated using five metrics: \textit{Adversarial Robustness} (AR), assessed by accuracy on adversarial samples; \textit{Benign Accuracy} (BA), assessed by accuracy on benign samples; \textit{Time}, which tracks the time required per training epoch; \textit{Parameter Size} (Paras), which quantifies the number of model parameters fine-tuned during training; \textit{GPU Memory} (Mem), the GPU required for adversarial training with a fixed batch size.

The results in \cref{moti0} show that adversarial fine-tuning with LoRA-FA reduces training time by up to 31.2\% compared to full-model fine-tuning, while requiring only 0.4\% of the parameters and saving about 70.8\% of GPU memory, making it highly efficient for edge devices. As model size increases, LoRA-based adversarial fine-tuning improves both AR and BA, with LoRA-FA achieving comparable performance at lower computational cost. This is consistent with prior findings~\cite{hu2021lora} that low-rank eigenvalues capture most of the model’s energy and preserve feature extraction.

\insight{1}{LoRA-based adversarial fine-tuning preserves or even improves defense effectiveness while drastically reducing computational overhead.}

However, a critical challenge arises when scaling from single-client to multi-client training, as it requires designing a collaborative adversarial defense framework using LoRA-FA to fully leverage its advantages.

\subsubsection{Impact of the Personalized Framework}

In multi-client environments, federated adversarial training has been widely adopted.
However, most existing approaches rely on training a global model, which becomes ineffective when data distributions across devices are heterogeneous. 
This limitation arises from attackers designing tailored attack strategies based on the data distribution of downstream tasks.
As a result, each client's model requires personalized defense mechanisms. 
This raises a crucial question: Can we decompose each client's model into two distinct components—one focused on learning generalized adversarial features, and the other on capturing personalized classification outcomes?

Given the powerful feature extraction capabilities of pre-trained model backbones, we believe that their feature extraction ability remains effective for adversarial examples after adversarial fine-tuning. 
To explore this, we conduct an experiment, where the training and testing data for each client are split according to the same Dirichlet distribution, thereby simulating a unique non-IID environment for each client. We utilize three different ViT models and implement a federated adversarial training strategy, applying TRADES for local fine-tuning through LoRA-FA, on the CIFAR-10~\cite{krizhevsky2009learning}, STL-10~\cite{coates2011analysis}, and GTSRB~\cite{stallkamp2011german} datasets. 
The federated adversarial training involve 5 local training epochs followed by parameter upload and cloud aggregation, repeated for 15 rounds.
For the model upload step, we compare two strategies: uploading whole parameters (W) versus uploading only the LoRA-FA parameters of the backbone while excluding the classifier parameters (L). The latter approach enables the training of a shared global backbone while allowing each client to personalize its classifier.

As shown in \cref{fig:moti1}, the personalized aggregation approach (L) achieves superior performance in both BA and AR metrics on average for each client. This is attributed to our framework's ability to learn generalized features while tailoring the classifier. In contrast, the whole parameter aggregation approach (W) performs poorly, as aggregation in a heterogeneous environment induces model drift, hindering convergence.

\insight{2}{Aggregating only the backbone (LoRA-FA) and personalizing the classifier preserves generalized feature extraction while enhancing the classifier’s sensitivity to heterogeneity, ensuring both global generalization and personalized defense.}

Despite the effectiveness of this personalized framework, a new training paradigm is urgently required to further enhance robust generalization and adaptability within the LoRA-based collaborative adversarial training framework.

\begin{figure}[!tb]
\centering
\subfloat[Adversarial robustness]{\includegraphics[width=1.65in]{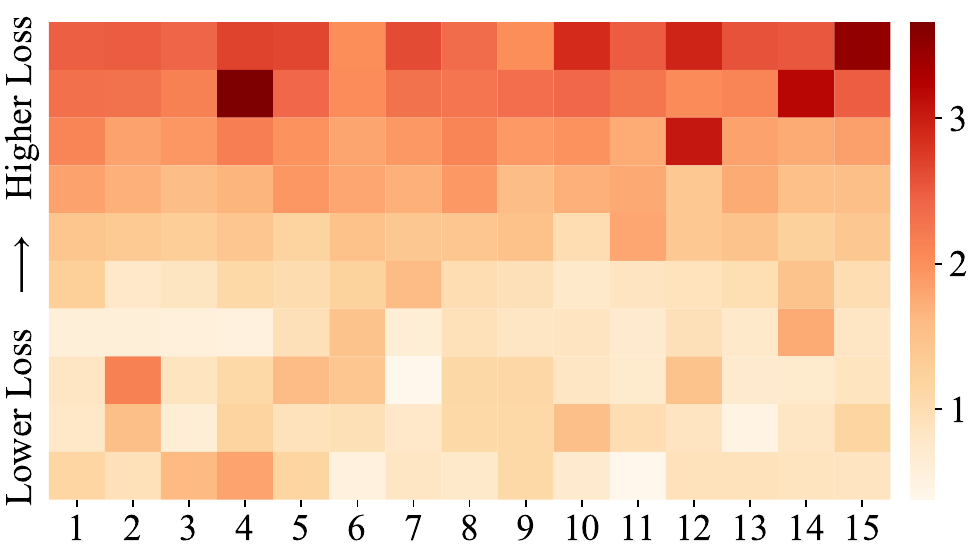}%
\label{fig:moti3_rob}}
\subfloat[Benign accuracy]{\includegraphics[width=1.65in]{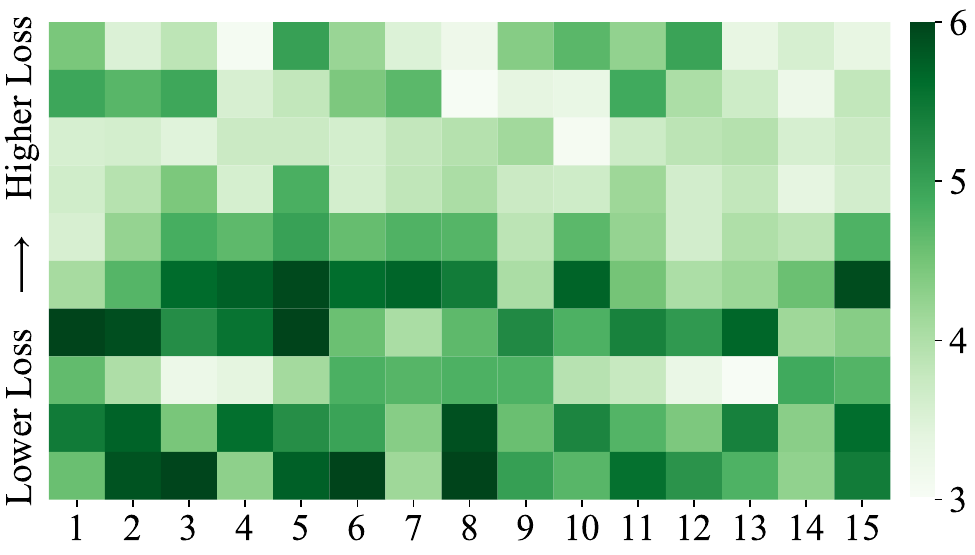}%
\label{fig:moti3_acc}}
\caption{Comparison of aggregation methods for adversarial training in heterogeneous environments}
\label{fig:moti3}
\vspace{-6mm}
\end{figure}

\subsubsection{Impact of the Aggregation Method}

Many studies show that data heterogeneity in federated adversarial training causes model drift on clients, with direct aggregation slowing convergence~\cite{zhou2022adversarial, chen2022calfat}. In the personalized federated adversarial framework, LoRA-FA parameters may likewise be affected by non-IID data, leading to drifts that weaken the adversarial generalization of the fine-tuned backbone.

We simulate a system with 15 clients whose heterogeneous data follow a Dirichlet distribution, and include a Byzantine client performing label-flipping during training. After 10 epochs of local adversarial training, we apply PCA algorithm to the LoRA-FA parameters, reducing them to 3 dimensions, as shown in \cref{fig:moti2_scatter}. The results reveal drifts in some local LoRA-FA models due to data heterogeneity, while the Byzantine client also drifts and overlaps with some honest heterogeneous models.
We design two aggregation strategies: one aggregates all models (All), while the other selectively aggregates only those whose PCA-compressed parameters are close, excluding models with significant drifts and Byzantine attacks (Sel). 
In this setting, naive drift-based aggregation may exclude honest heterogeneous models.
After the same federated adversarial fine-tuning, the averaged test results for each client, shown in \cref{fig:moti2}, demonstrate that only aggregating normal models yields higher AR and BA metrics. This motivates a Byzantine-aware aggregation strategy that preserves heterogeneity-aware adversarial training.

\insight{3}{Under data heterogeneity, both honest and Byzantine clients can exhibit drift, making naive aggregation insufficient for personalized adversarial training.}

However, drifted models may still encode client-specific knowledge arising from data heterogeneity, while similar drift patterns can also be induced by Byzantine updates, and indiscriminately discarding such models can accelerate convergence at the cost of losing valuable information for adversarial training.

\subsubsection{Limitations of the Robustness-Accuracy Trade-off}

In adversarial training within pre-trained models, the key challenge is to maximize benign accuracy while maintaining adversarial robustness.
The method in \cite{zhou2024securely} enhances benign accuracy by first computing the loss for each model layer using a set of adversarial samples. 
The top-$p$ layers with the lowest loss—indicating minimal sensitivity to robustness—are selected for unfreezing, while the others remain frozen. The model is then trained on benign data, enhancing benign accuracy without sacrificing robustness.

However, the method has a critical flaw: the $p$ selected layers may not remain optimal after joint selection due to layer interactions affecting overall performance. To verify this, we conduct the following experiment: for the robust models trained on the 15 clients, we measure the loss of each layer as in \cite{zhou2024securely}. 
We then select multiple combinations of $p$ layers, unfreezing them while freezing the others, and retrained the model using benign data.

\cref{fig:moti3} shows accuracy changes for adversarial and benign samples after fine-tuning. The x-axis represents clients, and the y-axis represents the sum of losses from each layer, arranged in ascending order.
In \cref{fig:moti3_rob}, the trend in robustness aligns with expectations: selecting $p$ layers with lower loss (i.e., lower robustness sensitivity) has a lesser impact on robustness, while layers with higher loss negatively affect it. However, an intriguing observation is that the combination yielding the lowest loss does not always produce the best results, as layer interactions can alter the final outcome. 
Meanwhile, \cref{fig:moti3_acc} shows significant variance in benign accuracy improvement, indicating the need for a trade-off: the selection with the lowest robustness loss does not always result in the greatest benign accuracy gain.

\insight{4}{Jointly selecting robustness-insensitive layers for fine-tuning effectively enhances benign accuracy, but the interactions among layers prevent achieving the optimal outcome.}

Therefore, it is crucial to design an effective and efficient fine-tuning method to identify the appropriate layers, aiming to minimize the reduction in robustness while maximizing benign accuracy improvement.

\begin{figure}[!tb]
    \centering
    \includegraphics[width=3.4  in]{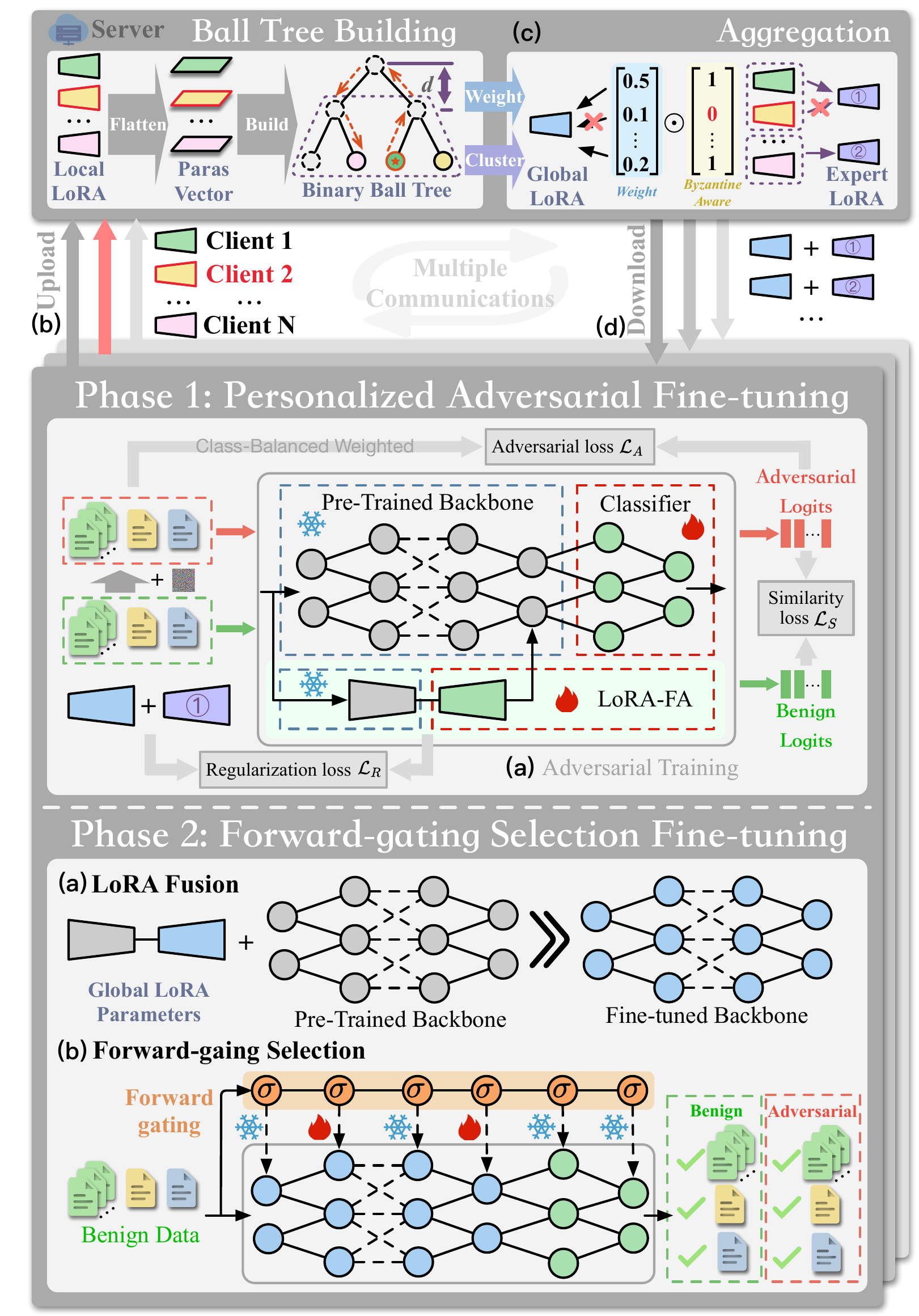}
    \caption{Overview of \textit{Lorica}. (1) Phase 1—Clients and the server collaborate in federated adversarial fine-tuning to obtain a generalized backbone and personalized classifier. (2) Phase 2—Clients fine-tune layers using a forward-gating algorithm, improving benign accuracy while preserving robustness.}
    \label{fig:Method}
    \vspace{-5mm}
\end{figure}

\section{Methodology}\label{sec3}

In this section, we present the framework of the \textit{Lorica} algorithm, offering a comprehensive explanation of its key components and the overall training process.

\subsection{Overview}

\cref{fig:Method} provides an overview of the framework and training process of the \textit{Lorica} algorithm, which is divided into two distinct phases.

In the Phase 1, personalized adversarial fine-tuning, the algorithm is implemented within a federated learning framework, where each client employs the LoRA-FA fine-tuning method to personalize adversarial training. 
%
Each client uploads its LoRA-FA parameters to the server, where a Byzantine-aware ball-tree-based aggregation method generates both a global model and expert models. Clients then download the respective models and perform personalized local fine-tuning using an MoE-inspired adaptive optimization strategy. This ensures robust global generalization while maintaining client-specific personalization.
%
Through this method, each client obtains a shared global backbone integrating the aggregated LoRA-FA modules, enabling robust feature extraction from adversarial samples. Simultaneously, clients retain personalized classifier modules to ensure targeted defense against adversarial attacks specific to their data distributions.

The Phase 2, cooperative game-based fine-tuning, aims to enhance the benign accuracy of models on standard downstream tasks while preserving their robustness.
For the personalized models trained by each client in Phase 1, we assume they already exhibit sufficient robustness against adversarial samples. However, their inference performance on local benign data can still be further improved. To achieve this, each client independently and in parallel fine-tunes its model using its local benign dataset.
%
In this phase, we propose a forward-gating network that adaptively selects which model layers to freeze or fine-tune during training. Guided by an optimization objective designed to balance benign accuracy and adversarial robustness, the network fine-tunes layers with the highest improvement potential, enhancing accuracy on benign data while minimizing robustness loss.

\subsection{Personalized Adversarial Fine-tuning}

We introduce the training details of Phase 1, including the optimization of the aggregation algorithm and the design of the novel LoRA-based training paradigm.

\subsubsection{Ball-tree-based Aggregation}

Similar to other distributed algorithms, after clients upload their models, the server is responsible for aggregating the received parameters. In this case, the server aggregates the uploaded LoRA-FA parameters to construct a global model. This aggregation process is critical for ensuring the generalization of the model while addressing the challenges posed by heterogeneous client data distributions.

We adopt a ball-tree-based~\cite{dolatshah2015ball} retrieval method to assign aggregation weights to each client model. The LoRA-FA model parameters received from each client are flattened into vectors, represented as
\begin{equation}
    \mathbf{G}=\{\mathcal{G}(w_{1}^{L}),\mathcal{G}(w_{2}^{L}),\ldots,\mathcal{G}(w_{N}^{L})\},
\end{equation}
where $\mathcal{G}(\cdot)$ is the flatten function.
We treat each $\mathcal{G}(w_{i}^{L})$ as a node and construct a ball tree for the entire vector $\mathbf{G}$. The ball tree algorithm is a spatial indexing method that organizes multidimensional data in a tree structure. Each non-leaf node represents a hypersphere enclosing a subset of data, facilitating efficient nearest neighbor searches by recursively partitioning the data.
For client $i$ , we can use the ball tree to identify the $k$ client model vectors $\{\mathcal{G}(w_{j_1}^L),\mathcal{G}(w_{j_2}^L),\ldots,\mathcal{G}(w_{j_k}^L)\}$ that are most similar to its parameters, and compute the distances between them
\begin{equation}
    d_{ij_m}=\|\mathcal{G}(w_i^L)-\mathcal{G}(w_{j_m}^L)\|_2 , \quad m=1,2,\ldots,k
\end{equation}
Then, gaussian weighting is applied to these distances to obtain the aggregation weight for each client model, denoted as 
\begin{equation}
    q_i=\frac{\sum_{m=1}^k\exp\left(-\frac{d_{ij_m}}{\sigma^2}\right)}{\sum_{i=1}^N\sum_{m=1}^k\exp\left(-\frac{d_{ij_m}}{\sigma^2}\right)}.
\end{equation}
The weighting scheme favors clients with stable data, centrally located in high-dimensional space and exhibiting stronger robust generalization, while assigning lower weights to those with more extreme, dispersed distributions.
%

Our aggregation computes weights over high-dimensional parameters. Naive nearest neighbor search has $O(N^2)$ complexity, but the ball tree algorithm reduces it to $O(N\log N)$, enabling efficient and scalable distance computation.
%
Leveraging the ball tree’s inherent hierarchical clustering and fast nearest-neighbor retrieval, we partition the tree at depth $d$ into $E$ clusters, each maintaining an expert model, where $e(i)$ denotes the index of the cluster to which client $i$ belongs.
To identify Byzantine clients under non-IID data distributions, we aim to detect the most distributionally deviant clients within each cluster.
Let the set of clients in the $e$-th cluster be denoted by $\mathbf{C}_e$. We define the robust cluster center as $\mu_e=\mathrm{median}\{\mathcal{G}(w_i^L)\mid i\in\mathbf{C}_e\}$, where the median is computed element-wise. The distance of each client update to the cluster center is then given by $\psi_i=\|\mathcal{G}(w_i^L)-\mu_e\|_2$. During aggregation, client updates can be filtered and reweighted according to these distances as
\begin{equation}
    \tilde{q}_i=q_i\cdot\mathbf{1}\{\psi_i\leq\tilde{\psi}_{e(i)}+\kappa\operatorname{MAD}_{e(i)}\},
\end{equation}
where $\tilde{\psi}_e=\mathrm{median}\{\psi_i\mid i\in\mathbf{C}_e\}$, $\mathrm{MAD}_e=\mathrm{median}\{|\psi_i-\tilde{\psi_e}|\mid i\in\mathbf{C}_e\}$. Finally, by combining \cref{equ:aggre}, the server can obtain the aggregated LoRA-FA model as
\begin{equation}\label{equ:aggre_weight}
    w_g^L = \sum_{i=1}^N\frac{\tilde{q_i}|D_i|w_i^L}{\sum_{i=1}^N\tilde{q_i}|D_i|}.
\end{equation}

Within each cluster, the LoRA-FA parameters are averaged to produce the expert $w_e^L$ by a robust trimmed mean strategy. The server transmits both the global LoRA-FA model and the corresponding expert.

\subsubsection{MoE-inspired Adaptive Optimization Loss}

Current adversarial training algorithms predominantly focus on designing loss functions that balance the relationship between benign data and adversarial samples, seeking an optimal trade-off between accuracy and robustness.
%
We propose a Mixture-of-Experts inspired method for personalized training. This approach adaptively optimizes the model for each client, tailoring the training process to unique data distributions and adversarial characteristics, enhancing adversarial training for each client's unique local data.

We address the class imbalance problem by utilizing the number of samples for each class. Specifically, on each client $i$, the class imbalance weight can be calculated based on its local dataset $\mathcal{D}_i$ as
\begin{equation}
    h_{i_c}^{Base}=\frac{1-\gamma}{1-\gamma^{n_{i_c}}},
\end{equation}
where $c$ represents a specific class, and $n_{i_c}$ denotes the number of samples belonging to class $c$ on client $i$. $\gamma$ is a hyperparameter, typically set in the range $[0.5,0.99]$. We assume there are $C$ total classes.
The method reduces weights for frequent classes while emphasizing underrepresented ones.

However, pre-trained models are primarily trained on benign data and may lack sufficient sensitivity to adversarial samples. Moreover, when client data distributions differ significantly from the pre-training data, directly applying the aforementioned weight scheme in adversarial fine-tuning can lead to convergence challenges.
To improve the model’s generalization in extracting adversarial features across classes, we introduce a dynamic smoothing weight scheme to stabilize the training process. The weight is defined as
\begin{equation}
    h_{i_c}^{Ada}(t)=\epsilon \cdot h_{i_c}^{Ada}(t-1)+(1-\epsilon)\cdot h_{i_c}^{Base}(t),
\end{equation}
where $\epsilon$ is a smoothing hyperparameter in the range $[0,1]$, and $t$ represents the local training epoch. The smoothed class weights are normalized as 
\begin{equation}
    h_{i_c}(t)=\frac{h_{i_c}^{Ada}(t)}{\sum_{c=1}^Ch_{i_c}^{Ada}(t)}.
\end{equation}
As local adversarial training progresses, the smoothed weights gradually approach the preset class weights. The corresponding adversarial training loss can be expressed as
\begin{equation}\label{equ:L_A}
    \mathcal{L}_{A}(w_{i},x^{adv},y)=\min\sum_{(x,y)\in\mathcal{D}_{i}}h_{i_y}\cdot\mathcal{L}_{CE}(\mathcal{F}(w_{i},x^{adv}),y),
\end{equation}
where $\mathcal{L}_{CE}(\cdot)$ denotes the cross-entropy loss to improve the robustness, $h_{i_y}$ represents the smoothed class weight for the class $y$ of the sample $x$ at the current training epoch.

Furthermore, it is crucial to ensure that adversarial training does not significantly compromise the model's accuracy on benign data, requiring the model to maintain sensitivity to benign inputs. 
This requirement is quantified by measuring the similarity between the model’s output vectors for benign and adversarial samples. To enforce it, we incorporate a KL-divergence loss term, defined as
\begin{equation}\label{equ:L_S}
    \begin{aligned}&\mathcal{L}_{S}(w_{i},x,x^{adv})=\mathcal{L}_{KL}(\mathcal{F}(w_{i},x),\mathcal{F}(w_{i},x^{adv}))\\&=\sum\mathrm{softmax}(\mathcal{F}(w_{i},x))\cdot\log\left(\frac{\mathrm{softmax}(\mathcal{F}(w_{i},x))}{\mathrm{softmax}(\mathcal{F}(w_{i},x^{adv}))}\right).\end{aligned}
\end{equation}

The optimization objective for the backbone module of pre-trained models is to enhance its generalization in extracting features from adversarial samples, while the classifier module focuses on achieving personalized classification performance. 
Inspired by the MoE framework, we propose a collaborative approach, guiding the backbone to shift toward clusters with similar distributions. Each client adjusts its model toward its local cluster, leveraging a fused reference model combining the global model $w_g^L$ and the expert model $w_{e,i}^L$:
\begin{equation}
    w_i^{ref} = \left(1-\eta\right)w_g^L + \eta w_{e,i}^L,
\end{equation}
where $\eta$ is a hyperparameter controlling the fusion ratio. The local training loss is then computed based on the distance between the client's parameters and the reference model:
\begin{equation}\label{equ:L_R}
    \mathcal{L}_R(w_i^L,w_i^{ref})=\left\|w_i^L-w_i^{ref}\right\|_2^2,
\end{equation}
Finally, by combining \cref{equ:L_A}, \cref{equ:L_S} and \cref{equ:L_R}, we weight the above loss terms to obtain the adversarial training loss as
\begin{equation}\label{equ:L}
    \mathcal{L}=\mathcal{L}_A+\lambda_1\mathcal{L}_S+\lambda_2\mathcal{L}_R.
\end{equation}
Each client performs local adversarial training using the objective in \cref{equ:L}, updating only its LoRA-FA and classifier modules. The LoRA-FA parameters are then periodically uploaded to the server for global aggregation.

\subsection{Convergence Analysis}

To theoretically establish the convergence upper bound of our method, we derive its convergence properties based on our adversarial training. Let $t \in {1,2,\dots,T_1}$ denote the index of global aggregation rounds at the server, and introduce an equivalent step size $\zeta > 0$, under which we can define as
\begin{equation}
    g_{i}^{t}\triangleq\frac{w_{g}^{L,t}-w_{i}^{L,t+1}}{\zeta},\quad\hat{g}^{t}\triangleq\sum_{i=1}^{N}\bar{q}_{i}^{t}g_{i}^{t}.
\end{equation}
The aggregation step can be expressed as $w_g^{L,t+1}=w_g^{L,t}-\zeta\left.\hat{g}^t\right.$. The local optimization can be defined by $\tilde{f}_i(w_i^L, w_i^C)$. Accordingly, the personalized LoRA training can be formulated as
\begin{equation}
    \Phi(w^L)\triangleq\frac{1}{N}\sum_{i=1}^N\min_{w^C}\tilde{f}_i(w_i^L,w_i^C).
\end{equation}
We make the following assumptions:

\textit{\textbf{Assumption 1.}} $\Phi$ is $\mathrm{L}$-smooth and lower bounded by $\Phi_{\mathrm{inf}}$, defined as
\begin{equation}
    \|\nabla\Phi(u)-\nabla\Phi(v)\|\leq \mathrm{L}\|u-v\|,\quad\Phi(w)\geq\Phi_{\inf}.
\end{equation}

\textit{\textbf{Assumption 2.}} We assume that the normalized weights and update directions contributed by honest clients are unbiased and have bounded variance, which can be expressed as:
\begin{equation}
    \bar{q}_{i,\mathbf{H}}^{t}\triangleq\frac{\tilde{q}_{i}^{t}|D_{i}|}{\sum_{j\in\mathbf{H}}\tilde{q}_{j}^{t}|D_{j}|},\quad g_{\mathbf{H}}^{t}\triangleq\sum_{i\in\mathbf{H}}\bar{q}_{i,\mathbf{H}}^{t}g_{i}^{t},
\end{equation}
where $\mathbb{E}[g_{\mathbf{H}}^t\mid w_g^{L,t}]=\nabla\Phi(w_g^{L,t})$, indicating unbiasedness, and $\mathbb{E}{\left[\|g_{\mathbf{H}}^t-\nabla\Phi(w_g^{L,t})\|^2\mid w_g^{L,t}\right]}\leq\pi^2$.

\textit{\textbf{Assumption 3.}} The filtered aggregation direction is unbiased with respect to the honest reference, i.e., $\mathbb{E}[\hat{g}^t\mid w_g^{L,t}]=\mathbb{E}[g_\mathbf{H}^t\mid w_g^{L,t}]$. The second-order residual moment has bounded as $\mathbb{E}\left[\|\hat{g}^t-g_\mathbf{H}^t\|^2\mid w_g^{L,t}\right]\leq b_t^2$.

We obtain the following theorem and its proof:

\textbf{\textit{Theorem 1.}} When $\zeta\leq1/\mathrm{L}$, we have
\begin{equation}
    \begin{aligned}
    \frac{1}{T_1}\sum_{t=0}^{T_1-1}\mathbb{E}&\|\nabla\Phi(w_g^{L,t})\|^2 \leq \frac{2\left(\Phi(w_g^{L,0})-\Phi_{\inf}\right)}{\zeta T_1} \\
    & + \mathrm{L}\zeta\left(\pi^2 + \frac{1}{T_1}\sum_{t=0}^{T_1-1}b_t^2\right).
    \end{aligned}
\end{equation}

\textbf{\textit{Proof 1.}}
By the descent lemma in Assumption 1, for any $t$, we have
\begin{equation}
    \begin{aligned}
        \Phi(w_g^{L,t+1})&\leq\langle\nabla\Phi(w_g^{L,t}),w_g^{L,t+1}-w_g^{L,t}\rangle+\\
        &\Phi(w_g^{L,t})+\frac{\mathrm{L}}{2}\|w_g^{L,t+1}-w_g^{L,t}\|^2.
    \end{aligned}
\end{equation}
By $w_g^{L,t+1}=w_g^{L,t}-\zeta\hat{g}^t$ we get
\begin{equation}
    \Phi(w_g^{L,t+1})\leq\Phi(w_g^{L,t})-\zeta\langle\nabla\Phi(w_g^{L,t}),\hat{g}^t\rangle+\frac{\mathrm{L}\zeta^2}{2}\|\hat{g}^t\|^2.
\end{equation}
Let $u^t=\hat{g}^t-\nabla\Phi(w_g^{L,t})$, we have $\hat{g}^t=\nabla\Phi(w_g^{L,t})+u^t$ and $\|\hat{g}^t\|^2=\|\nabla\Phi(w_g^{L,t})\|^2+2\langle\nabla\Phi(w_g^{L,t}),u^t\rangle+\|u^t\|^2$. Under Assumptions 2–3, we get $\mathbb{E}[u^t\mid w_g^{L,t}]=0$, and thus
\begin{equation}
    \mathbb{E}[\langle\nabla\Phi(w_g^{L,t}),\hat{g}^t\rangle\mid w_g^{L,t}]=\|\nabla\Phi(w_g^{L,t})\|^2,
\end{equation}
\begin{equation}
    \mathbb{E}[\|\hat{g}^t\|^2\mid w_g^{L,t}]=\|\nabla\Phi(w_g^{L,t})\|^2+\mathbb{E}[\|u^t\|^2\mid w_g^{L,t}].
\end{equation}
Taking expectation on the above expression and substituting it back into the preceding descent inequality yields
\begin{equation}
\begin{aligned}
    \mathbb{E}[\Phi(w_g^{L,t+1})]\leq\mathbb{E}[\Phi(w_g^{L,t})]-&\left(\zeta-\frac{\mathrm{L}\zeta^2}{2}\right)\mathbb{E}\|\nabla\Phi(w_g^{L,t})\|^2\\
    &+\frac{\mathrm{L}\zeta^2}{2}\mathbb{E}\|u^t\|^2
\end{aligned}
\end{equation}
Furthermore, since $u^t=(\hat{g}^t-g_{\mathbf{H}}^t)+(g_{\mathbf{H}}^t-\nabla\Phi(w_g^{L,t}))$, and using the inequality $\|a+b\|^2\leq2\|a\|^2+2\|b\|^2$, Assumptions 2-3 imply $\mathbb{E}\|u^t\|^2\leq2b_t^2+2\pi^2$.
Finally, choosing $\zeta\leq1/\mathrm{L}$ yeilds $\zeta-\frac{\mathrm{L}\zeta^2}{2}\geq\frac{\zeta}{2}$. Summing over $t=0,\ldots,T_1-1$ and applying a telescoping sum together with $\Phi(w)\geq\Phi_{\inf}$ completes the proof of the theorem.

\subsection{Forward-gating Selection Fine-tuning}

After Phase 1, each client fuses the global LoRA-FA with the pre-trained model to obtain a robustly generalized backbone. Adding a personalized classifier enables effective adversarial defense. However, we believe that benign accuracy can be further improved without sacrificing robustness. 
%


In Phase 2, each client independently trains using its own data in parallel, further fine-tuning its personalized model.
Our goal is to select $p$ layers from the $L$ layers of the robust model for training on benign data, while freezing the remaining layers, to maximize accuracy improvements and minimize robustness degradation. 
Previous studies jointly compute the robustness and accuracy losses for each layer to determine the corresponding layer selection~\cite{qi2025sylva, zhou2024securely}, but this method involves high computational complexity.

As shown in \cref{fig:Method}, for each client $i$, we first fuse the locally trained weights $w_i^L$ obtained from Phase 1 with its backbone model. After attaching a personalized classifier, we obtain the fine-tuned model $w_i^{{fus}}$. Subsequently, we introduce a forward gating network $Z_i=\{z_{i,r}\}$ into the bypass of the forward propagation, where each gate is controlled by a sigmoid function and corresponds to one layer $r$ of the network. The forward propagation of each layer can then be formulated as
\begin{equation}\label{equ:w_eff}
    w_{i,r}^{eff} = w_{i,r}^{{fus}}+z_{i,r}\Delta w_{i,r},
\end{equation}
where $\Delta w_{i,r}$ denotes the parameter updates during training. When $z_{i,r}=0$, the parameters of this layer are frozen and remain identical to those obtained after Phase 1. Conversely, when $z_{i,r}=1$, the layer is unfrozen and participates in training.

During model optimization, we adopt a two-stage alternating update scheme consisting of an inner and an outer loop.
In the inner loop, the gating network $Z_i$ is frozen while only the fused model $w_i^{{fus}}$ is updated using a small amount of data. This stage requires only a few update steps, allowing the gradients to be accumulated and stored within the gating network.
Subsequently, in the outer loop, the model is fixed and the gating network is updated. The optimization objective at this stage is defined as
\begin{equation}\label{equ:j}
    J=\mathcal{L}_{acc}(w_i^{{eff}})-\beta\mathcal{L}_{rob}(w_i^{{eff}})+\lambda_3{\left[\sum_r z_{i,r}-B\right]}_+,
\end{equation}
where $B$ denotes the total training budget, and $[\cdot]_{+}$ represents a penalty term that is activated only when the budget is exceeded and remains zero otherwise.
This objective encourages the gating network to identify a subset of layers that reduce robustness loss but increase accuracy loss, selectively opening them for training while keeping the total number of active layers within the budget.
After each outer-loop update, the model parameters are reset to $w_i^{fus}$ and further trained for multiple iterations.

Through the alternating optimization, the gating network gradually converges to the target $J$. 
In the final outer-loop iteration, a top-$B$ projection is applied to retain the $B$ largest gates, forming the selected layer set $\mathbf{R}=\{r_1,r_2,\dots,r_B\}$, where the corresponding layers are opened for training while the others remain frozen. The model is then retrained on clean data for final fine-tuning.

\subsection{The Workflow of \textit{Lorica}}

\begin{algorithm}[!tb]
    \caption{\textit{Lorica}'s Adversarial Training Process}\label{algorithm}
    \begin{algorithmic}[1]\label{alg:1}
        \STATE \textbf{Input}: Aggregation epoch $T_1$, local training epoch $T_2$, local benign data $\mathcal{D}_i$, global model $w_g(w_g^L, w_g^P, w_g^C)$, and local model $w_i(w_i^L, w_i^P, w_i^C)$, where $i\in \{1,2,\dots, N\}$.
        \STATE \textbf{Output}: Personalized, robust, high-accuracy client models.
        \STATE \textbf{Initialize}: Server's model $w_g$ and clients' models $w_i$.
        \FOR{$1$ \TO $T_1$ } 
            \STATE \textbf{for} client $i\in \{1,2,\dots,N\}$ \textbf{parallel do} 
            \INDSTATE Downloads the global LoRA-FA as $w_i^L$ and expert model $w_{e,i}^L$. Combine $w_i^L$, $w_i^P$ and $w_i^C$ to form $w_i$;
            \INDSTATE \textbf{for} 1 \textbf{to} $T_2$ \textbf{do}
                \INDSTATE[2] \textbf{for} $x\in \mathcal{D}_i$ \textbf{do}
                    \INDSTATE[3] $x_{adv} = \text{PGD}(w_i, x)$;
                    \INDSTATE[3] Train and calculate loss $\mathcal{L}$ with \cref{equ:L};
                    \INDSTATE[3] $w_i^L, w_i^C \gets \text{SGD}(w_i, \mathcal{L})$;
                \INDSTATE[2] \textbf{end for}
            \INDSTATE \textbf{end for}
            \INDSTATE Each client upload LoRA-FA model $w_i^L$ to server;
            \STATE \textbf{end for}
            \STATE Server performs Byzantine-aware aggregate LoRA-FA model with \cref{equ:aggre_weight};
        \ENDFOR
        \STATE \textbf{for} client $i\in \{1,2,\dots,N\}$ \textbf{parallel do} 
            \INDSTATE Fuse $w_i^L$, $w_i^P$, and $w_i^C$ to form the model $w_i^{fus}$, initialize forward-gating network $Z_i$;
            \INDSTATE \textbf{for} 1 \textbf{to} $T_3$ \textbf{do}
                \INDSTATE[2] \textbf{for} 1 \textbf{to} $T_4$ \textbf{do}
                    \INDSTATE[3] Freeze $Z_i$ and train the model by \cref{equ:w_eff};
                \INDSTATE[2] \textbf{end for}
                \INDSTATE[2] Freeze the model and optimize $Z_i$ by \cref{equ:j};
            \INDSTATE \textbf{end for}
            \INDSTATE Select top-$B$ largest gate as $\mathbf{R}=\{r_1,r_2,\dots,r_B\}$;
            \INDSTATE Freeze layers except $\mathbf{P}$ and train model $w_i^{fus}$ by $\mathcal{D}_i$;
        \STATE \textbf{end for}
    \end{algorithmic}
\end{algorithm}

\cref{alg:1} outlines the \textit{Lorica} defense algorithm. Each client $i$ holds its private benign dataset $\mathcal{D}_i$. A global model $w_g$ and a local model $w_i$, both utilizing LoRA-FA for fine-tuning, are constructed and deployed.


In Phase 1, personalized adversarial fine-tuning is conducted as follows. 
The server initializes the global model $w_g$, and clients download the LoRA-FA $w_g^L$ and expert model $w_{e,i}^L$ to build their local models $w_i$. 
Each client generates adversarial examples from its local dataset $\mathcal{D}_i$ using the PGD method and trains locally for $T_2$ epochs with \cref{equ:L}, updating only its LoRA-FA $w_i^L$ and classifier $w_i^C$. 
After local training, clients upload $w_i^L$ to the server, which aggregates them via \cref{equ:aggre_weight} to update the global LoRA-FA $w_g^L$ and expert model $w_{e,i}^L$. 
This process repeats for $T_1$ rounds.

After completing Phase 1, each client possesses the shared global LoRA-FA model $w_i^L$, the frozen pre-trained backbone $w_i^P$, and a personalized classifier $w_i^C$. 
Phase 2 involves forward-gating selection fine-tuning. 
First, the model is fused and augmented with a bypass gating network, yielding $w_i^{\mathrm{fus}}$ and $Z_i$. 
Then, the gating network is iteratively optimized according to \cref{equ:w_eff} and \cref{equ:j}, enabling layer-wise selection. 
Finally, the top-$B$ largest gates are selected as $\mathbf{P}$. All other layers are frozen. The fused model $w_i^{\mathrm{fus}}$ is further trained.

Through this workflow, each client obtains a personalized model suited to its data distribution, achieving high benign accuracy while maintaining strong adversarial robustness.

\begin{table}[!tb]
\centering
\caption{Performance comparison of \textit{Lorica} and baseline under different datasets and attack algorithms (ViT-B/16)}\label{tab:result_vit_b}
\scalebox{0.82}{
\renewcommand{\arraystretch}{0.8}  
\begin{tabular}{@{}c|c|ccccc@{}}
\toprule
Datasets                   & Baselines            & Benign(BA) & FGSM  & PGD   & SparseFool & PAP   \\ \midrule
\multirow{11}{*}{CIFAR-10}  & FA-PGD-AT   & 39.25      & 45.89 & 42.50 & 48.02      & 44.12 \\
                           & FA-TRADES   & 50.12      & 45.04 & 43.39 & 45.71      & 43.56 \\
                           & FA-Gen-AF   & 53.42      & 46.95 & 45.73 & 45.53      & 45.39 \\
                           & FP-PGD-AT   & 41.04      & 45.28 & 43.63 & 48.51      & 45.43 \\
                           & FP-TRADES   & 52.85      & 47.45 & 45.42 & 46.02      & 44.46 \\
                           & FP-Gen-AF   & 54.70      & 48.52 & 46.21 & 49.88      & 47.12 \\
                           & DBFAT       & 53.56      & 45.63 & 44.94 & 48.93      & 48.41 \\
                           & Per-Adv     & 48.28      & 44.52 & 42.61 & 44.31      & 44.40 \\
                           & Per-LoRA    & 53.02      & 47.21 & 44.32 & 46.71      & 44.35 \\
                           & Sylva       & 59.03      & 52.88 & 55.03 & 52.68      & 51.89 \\
                           & \cellcolor[HTML]{DCDCDC}Lorica(Ours) & \cellcolor[HTML]{DCDCDC}\textbf{60.83}      & \cellcolor[HTML]{DCDCDC}\textbf{54.26} & \cellcolor[HTML]{DCDCDC}\textbf{57.52} & \cellcolor[HTML]{DCDCDC}\textbf{54.05}      & \cellcolor[HTML]{DCDCDC}\textbf{53.18} \\ \midrule
\multirow{11}{*}{STL-10}    & FA-PGD-AT   & 34.53      & 39.85 & 38.61 & 40.26      & 39.88 \\
                           & FA-TRADES   & 42.65      & 37.65 & 36.50 & 39.80      & 39.20 \\
                           & FA-Gen-AF   & 44.72      & 38.57 & 37.18 & 39.65      & 40.15 \\
                           & FP-PGD-AT   & 36.48      & 38.58 & 37.21 & 41.32      & 39.74 \\
                           & FP-TRADES   & 43.68      & 39.10 & 37.21 & 38.92      & 38.85 \\
                           & FP-Gen-AF   & 45.83      & 41.01 & 38.53 & 41.83      & 41.23 \\
                           & DBFAT       & 45.55      & 41.05 & 38.01 & 41.60      & 40.75 \\
                           & Per-Adv     & 36.71      & 40.20 & 38.87 & 39.93      & 39.67 \\
                           & Per-LoRA    & 44.37      & 40.84 & 38.11 & 41.09      & 40.66 \\
                           & Sylva       & 49.11      & 43.78 & 41.48 & 44.05      & 44.20 \\
                           & \cellcolor[HTML]{DCDCDC}Lorica(Ours) & \cellcolor[HTML]{DCDCDC}\textbf{52.13}     & \cellcolor[HTML]{DCDCDC}\textbf{44.25} & \cellcolor[HTML]{DCDCDC}\textbf{44.07} & \cellcolor[HTML]{DCDCDC}\textbf{45.46}      & \cellcolor[HTML]{DCDCDC}\textbf{44.91} \\ \midrule
\multirow{11}{*}{GTSRB}     & FA-PGD-AT   & 60.17      & 68.34 & 68.98 & 71.67      & 71.52 \\
                           & FA-TRADES   & 74.38      & 67.88 & 68.68 & 70.65      & 72.33 \\
                           & FA-Gen-AF   & 76.45      & 68.91 & 67.89 & 71.72      & 71.48 \\
                           & FP-PGD-AT   & 62.53      & 68.85 & 68.92 & 71.93      & 71.83 \\
                           & FP-TRADES   & 76.43      & 69.37 & 69.85 & 70.88      & 71.49 \\
                           & FP-Gen-AF   & 79.55      & 69.51 & 69.90 & 71.56      & 73.15 \\
                           & DBFAT       & 79.40      & 68.58 & 68.72 & 70.70      & 73.45 \\
                           & Per-Adv     & 63.70      & 68.03 & 67.83 & 69.68      & 70.09 \\
                           & Per-LoRA    & 74.84      & 68.70 & 68.34 & 71.01      & 70.78 \\
                           & Sylva       & 80.49      & 72.63 & 72.40 & 73.78      & 74.30 \\
                           & \cellcolor[HTML]{DCDCDC}Lorica(Ours) & \cellcolor[HTML]{DCDCDC}\textbf{81.54}     & \cellcolor[HTML]{DCDCDC}\textbf{74.05} & \cellcolor[HTML]{DCDCDC}\textbf{73.92} & \cellcolor[HTML]{DCDCDC}\textbf{74.64}     & \cellcolor[HTML]{DCDCDC}\textbf{74.80} \\ \midrule
\multirow{11}{*}{CIFAR-100} & FA-PGD-AT   & 12.45      & 25.82 & 25.27 & 27.36      & 26.59 \\
                           & FA-TRADES   & 24.08      & 24.88 & 23.35 & 25.20      & 25.44 \\
                           & FA-Gen-AF   & 23.65      & 25.33 & 24.20 & 26.16      & 25.47 \\
                           & FP-PGD-AT   & 13.71      & 25.70 & 25.25 & 28.00      & 26.79 \\
                           & FP-TRADES   & 24.61      & 23.83 & 25.32 & 26.67      & 25.69 \\
                           & FP-Gen-AF   & 25.07      & 25.45 & 25.34 & 27.46      & 26.63 \\
                           & DBFAT       & 25.02      & 25.11 & 25.29 & 27.11      & 28.33 \\
                           & Per-Adv     & 15.31      & 24.37 & 24.47 & 25.02      & 26.80 \\
                           & Per-LoRA    & 24.85      & 25.38 & 24.92 & 27.13      & 27.06 \\
                           & Sylva       & 26.95      & 27.63 & 26.23 & 28.95      & 28.70 \\
                           & \cellcolor[HTML]{DCDCDC}Lorica(Ours) & \cellcolor[HTML]{DCDCDC}\textbf{27.93}     & \cellcolor[HTML]{DCDCDC}\textbf{28.45} & \cellcolor[HTML]{DCDCDC}\textbf{27.77} & \cellcolor[HTML]{DCDCDC}\textbf{29.43}      & \cellcolor[HTML]{DCDCDC}\textbf{29.15}\\ \bottomrule
\end{tabular}
}
\vspace{-4mm}
\end{table}

\section{Experiments}\label{sec4}

We evaluate \textit{Lorica} across diverse distributed settings and attack scenarios, comparing it with baselines to assess personalized defense performance and adaptability.
\textbf{RQ1}: How does \textit{Lorica} perform against different attack algorithms in terms of robustness and accuracy?  
\textbf{RQ2}: How does \textit{Lorica} handle varying heterogeneous data distributions across diverse clients?
\textbf{RQ3}: How does \textit{Lorica} achieve efficiency in communication and computation in distributed environments?
\textbf{RQ4}: How effective is \textit{Lorica} in defending against Byzantine attacks during adversarial training?
\textbf{RQ5}: How do \textit{Lorica}'s hyperparameters influence its overall robustness and accuracy performance?

\subsection{Experimental Setup}

We provide a brief overview of the experimental setup, with detailed specifics available in Appendix A.

\textbf{Datasets and models}. This study leverages datasets widely recognized in adversarial training, including CIFAR-10~\cite{krizhevsky2009learning}, STL-10~\cite{coates2011analysis}, and GTSRB~\cite{stallkamp2011german}, as introduced in \cref{sec2}. To better align with tasks involving large models, the more complex CIFAR-100 dataset is also incorporated. For model selection, we utilize large vision models based on transformers, specifically ViTs~\cite{dosovitskiy2020image} pre-trained on the ImageNet dataset~\cite{russakovsky2015imagenet}. These models are categorized into three scales—ViT-T/16, ViT-B/16, and ViT-L/16.

\begin{figure}[!tb]
\centering
\subfloat[CIFAR-10]{\includegraphics[width=1.65in]{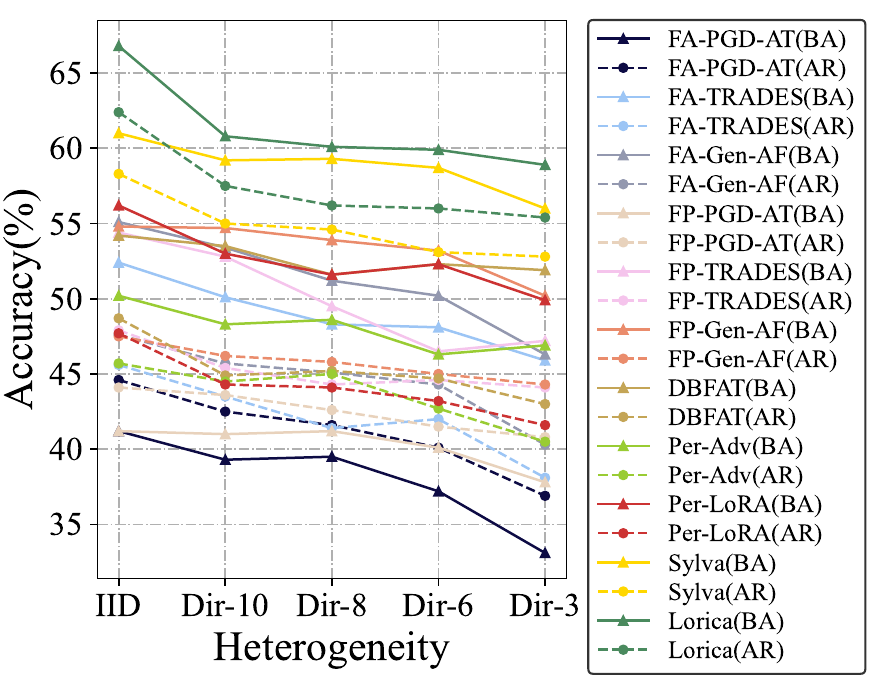}%
\label{fig:noniid_cifar-10_vit-b}}
\subfloat[STL-10]{\includegraphics[width=1.65in]{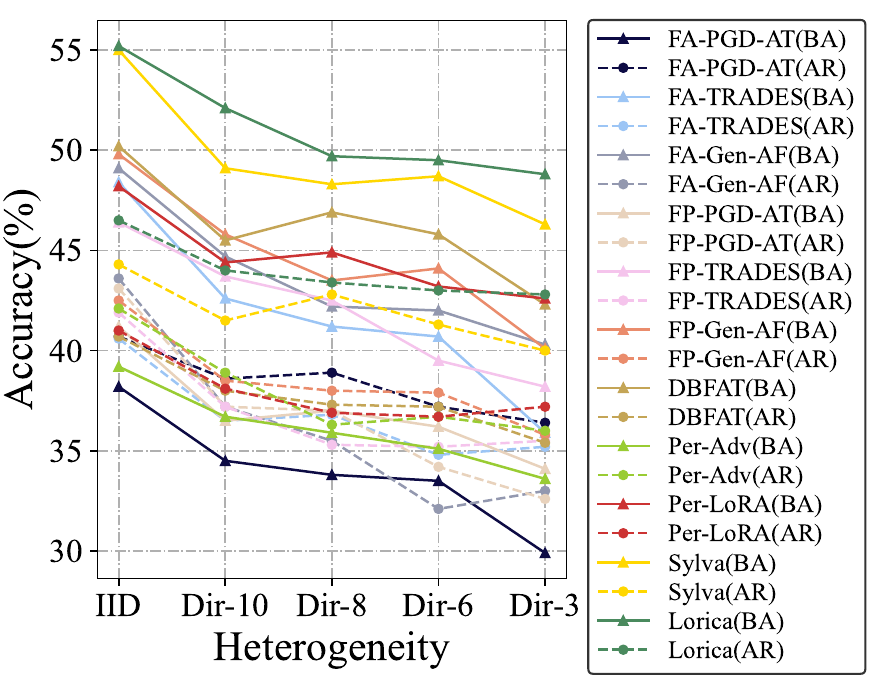}%
\label{fig:noniid_stl-10_vit-b}}

\vspace{-3mm} 
\subfloat[GTSRB]{\includegraphics[width=1.65in]{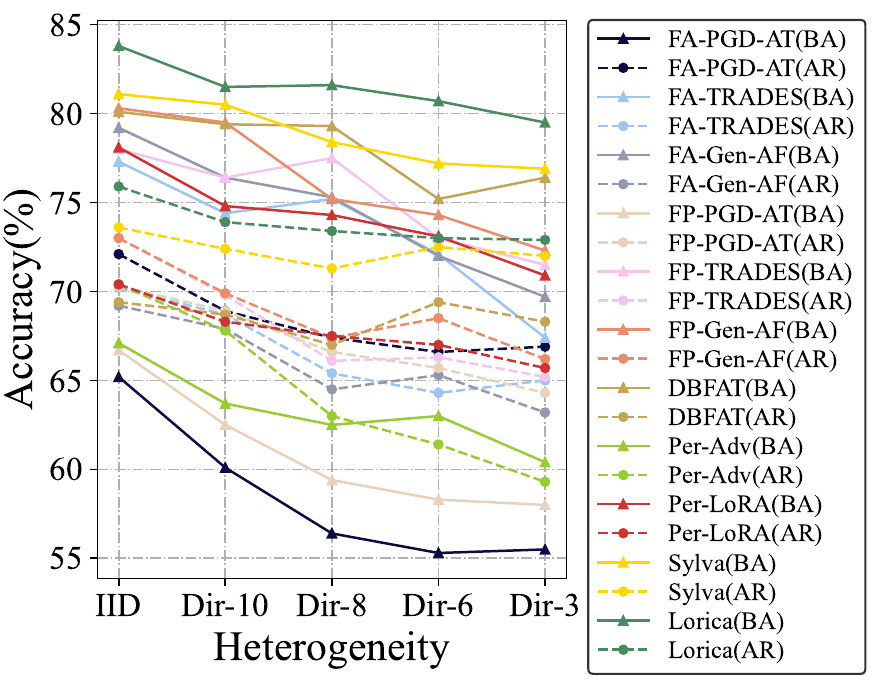}%
\label{fig:noniid_gtsrb_vit-b}}
\subfloat[CIFAR-100]{\includegraphics[width=1.65in]{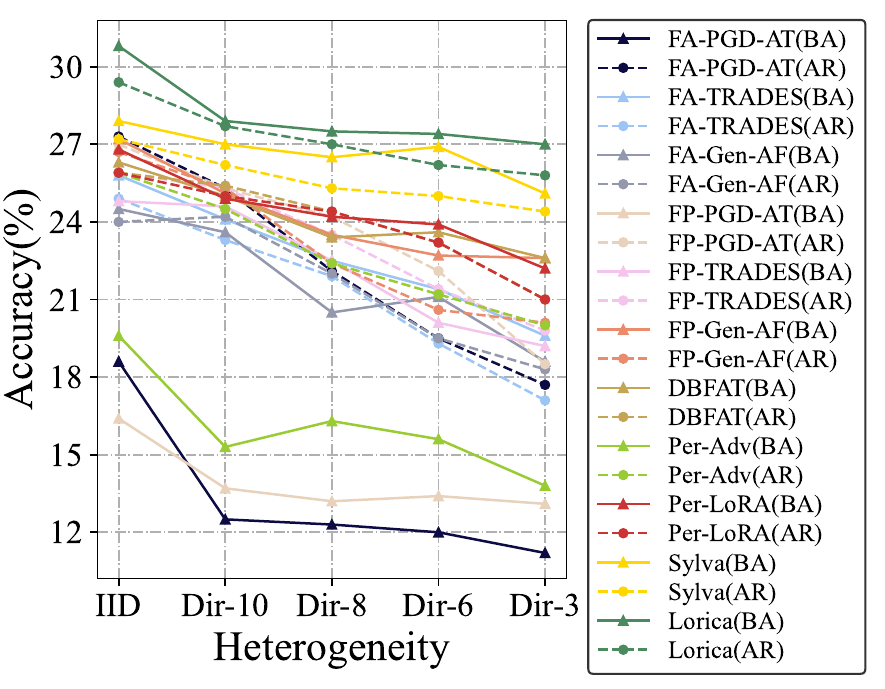}%
\label{fig:noniid_cifar-100_vit-b}}
\caption{Comparison of adversarial training performance under different heterogeneity levels (ViT-B/16)}
\label{fig:noniid_vit-b}
\vspace{-6mm}
\end{figure}

\textbf{Attacks}. We simulate both white-box and gray-box attacks using recent and widely adopted image attack algorithms. For white-box attacks, we employ the classic FGSM algorithm~\cite{goodfellow2014explaining} and its variants, alongside the more robust PGD algorithm~\cite{madry2017towards}. To explore subtle perturbations, we include the SparseFool attack~\cite{modas2019sparsefool}, which modifies only a few pixels. For gray-box attacks, we utilize the PAP algorithm~\cite{ban2022pre}, which generates universal perturbations tailored for pre-trained models and remains effective against fine-tuned models without requiring knowledge of downstream tasks.
We consider Byzantine attackers with partial client visibility rather than global knowledge, including label-flipping attacks~\cite{rosenfeld2020certified} at the data level and model poisoning using the MPAF~\cite{cao2022mpaf}.

\textbf{Baselines}. We use several standard defense algorithms for adversarial training: PGD-AT~\cite{madry2017towards}, which employs the PGD algorithm to generate adversarial samples and enhance model robustness; TRADES~\cite{zhang2019theoretically}, which optimizes the loss to balance accuracy on benign data and robustness against adversarial attacks; and Gen-AF~\cite{zhou2024securely}, a fine-tuning approach for adversarial training on pre-trained models, maintaining robustness while improving accuracy.
In the distributed scenario, we employ FedAvg~\cite{mcmahan2017communication} and FedProx~\cite{li2020federated}, with FedProx addressing client data heterogeneity. We integrate them with various defense algorithms, establishing them as baseline approaches for comparison. Additionally, we incorporate DBFAT~\cite{zhang2023delving}, designed specifically for non-IID settings, into our baseline approaches.
We also include two personalized federated learning algorithms for comparison. \cite{allouah2024fine} proposes addressing Byzantine attacks in distributed training by customizing the loss function for each client, which we refer to as Per-Adv. \cite{wagner2024personalized} introduces a method that computes LoRA-based trust weights among clients during large model fine-tuning to achieve personalized updates, which we denote as Per-LoRA.
It is worth mentioning that we incorporate the TRADES loss into both personalized algorithms, thereby improving their alignment with the threat model and experimental tasks defined in this study.
For Byzantine robustness, we further compare against widely adopted aggregation rules, including Median and Trimmed Mean~\cite{yin2018byzantine}.
In addition, as expected, we also compare our method with our previous work Sylva~\cite{qi2025sylva}.

\textbf{Implementation}. For data heterogeneity, client data is partitioned using a Dirichlet distribution with 15 clients and a Dirichlet parameter of 10. In Phase 1, $\gamma$ and $\epsilon$ are set to 0.9, $\lambda_1$ to 20, $\lambda_2$ to 0.001, and $k$, the number of clients for weight aggregation, is set to 5. We set the malicious client ratio $\rho$ to 5–15\% only when evaluating Byzantine attacks.
In Phase 2, $\beta$ is set to 5, $B$ to 300, and $p$, the proportion of layers selected for training, to 3\% of the total layers. The learning rate is 0.005. FAT uses 5 local epochs, with ViT-T performing 30 cloud aggregation rounds, while ViT-B and ViT-L performing 50. Experiments are conducted on 4 NVIDIA A100 GPUs using the Ray framework for multi-process client simulation.
%
To evaluate \textit{Lorica}'s efficiency in real-world settings, we conduct experiments on five representative devices, including three consumer GPUs (RTX 4090, 3090, and 2080Ti) and two edge platforms for autonomous driving: the Apollo D-KIT Advanced (RTX 3060) and Jetson AGX Orin.

\subsection{Performance of Adversarial Training}

We perform adversarial training in a distributed scenario with 15 clients, following default settings. For each client, we evaluate performance by calculating the average accuracy on benign data (BA) and adversarial robustness (AR) across all clients under various attack algorithms. As baselines, we integrate local adversarial training with FedAvg (FA) and FedProx (FP). 

We evaluate on four datasets, reporting average performance across all clients (\cref{tab:result_vit_b}). In heterogeneous settings, non-IID-aware baselines like FedProx variants and DBFAT show relatively better performance.
\textit{Lorica} consistently outperforms across diverse datasets, excelling on both benign and adversarial samples. On datasets like CIFAR-10, it shows substantial gains over both traditional federated adversarial training and personalized methods.
Specifically, \textit{Lorica} leads to an accuracy increase of up to 55.0\% and a robustness enhancement of 35.3\%, demonstrating its superior capability in enhancing both performance and resilience against adversarial attacks.
When compared to the baselines of the two personalized algorithms, \textit{Lorica} continues to deliver significant improvements, achieving an average increase of 50.2\% in benign accuracy and 18.7\% in robustness. This performance boost can be attributed to \textit{Lorica}'s greater adaptability, as it effectively balances generalization across tasks with maximizing personalization for individual clients, thereby enhancing both accuracy and robustness.
Moreover, benefiting from the optimizations of the \textit{Lorica} modules, both accuracy and robustness show consistent improvements over \textit{Sylva}.
While the improvements on more complex datasets like CIFAR-100 are comparatively smaller, they remain significant, highlighting \textit{Lorica}'s adaptability and effectiveness in varied scenarios.
Furthermore, we evaluate \textit{Lorica} on models of varying scales, with detailed results provided in Appendix B. \textit{Lorica} consistently delivers strong performance across all model scales. 

\answer{1}{By leveraging its personalized design, \textit{Lorica} enables each client's model to effectively adapt to its specific data distribution, achieving superior accuracy and robustness even in highly heterogeneous environments.}

\begin{table*}[]
\caption{Efficiency comparison of \textit{Lorica} on different edge devices}\label{tab:result_efficency}
\renewcommand{\arraystretch}{0.88}
\scalebox{0.83}{
\begin{tabular}{@{}cc|c|ccccc|ccccc@{}}
\toprule
 &
   &
  \multirow{2}{*}{\begin{tabular}[c]{@{}c@{}}Mem$\downarrow$\\ (G)\end{tabular}} &
  \multicolumn{5}{c|}{Time$\downarrow $ ($\times 10^3$s)} &
  \multicolumn{5}{c}{Com$\downarrow $ (s)} \\ \cmidrule(l){4-13} 
                                            &          &      & RTX 4090 & RTX 3090 & RTX 2080-Ti & RTX 3060 & AGX Orin & RTX 4090 & RTX 3090 & RTX 2080-Ti & RTX 3060 & AGX Orin \\ \midrule
\multicolumn{1}{c|}{\multirow{4}{*}{ViT-T}} & FAT      & 1.04  & 0.35 & 0.45 & 0.47    & 0.49 & 0.99 & 0.7  & 1.3  & 1.5    & 1.0 & 1.2 \\
\multicolumn{1}{c|}{}                       & Per-LoRA & 0.89  & 0.31 & 0.32 & 0.33    & 0.32 & 0.82 & 0.9  & 0.7  & 1.1    & 1.1 & 1.0 \\
\multicolumn{1}{c|}{}                       & Sylva    & 0.89  & 0.30 & 0.32 & 0.33    & 0.32 & 0.82 & 0.8  & 0.6  & 1.2    & 0.9 & 0.9 \\
\multicolumn{1}{c|}{}                       & \cellcolor[HTML]{DCDCDC}Lorica    & \cellcolor[HTML]{DCDCDC}\textbf{0.63}  & \cellcolor[HTML]{DCDCDC}\textbf{0.28} & \cellcolor[HTML]{DCDCDC}\textbf{0.30} & \cellcolor[HTML]{DCDCDC}\textbf{0.31}    & \cellcolor[HTML]{DCDCDC}\textbf{0.31} & \cellcolor[HTML]{DCDCDC}\textbf{0.77} & \cellcolor[HTML]{DCDCDC}\textbf{0.7}  & \cellcolor[HTML]{DCDCDC}\textbf{0.5}  & \cellcolor[HTML]{DCDCDC}\textbf{1.1}    & \cellcolor[HTML]{DCDCDC}\textbf{0.7} & \cellcolor[HTML]{DCDCDC}\textbf{0.8} \\ \midrule
\multicolumn{1}{c|}{\multirow{4}{*}{ViT-B}} & FAT      & 4.86  & 0.40 & 0.47 & 0.49    & 0.53 & 1.13 & 15.0 & 13.5 & 13.8   & 14.1 & 13.8 \\
\multicolumn{1}{c|}{}                       & Per-LoRA & 3.14  & 0.34 & 0.39 & 0.42    & 0.45 & 0.92 & 1.3  & 1.4  & 1.4    & 1.2 & 1.5 \\
\multicolumn{1}{c|}{}                       & Sylva    & 3.14  & 0.33 & 0.37 & 0.41    & 0.45 & 0.91 & 0.8  & 0.9  & 0.8    & 0.7 & 0.8 \\
\multicolumn{1}{c|}{}                       & \cellcolor[HTML]{DCDCDC}Lorica    & \cellcolor[HTML]{DCDCDC}\textbf{2.73}  & \cellcolor[HTML]{DCDCDC}\textbf{0.29} & \cellcolor[HTML]{DCDCDC}\textbf{0.34} & \cellcolor[HTML]{DCDCDC}\textbf{0.37}    & \cellcolor[HTML]{DCDCDC}\textbf{0.39} & \cellcolor[HTML]{DCDCDC}\textbf{0.80} & \cellcolor[HTML]{DCDCDC}\textbf{0.8}  & \cellcolor[HTML]{DCDCDC}\textbf{0.7}  & \cellcolor[HTML]{DCDCDC}\textbf{0.8}    & \cellcolor[HTML]{DCDCDC}\textbf{0.6} & \cellcolor[HTML]{DCDCDC}\textbf{0.7} \\ \midrule
\multicolumn{1}{c|}{\multirow{4}{*}{ViT-L}} & FAT      & 11.43 & 0.75 & 0.93 & 1.78    & 1.87 & 2.41 & 49.9 & 47.6 & 49.1   & 48.7 & 48.2 \\
\multicolumn{1}{c|}{}                       & Per-LoRA & 6.63  & 0.57 & 0.88 & 1.29    & 1.39 & 1.71 & 3.1  & 2.9  & 3.2    & 3.4 & 3.0 \\
\multicolumn{1}{c|}{}                       & Sylva    & 6.63  & 0.55 & 0.87 & 1.28    & 1.39 & 1.70 & 1.1  & 1.0  & 1.2    & 1.2 & 1.1 \\
\multicolumn{1}{c|}{}                       & \cellcolor[HTML]{DCDCDC}Lorica    & \cellcolor[HTML]{DCDCDC}\textbf{5.37}  & \cellcolor[HTML]{DCDCDC}\textbf{0.43} & \cellcolor[HTML]{DCDCDC}\textbf{0.65} & \cellcolor[HTML]{DCDCDC}\textbf{0.92}    & \cellcolor[HTML]{DCDCDC}\textbf{1.01} & \cellcolor[HTML]{DCDCDC}\textbf{1.38} & \cellcolor[HTML]{DCDCDC}\textbf{0.8}  & \cellcolor[HTML]{DCDCDC}\textbf{0.7}  & \cellcolor[HTML]{DCDCDC}\textbf{0.8}    & \cellcolor[HTML]{DCDCDC}\textbf{0.9} & \cellcolor[HTML]{DCDCDC}\textbf{0.9} \\ \bottomrule
\end{tabular}
}
\vspace{-2mm}
\end{table*}

\subsection{Performance of Different Heterogeneous Distributions}\label{Sec4:RQ2}

To evaluate the impact of varying levels of data heterogeneity on adversarial defense performance, we conduct experiments in five distinct scenarios: one with independent identical (IID) distribution and four with increasing heterogeneity levels determined by Dirichlet distribution parameters. Lower Dirichlet parameters indicate greater data heterogeneity, as detailed in Appendix A.4. For the attack algorithms, we use PGD and measure the benign accuracy and adversarial robustness metrics. The performance of \textit{Lorica} and baseline models is assessed under these conditions across four datasets, with the results summarized in \cref{fig:noniid_vit-b}.

The results clearly show that as data heterogeneity increases, the overall performance of all algorithms declines, underscoring the significant influence of heterogeneity on training outcomes. Among the baselines, FedProx-based algorithms consistently outperform FedAvg-based ones, benefiting from their regularization mechanism that better accommodates heterogeneous data distributions. 
Although the two personalized baselines show good performance, they do not fully exploit the generalization potential of the model's backbone, nor are their loss functions adequately aligned with the threat model for adversarial training. Consequently, they still fall behind \textit{Lorica} in terms of effectiveness.
Notably, \textit{Lorica} consistently outperforms other methods and exhibits greater resilience to increasing heterogeneity.
As heterogeneity increases, \textit{Lorica} outperforms \textit{Sylva} more significantly. This improvement arises from \textit{Lorica}’s module design, which emphasizes personalization and thus excels in highly heterogeneous settings.
%
%
All experiments above are conducted using the ViT-B architecture, and results for other model scales are provided in Appendix C.

\answer{2}{\textit{Lorica}'s personalized framework enhances robustness by minimizing data heterogeneity, ensuring strong performance across varying heterogeneous environments.}

\begin{figure}[!tb]
\centering
\subfloat[Label-flipping]{\includegraphics[width=1.65in]{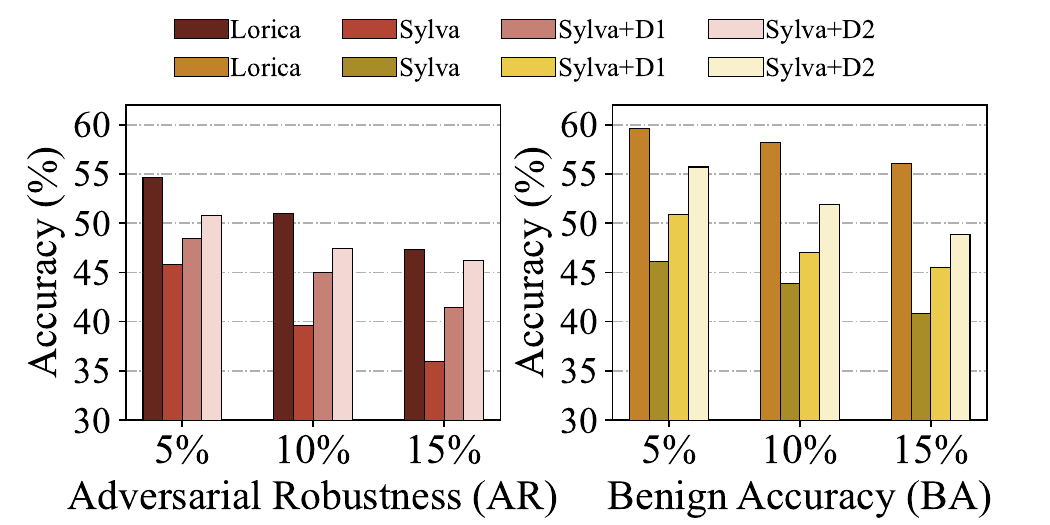}%
\label{fig:byzantine_1_cifar10}}
\subfloat[MPAF]{\includegraphics[width=1.65in]{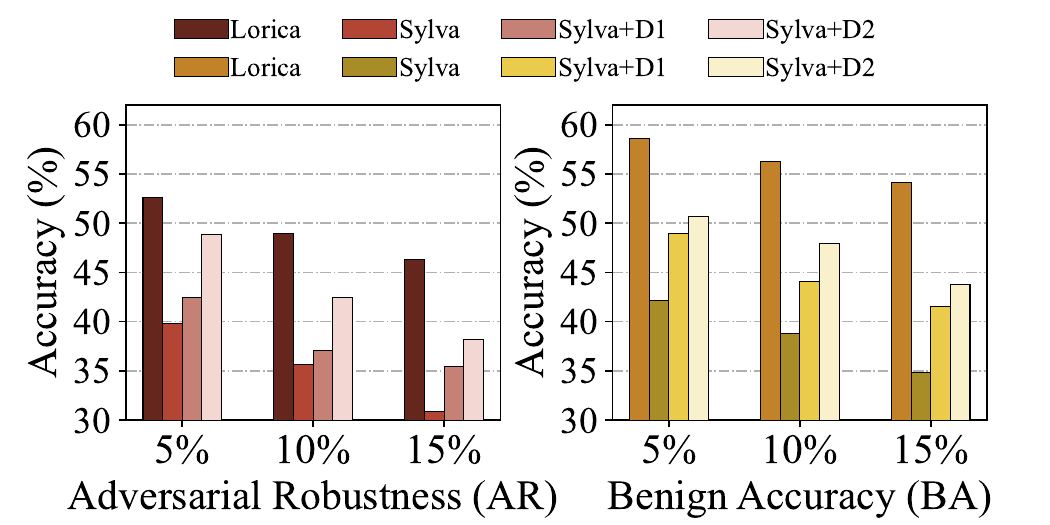}%
\label{fig:byzantine_2_cifar10}}
\caption{Defense performance against Byzantine attacks on CIFAR-10}
\label{fig:byzantine_cifar10}
\vspace{-5mm}
\end{figure}

\begin{figure}[!tb]
\centering
\subfloat[Label-flipping]{\includegraphics[width=1.65in]{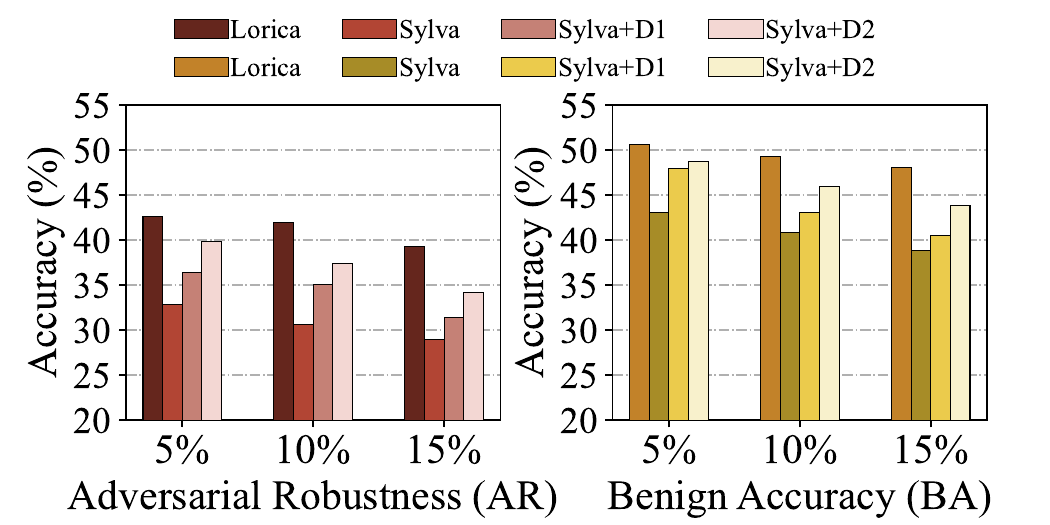}%
\label{fig:byzantine_1_stl10}}
\subfloat[MPAF]{\includegraphics[width=1.65in]{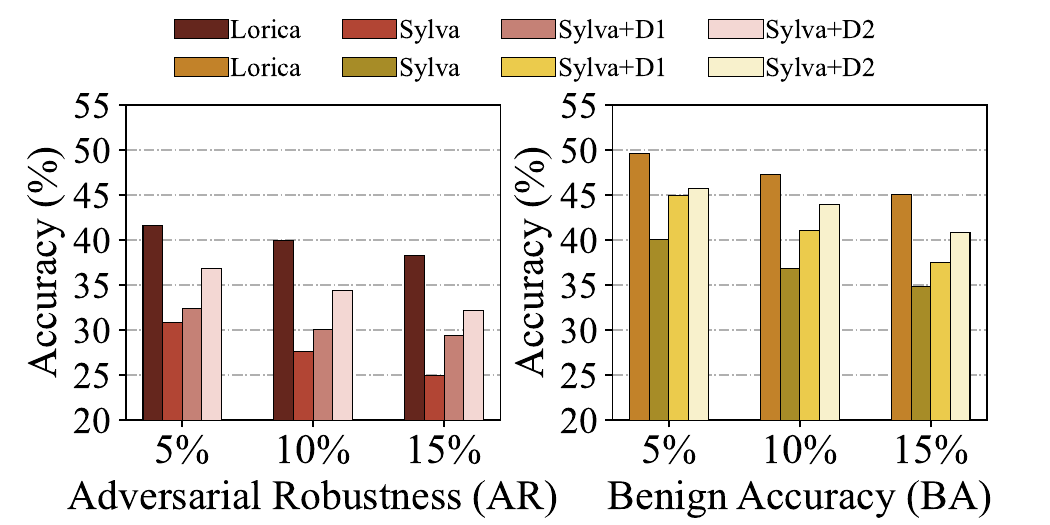}%
\label{fig:byzantine_2_stl10}}
\caption{Defense performance against Byzantine attacks on STL-10}
\label{fig:byzantine_stl10}
\vspace{-5mm}
\end{figure}

\begin{figure}[!tb]
\centering
\subfloat[CIFAR-10]{\includegraphics[width=1.65in]{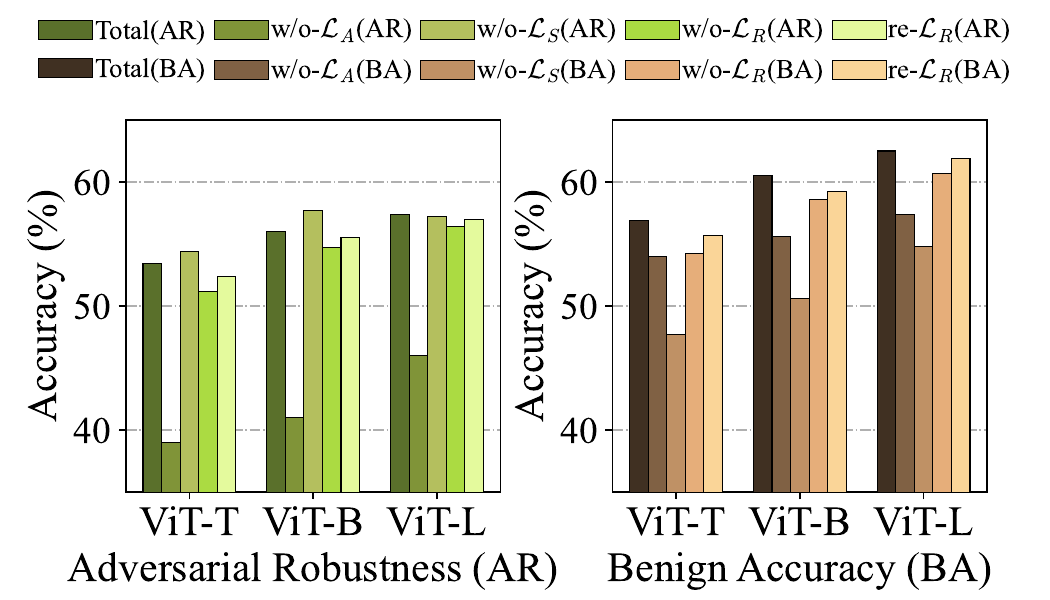}%
\label{fig:ablation_loss_cifar10}}
\subfloat[STL-10]{\includegraphics[width=1.65in]{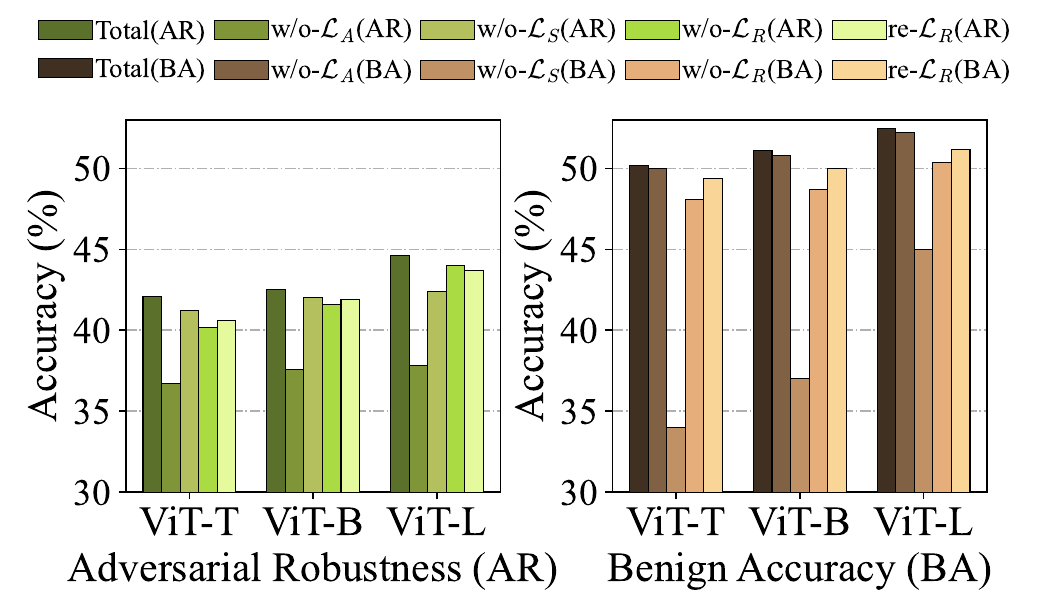}%
\label{fig:ablation_loss_stl10}}
\caption{Impact of different loss modules on adversarial robustness and benign accuracy on CIFAR-10 and STL-10}
\label{fig:ablation_loss}
\vspace{-5mm}
\end{figure}

\subsection{Comparison of Adversarial Training Efficiency}\label{Sec4:RQ3}

Given that \textit{Lorica} is designed for deployment in distributed systems with multiple edge devices, evaluating its efficiency in real-world environments is essential. To simulate a real distributed system, we employ three widely used GPUs—NVIDIA GeForce RTX 4090, 3090, and 2080-Ti—on edge devices, while leveraging a high-speed subnet server as the cloud. 
To align with real-world scenarios, we also explore the deployment of \textit{Lorica} in autonomous driving, including tasks like road sign recognition, as described in our threat model. We test \textit{Lorica} on both the autonomous driving simulation device, specifically the RTX 3060 in the Apollo D-KIT Advanced, and the widely used Jetson AGX Orin in real vehicles.
%
In contrast, real-world vehicles are typically equipped with 2-4 Jetson AGX Orin units, allowing parallel deployment to approach the computational power of the RTX 3090 per vehicle.
The detailed device comparison can be found in Appendix.

In this setup, we assess \textit{Lorica}'s efficiency by measuring its memory overhead (Mem), per-epoch training time (Time), and model upload communication time (Com) on edge devices. These metrics are compared against FA-TRADES (FAT) to provide a comprehensive evaluation. The results are shown in \cref{tab:result_efficency}.
Compared to traditional FAT algorithms, \textit{Lorica} employs an efficient parameter fine-tuning approach for pre-trained models, offering significant reductions in memory usage, training time, and communication overhead. These benefits become more pronounced as model size increases. Specifically, \textit{Lorica} reduces memory usage by up to 53.1\%, training time by 42.7\% on high-performance GPUs like the NVIDIA 4090, and up to 48.3\% on resource-constrained devices such as the 2080-Ti. 
In terms of communication, \textit{Lorica} achieves a 50$\times$ reduction in communication time, even with larger models and high-speed local networks, making it ideal for real-world distributed systems.
Notably, \textit{Lorica} outperforms \textit{Sylva} in both training time and memory efficiency by leveraging LoRA-FA to train only half of the parameters. In addition, uploading only LoRA-FA modules further reduces communication overhead.
Compared to Per-LoRA, \textit{Lorica} requires less memory due to the LoRA-FA module. \textit{Lorica} aggregates only the backbone’s LoRA-FA parameters, whereas Per-LoRA also uploads classifier parameters, yielding up to 75.0\% communication savings. Moreover, Per-LoRA's trust-weight computation increases complexity, while \textit{Lorica}’s ball-tree aggregation simplifies it, resulting in slightly faster training.

On the other hand, \textit{Lorica} not only improves performance on high-end GPUs but also reduces training time and communication overhead on real edge devices. Experimental results closely match the computational power estimates derived from our theoretical analysis based on GPU parameters. Notably, for the Jetson AGX Orin, we only use a single unit for the experiments, but real-world vehicles typically use multiple units. Model or data parallelism could potentially double the training speed. With 32GB of memory, the Orin can run \textit{Lorica} alongside other autonomous driving tasks. Overall, \textit{Lorica} outperforms baseline algorithms in real-world scenarios and is viable for practical deployment.
In addition, we also test the communication efficiency when training with different numbers of clients, and the results are in Appendix D.

\answer{3}{\textit{Lorica} significantly enhances distributed adversarial training by reducing memory usage, training time, and communication overhead, with its advantages growing as model size increases, making it highly efficient for real-world scenarios.}

\begin{figure}[!tb]
\centering
\subfloat[CIFAR-10]{\includegraphics[width=1.65in]{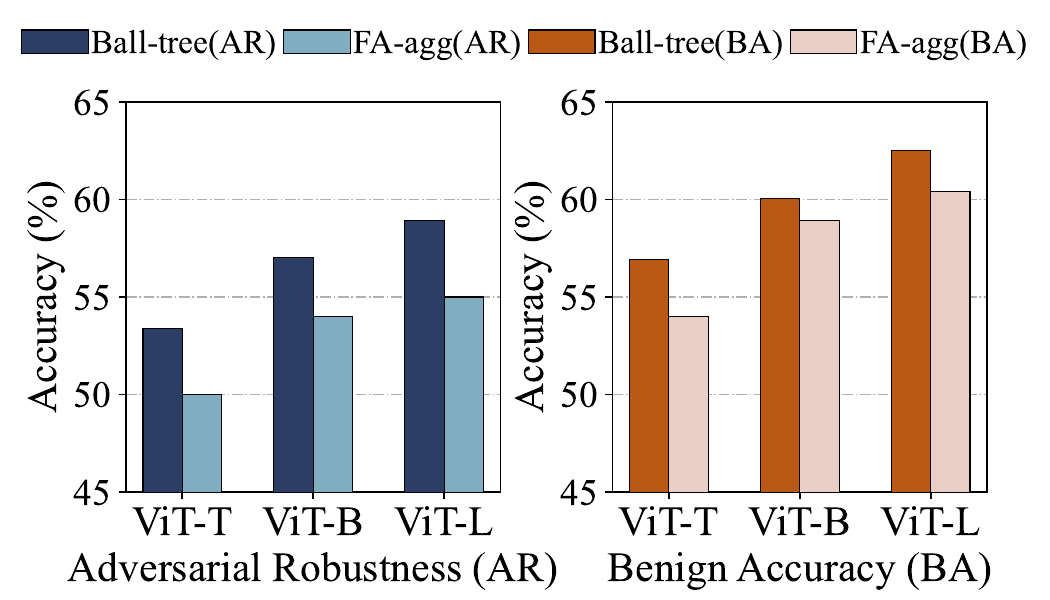}%
\label{fig:ablation_aggre_cifar10}}
\subfloat[STL-10]{\includegraphics[width=1.65in]{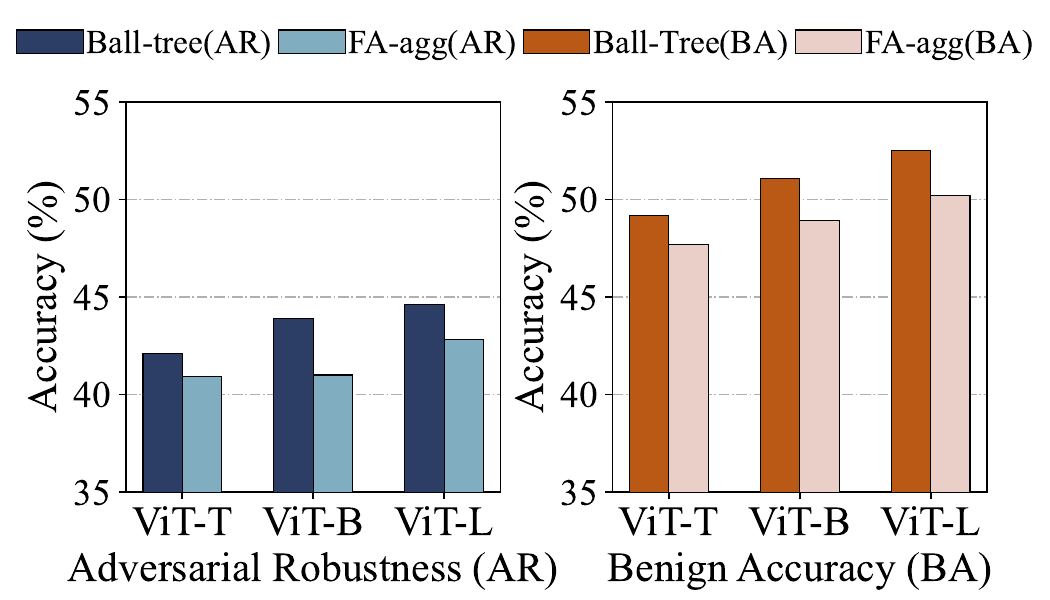}%
\label{fig:ablation_aggre_stl10}}
\caption{Impact of ball-tree-based aggregation algorithm on CIFAR-10 and STL-10}
\label{fig:ablation_aggre}
\vspace{-5mm}
\end{figure}

\begin{figure}[!tb]
\centering
\subfloat[CIFAR-10]{\includegraphics[width=1.65in]{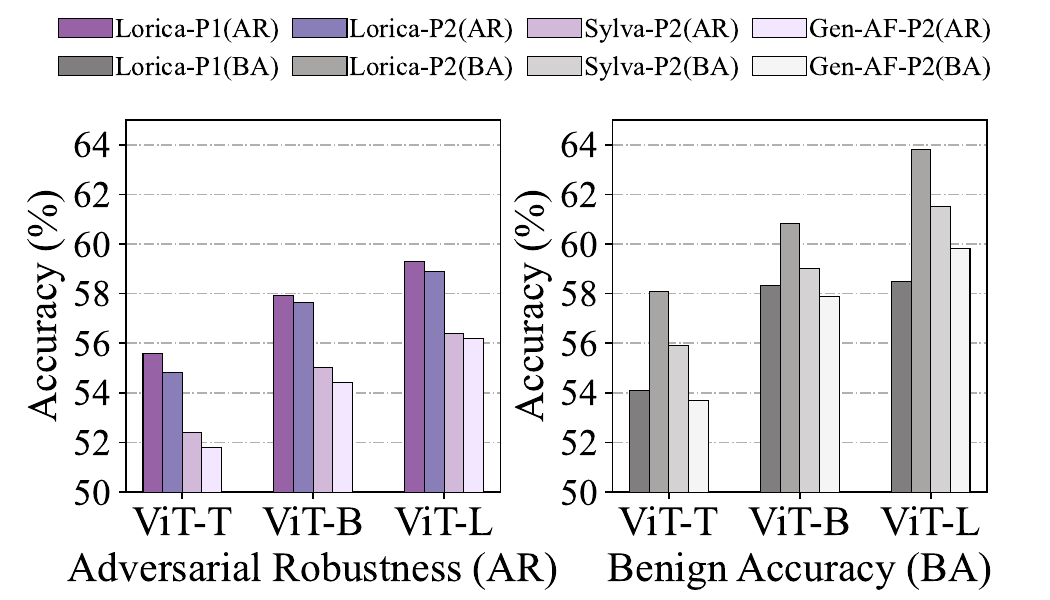}%
\label{fig:ablation_phase_cifar10}}
\subfloat[STL-10]{\includegraphics[width=1.65in]{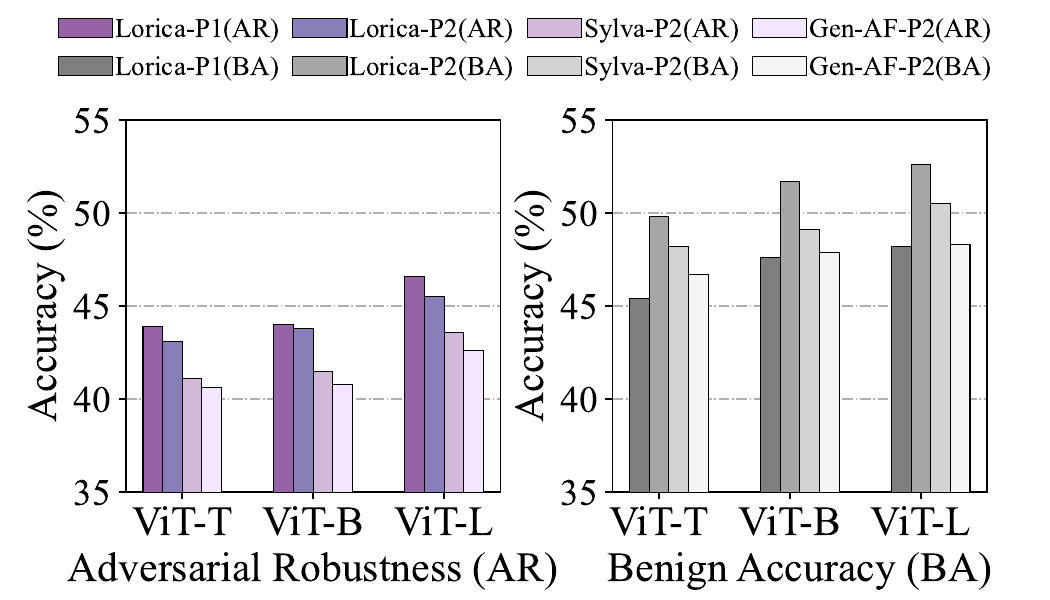}%
\label{fig:ablation_phase_stl10}}
\caption{Impact of different phases on the trade-off between robustness and accuracy for CIFAR-10 and STL-10}
\label{fig:ablation_phase}
\vspace{-4mm}
\end{figure}

\subsection{Defense performance under Byzantine attacks}\label{Sec4:RQ4}

To evaluate robustness against Byzantine attacks, we randomly select 5\%–15\% of the participating devices as attackers under the default setting. These malicious devices perform attacks via label flipping and MPAF, aiming to corrupt both local training data and model updates. We compare \textit{Lorica} with Sylva, which does not natively incorporate Byzantine defense mechanisms. To ensure a fair comparison, Sylva is augmented with two widely adopted aggregation-based defenses: Median (D1) and Trimmed Mean (D2).
As shown in \cref{fig:byzantine_cifar10} and \cref{fig:byzantine_stl10}, the Byzantine defense experiments on CIFAR-10 and STL-10 indicate that \textit{Lorica}, leveraging its explicit malicious client detection capability, consistently achieves higher robustness and accuracy. In contrast, Sylva combined with conventional Byzantine defenses exhibits relatively degraded performance, particularly as the proportion of Byzantine clients increases.
Additional results on other datasets are provided in Appendix E for completeness.

\answer{4}{\textit{Lorica} effectively mitigates Byzantine attacks during adversarial training, consistently achieving superior robustness and accuracy compared to conventional defense schemes.}

\subsection{Ablation Experiments}\label{Sec4:RQ5}

\subsubsection{Impact of Different Modules}

We evaluate the impact of our loss function, as illustrated in \cref{fig:ablation_loss}, by testing the total loss and assessing adversarial robustness and benign accuracy without the $\mathcal{L}_A$, $\mathcal{L}_S$, and $\mathcal{L}_R$ components individually, and we further replace our $\mathcal{L}_R$ with \textit{Sylva}'s $\mathcal{L}_R$.
The results clearly demonstrate that each component plays a positive role in enhancing overall performance. Among these, $\mathcal{L}_A$ and $\mathcal{L}_S$ prove to be particularly critical, as their absence leads to a substantial decline in both robustness and accuracy, underscoring their importance in the design of the loss function. Furthermore, our redesigned $\mathcal{L}_R$ outperforms that of \textit{Sylva}, evidencing the effectiveness of expert-guided personalization.
We next evaluate the impact of the ball tree aggregation algorithm employed in Phase 1 of \textit{Lorica} by comparing it with the traditional aggregation method used in FedAvg (FA-agg). As shown in \cref{fig:ablation_aggre}, the proposed algorithm effectively mitigates model drift caused by data heterogeneity, resulting in superior performance in both adversarial robustness and benign accuracy. 
We evaluate the effectiveness of the two-phase design on model robustness and benign accuracy. Phase 1 (P1) adversarial training generates a highly robust model, establishing a strong foundation for further refinement. In Phase 2 (P2), we compare \textit{Lorica}'s algorithm with \textit{Sylva} and Gen-AF's Stage 2 method by training the robust models with their respective approaches. 
%
As shown in \cref{fig:ablation_phase}, Phase 2 notably improves benign accuracy with minimal robustness loss, outperforming both Gen-AF and \textit{Sylva}. This demonstrates that \textit{Lorica}’s two-phase framework, particularly the forward-gating selection fine-tuning, more effectively identifies layers that balance robustness and accuracy.
%
Additional results for GTSRB and CIFAR-100 are provided in Appendix F for further validation and analysis.

\begin{table}[!tb]
\centering
\caption{The Impact of Hyperparameters in CIFAR-10}\label{tab:result_hyper}
\scalebox{0.95}{
\begin{tabular}{@{}c|c|cc|c|c|cc@{}}
\toprule
Para                  & Value & AR    & BA    & Para                 & Value & AR    & BA    \\ \midrule
\multirow{4}{*}{$\gamma$}    & 0.9 & 57.52 & 60.83 & \multirow{4}{*}{$\beta$}                                                & 20  & 57.17 & 57.43 \\
                             & 0.7 & 56.65 & 59.61 &                                                                         & 10  & 56.50 & 58.42 \\
                             & 0.5 & 56.88 & 58.54 &                                                                         & 5   & 57.52 & 60.83 \\
                             & 0.3 & 55.20 & 57.39 &                                                                         & 1   & 51.35 & 61.47 \\ \midrule
\multirow{4}{*}{$\epsilon$}  & 0.9 & 57.52 & 60.83 & \multirow{4}{*}{$B$}                                                    & 5\%  & 55.12 & 55.24 \\
                             & 0.7 & 57.10 & 60.15 &                                                                         & 10\% & 57.52 & 60.83 \\
                             & 0.5 & 56.48 & 59.50 &                                                                         & 20\% & 55.08 & 59.10 \\
                             & 0.3 & 57.22 & 59.42 &                                                                         & 30\% & 55.21 & 59.85 \\ \midrule
\multirow{4}{*}{$r$}         & 8   & 57.13 & 59.81 & \multirow{4}{*}{\begin{tabular}[c]{@{}c@{}}PGD\\ Strength\end{tabular}} & 3   & 50.66 & 61.54 \\
                             & 4   & 57.52 & 60.83 &                                                                         & 5   & 54.59 & 61.83 \\
                             & 2   & 57.58 & 59.53 &                                                                         & 10  & 57.52 & 60.83 \\
                             & 1   & 55.56 & 59.25 &                                                                         & 15  & 59.12 & 57.37 \\ \bottomrule
\end{tabular}
}
\vspace{-4mm}
\end{table}

\subsubsection{Impact of Different Hyperparameters}

To investigate the impact of hyperparameters on \textit{Lorica}, we conduct a comprehensive set of experiments on the ViT-B model using the CIFAR-10 dataset. This analysis evaluates the effects of key hyperparameters, including $\gamma$ and $\epsilon$ in the Phase 1 loss function, $\beta$ and $B$ in Phase 2, the hidden space dimension $r$ for LoRA-FA, and the PGD strength employed in adversarial training. \cref{tab:result_hyper} presents the corresponding results for adversarial robustness and benign accuracy, providing a detailed comparison across different hyperparameter settings.
The findings demonstrate that the default hyperparameter settings deliver strong overall performance, balancing robustness and accuracy effectively. 
%
Notably, in Phase 2, selecting different proportions of model layers as the training budget $B$ leads to suboptimal results when the ratio is either too high or too low, suggesting that the fine-tuning layer proportion should remain moderate.
Furthermore, the hidden space dimension $r$ for LoRA-FA shows minimal impact on the final results; even with $r=1$, performance remains excellent. This observation aligns with LoRA-FA's established efficiency and adaptability in traditional fine-tuning settings, further validating its applicability in adversarial training scenarios.
Additionally, we evaluate the effects of these hyperparameters on three other datasets, as well as on scenarios with varying numbers of clients. The detailed results are provided in Appendix F.

\answer{5}{\textit{Lorica} excels in balancing robustness and accuracy through its modular design, efficient aggregation, and adaptive hyperparameter tuning, validated across diverse datasets and scenarios.}

\section{Discussion}\label{sec5}

This section discusses key directions for further advancing \textit{Lorica}, focusing on balancing efficiency and expressiveness, stabilizing expert-guided aggregation, and strengthening robustness against more adaptive Byzantine adversaries.

\textbf{Expressivity–efficiency boundary of LoRA-FA.}
By adopting LoRA-FA, \textit{Lorica} significantly reduces computation and communication overhead, but this efficiency comes with a constrained trainable subspace. Under severe domain shifts or highly heterogeneous client distributions, limited adaptation capacity restrict personalization and robustness gains. Future work could explore adaptive rank or layer-wise budget allocation to dynamically balance efficiency and expressive power.

\textbf{Reliability and load balancing in expert-guided aggregation.}
The MoE-inspired expert-guided aggregation improves personalization by leveraging specialization, yet its effectiveness depends on reliable and balanced expert utilization. In skewed settings, routing may concentrate on a few experts, reducing diversity and potentially amplifying bias. Incorporating lightweight load-balancing or reliability-aware weighting mechanisms could further stabilize aggregation without sacrificing efficiency.

\textbf{Residual risks under stronger Byzantine adversaries.}
While \textit{Lorica} demonstrates robustness against representative Byzantine attacks, stronger adaptive or colluding adversaries may still evade detection by mimicking benign updates or exploiting aggregation dynamics. Thus, Byzantine defense should be viewed as a defense-in-depth objective, where robust aggregation can be complemented by low-overhead anomaly or reputation-based mechanisms to further enhance resilience.

\section{Related works}\label{sec6}

\subsection{Adversarial Attack}

Recent studies have crafted misclassifying adversarial examples through subtle perturbations. Goodfellow et al. proposed FGSM~\cite{goodfellow2014explaining}, highlighting neural networks' vulnerability due to their linearity and introducing a simple adversarial training approach. Tramèr et al. improved robustness against black-box attacks with RFGSM~\cite{tramer2017ensemble}, which integrates perturbations from multiple models. Madry et al. introduced PGD~\cite{madry2017towards}, providing robust optimization with security guarantees. Liu et al. developed EOTPGD~\cite{liu2018adv}, a Bayesian framework incorporating randomness for greater robustness. Croce et al. proposed APGD~\cite{croce2020reliable}, a parameter-free ensemble attack for evaluating defense strategies.
Su and Modas et al. introduced the One Pixel Attack~\cite{su2019one} and Sparsefool~\cite{modas2019sparsefool}, optimizing minimal pixel changes to reduce computational overhead. Ban et al. presented PAP~\cite{ban2022pre}, leveraging Low-Level Layer Lifting Attacks (L4A) to manipulate neuron activations in pre-trained models. Zhou et al. introduced AdvCLIP~\cite{zhou2023advclip}, generating downstream-agnostic adversarial examples with cross-modal pre-trained encoders.

\subsection{Adversarial Defence in Federated Learning}

Federated learning (FL) enables distributed model training while preserving user privacy~\cite{mcmahan2017communication}, but faces challenges due to statistical heterogeneity in device data distributions~\cite{qi2023hwamei, luo2021no}. 
%
To overcome the limitations of global models in non-IID scenarios, personalized federated learning approaches have been proposed to train client-specific models~\cite{wang2024towards}.
Many works leverage parameter-efficient methods in FL for LLMs~\cite{qi2024tomtit, guo2023promptfl}, including adapter tuning, prompt learning, and localized adjustments.
With the continuous development of edge computing, many works have implemented attacks and defenses based on the FL framework~\cite{yazdinejad2024robust}.
%
He et al. proposed PPBR~\cite{he2025ppbr}, a privacy-preserving and Byzantine-robust hierarchical federated learning framework for non-IID settings.
Li et al. provided a convergence proof for this federated adversarial learning approach~\cite{li2023federated}. Zhou et al. addressed the aggregation error in FL, breaking it down into bias and variance~\cite{zhou2022adversarial}. Additionally, Chen et al. explored certified defenses against adversarial examples in FL, providing an alternative direction for enhancing robustness~\cite{chen2022calfat}.
Zhang et al. proposed DBFAT~\cite{zhang2023delving}, which adjusts sample weights based on PGD steps for generating adversarial examples. 
Zhang et al. propose FedEdge~\cite{zhang2025cost}, a cost-efficient and Byzantine-robust FL framework for edge computing with adaptive compression and dynamic filtering.
Zuo et al. propose RAGA~\cite{zuo2025federated}, a geometric-median-based Byzantine-robust FL algorithm with convergence guarantees under heterogeneous data.

\section{Conclusion}\label{sec7}

This paper introduces \textit{Lorica}, the first adversarial defense algorithm for pre-trained models in distributed settings. \textit{Lorica} ensures robust and accurate models even with imbalanced data. Through a two-phase training framework—utilizing a novel loss function with personalized federated fine-tuning for enhanced robustness (Phase 1) and a forward-gating selection fine-tuning strategy for balanced robustness and accuracy (Phase 2)—\textit{Lorica} demonstrably outperforms state-of-the-art adversarial defenses in both metrics. Crucially, it achieves this with minimal computational overhead, making it highly suitable for edge deployment.

\bibliographystyle{IEEEtran}
\bibliography{cite}

\vfill

\newpage

{
\appendices

\section{Details of Experimental Setup}\label{App A}

\subsection{Datasets and Models}

In this experiment, we select several datasets that are commonly employed in adversarial training research. These datasets encompass a variety of characteristics, including differing numbers of categories, distinct application scenarios, and varying dataset scales. A detailed overview of each dataset is presented below:

\begin{itemize}
    \item CIFAR-10~\cite{krizhevsky2009learning}: A dataset of 60,000 32$\times$32 color images across 10 classes, with 50,000 for training and 10,000 for testing.
    \item STL-10~\cite{coates2011analysis}: A dataset with 10 object classes, consisting of 5,000 training and 8,000 test images at 96$\times$96 resolution, commonly used for unsupervised and semi-supervised learning benchmarks.
    \item GTSRB~\cite{stallkamp2011german}: A traffic sign recognition dataset with over 50,000 images spanning 43 classes, serving as a benchmark for intelligent transportation systems.
    \item CIFAR-100~\cite{krizhevsky2009learning}: Similar to CIFAR-10 but with 100 classes, comprising 60,000 images (600 per class), offering a greater challenge in image classification.
\end{itemize}

The Vision Transformer (ViT) is a transformer-based model for image classification, treating images as sequences of patches. We use three pre-trained variants: ViT-Tiny (ViT-T/16), ViT-Base (ViT-B/16), and ViT-Large (ViT-L/16), differing in scale and complexity. Their configurations, summarized in \cref{table:vit}, enable performance evaluation under varying computational demands.

\subsection{Attacks}

Building on the threat model outlined in \cref{sec2}, attacker methods can be broadly divided into white-box and gray-box attacks. In our experiments, we focus on four representative attack methods, detailed as follows:
\begin{itemize}
    \item FGSM~\cite{goodfellow2014explaining}: A white-box attack that perturbs inputs along the gradient direction of the loss function, efficiently generating small but impactful adversarial examples.
    \item PGD~\cite{madry2017towards}: An iterative extension of FGSM, applying multiple small perturbations projected back into the allowed space for a more thorough adversarial optimization.
    \item SparseFool~\cite{modas2019sparsefool}: Focuses on creating sparse perturbations by minimizing pixel changes, offering effective attacks with high computational efficiency for high-dimensional data.
    \item PAP~\cite{ban2022pre}: A gray-box attack targeting pre-trained models, effective even against fine-tuned models by leveraging low-level layer manipulations and noise augmentation for strong transferability.
\end{itemize}

The aforementioned attack algorithms are applied to a range of datasets, with one representative image selected from each dataset for visualization purposes. Results on ViT-B are presented in \cref{fig:attacks}. These algorithms are straightforward to implement and effectively satisfy the inconspicuousness criteria outlined in the threat model, ensuring minimal perceptual differences while achieving their intended effects.

For Byzantine attacks, we consider two representative strategies:
\begin{itemize}
    \item Label-flipping~\cite{rosenfeld2020certified}: A data-level attacks, where malicious clients perform label-flipping during local training.
    \item MPAF~\cite{cao2022mpaf}: A model-level attacks, where malicious clients poison the global model by submitting manipulated local updates.
\end{itemize}
Attackers can compromise a subset of clients as Byzantine attackers and apply the above attacks to disrupt the normal federated adversarial training process.

\begin{table}[!tb]
\centering
\caption{Details of vision transformer models}\label{table:vit}
\setlength{\tabcolsep}{4pt}        
\scalebox{0.95}{
\begin{tabular}{@{}c|cccc@{}}
\toprule
Model    & Layers & Hidden Size & Intermediate Size & Heads \\ \midrule
ViT-T/16 & 12     & 192         & 768               & 3     \\
ViT-B/16 & 12     & 768         & 3072              & 12    \\
ViT-L/16 & 24     & 1024        & 4096              & 16    \\ \bottomrule
\end{tabular}
}
\end{table}

\begin{figure}[!tb]
    \centering
    \includegraphics[width=2.2  in]{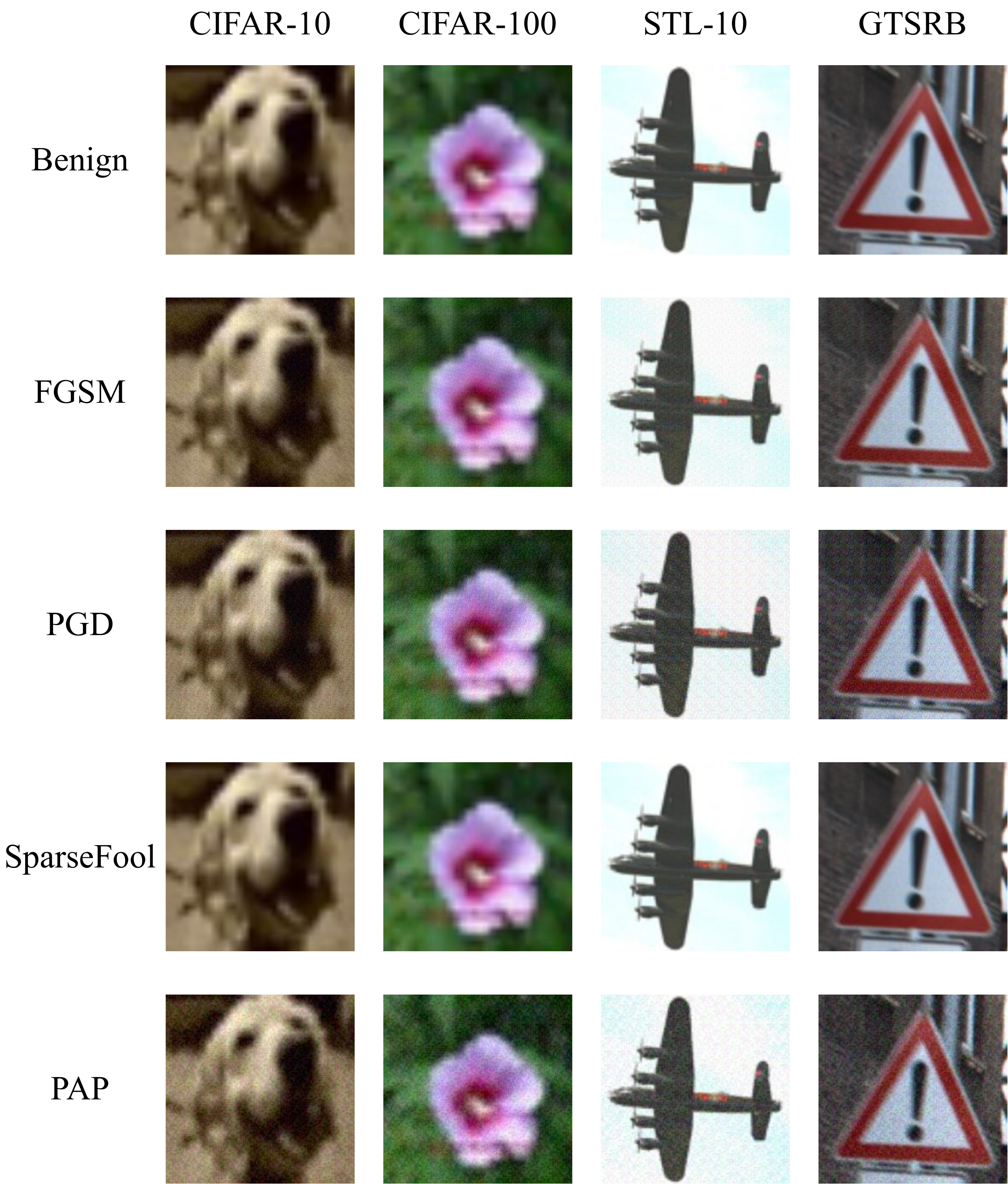}
    \caption{Visualization under different attack algorithms}
    \label{fig:attacks}
    \vspace{-5mm}
\end{figure}

\subsection{Baselines}

To demonstrate the superiority of \textit{Lorica}, we first examine commonly used and novel defense algorithms. For local defense, the following three algorithms are employed:

\begin{itemize}
    \item PGT-AT~\cite{madry2017towards}: PGD-AT employs the PGD algorithm to generate adversarial examples during training, thereby enhancing model robustness. By repeatedly exposing the model to adversarial perturbations, it improves the model's ability to resist various attacks effectively.
    \item TRADES~\cite{zhang2019theoretically}: TRADES achieves a balance between accuracy on benign data and robustness against adversarial attacks. It minimizes a carefully designed loss function that combines the standard classification loss with a penalty term measuring output discrepancies between benign and adversarial examples.
    \item Gen-AF~\cite{zhou2024securely}: Gen-AF improves model robustness through a two-stage fine-tuning approach for pre-trained models. It first optimizes the encoder using genetic regularization, followed by the selection and refinement of robust layers, leading to enhanced accuracy and robustness across diverse datasets.
\end{itemize}

In the distributed multi-client scenario addressed in this paper, we integrate commonly used federated learning algorithms with the aforementioned local defense strategies to enable effective deployment. The specific federated learning algorithms employed are as follows:

\begin{itemize}
    \item FedAvg~\cite{mcmahan2017communication}: FedAvg is a federated learning algorithm where each client trains a local model and shares updates with the server, which computes a weighted average to update the global model.
    \item FedProx~\cite{li2020federated}: FedProx extends FedAvg by adding a proximal term to the loss function, mitigating the impact of heterogeneous data across clients. This regularization ensures that local models remain close to the global model, improving convergence and robustness.
    \item DBFAT~\cite{zhang2023delving}: DBFAT enhances adversarial robustness in federated learning through local re-weighting and global regularization, improving both accuracy and robustness, especially in non-IID settings.
    \item Per-Adv~\cite{allouah2024fine}: Per-Adv enhances robustness against Byzantine attacks by introducing a personalized loss function, wherein each client interacts with others to collaboratively construct a robust interpolated objective.
    \item Per-LoRA~\cite{wagner2024personalized}: Per-LoRA leverages LoRA-based fine-tuning to perform local updates of large models, while additionally computing trust weights from each client’s LoRA parameters, which are used to guide the personalized update of the local model.
    \item Sylva~\cite{qi2025sylva}: Sylva is the previous version of \textit{Lorica}, which first introduced a two-phase federated fine-tuning framework and employed conventional LoRA-based tuning to enhance model robustness.
\end{itemize}
Specifically, we integrate the FedAvg and FedProx algorithms with the three aforementioned local defense algorithms, yielding the following configurations: FA-PGD-AT, FA-TRADES, and FA-Gen-AF, as well as FP-PGD-AT, FP-TRADES, and FP-Gen-AF. These combinations leverage the strengths of both federated optimization strategies and local defense mechanisms to enhance model robustness.
For the two personalized algorithms, Per-Adv and Per-LoRA, we apply appropriate optimizations to adapt them to our specific tasks. We utilize the TRADES loss function for both algorithms to facilitate adversarial training. In the case of Per-Adv, we assume an adversary without Byzantine attacks to better align with the threat model outlined in this study. For Per-LoRA, we employ strategy 2 proposed by the authors, to compute the trust weights. 

We adopt two widely used Byzantine defense schemes~\cite{yin2018byzantine}:
\begin{itemize}
    \item Median: Server aggregates client updates by taking the coordinate-wise median, effectively suppressing extreme outliers caused by Byzantine clients.
    \item Trimmed Mean: Server mitigates Byzantine effects by discarding a fraction of the largest and smallest updates before averaging, improving robustness under partial client corruption.
\end{itemize}

All other configurations of the personalization algorithm follow those in Sylva.

\begin{figure}[!tb]
\centering
\subfloat[IID]{\includegraphics[width=1.65in]{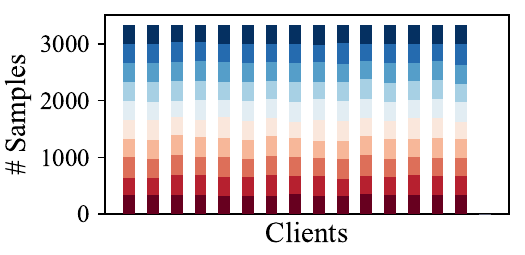}%
\label{fig:dir-1.0}}
\subfloat[Dir-10]{\includegraphics[width=1.65in]{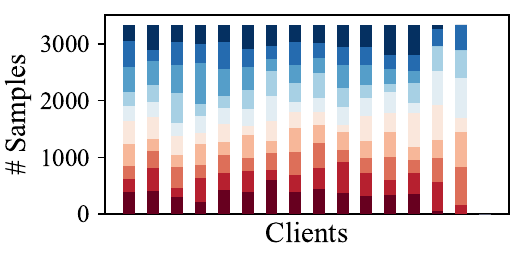}%
\label{fig:dir-0.8}}

\vspace{-3mm} 
\subfloat[Dir-6]{\includegraphics[width=1.65in]{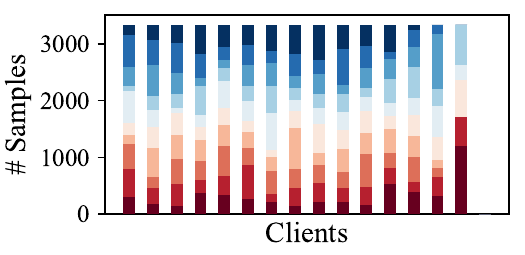}%
\label{fig:dir-0.5}}
\subfloat[Dir-3]{\includegraphics[width=1.65in]{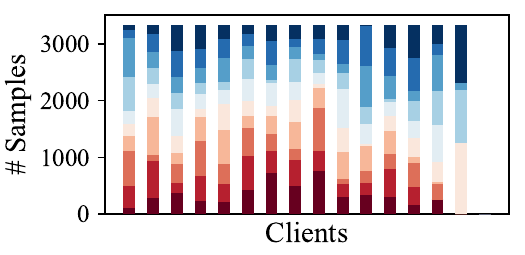}%
\label{fig:dir-0.2}}
\caption{Visualization of Dirichlet distribution}
\label{fig:dir}
\vspace{-5mm}
\end{figure}

\begin{table}[!tb]
\centering
\caption{Performance comparison of \textit{Lorica} and baseline under different datasets and attack algorithms (ViT-T/16)}\label{tab:result_vit_t}
\scalebox{0.82}{
\renewcommand{\arraystretch}{0.8}  
\begin{tabular}{@{}c|c|ccccc@{}}
\toprule
Datasets                           &  Baselines           & Benign(BA) & FGSM  & PGD   & SparseFool & PAP   \\ \midrule
\multirow{11}{*}{CIFAR-10}  & FA-PGD-AT   & 37.54      & 41.82 & 40.95 & 46.10      & 42.23 \\
                           & FA-TRADES   & 48.24      & 43.10 & 42.58 & 44.15      & 41.48 \\
                           & FA-Gen-AF   & 50.12      & 44.68 & 42.74 & 43.59      & 42.61 \\
                           & FP-PGD-AT   & 39.91      & 42.35 & 41.62 & 45.86      & 42.48 \\
                           & FP-TRADES   & 49.58      & 44.90 & 42.69 & 43.53      & 42.10 \\
                           & FP-Gen-AF   & 51.73      & 45.50 & 43.91 & 45.57      & 44.91 \\
                           & DBFAT       & 51.45      & 43.62 & 42.80 & 45.35      & 44.50 \\
                           & Per-Adv     & 39.96      & 41.93 & 41.32 & 46.50      & 42.77 \\
                           & Per-LoRA    & 50.89      & 45.90 & 43.03 & 45.21      & 44.30 \\
                           & Sylva       & 55.92      & 50.02 & 52.45 & 50.25      & 49.58 \\
                           & \cellcolor[HTML]{DCDCDC}Lorica(Ours) & \cellcolor[HTML]{DCDCDC}\textbf{57.46}     & \cellcolor[HTML]{DCDCDC}\textbf{52.32} & \cellcolor[HTML]{DCDCDC}\textbf{53.19} & \cellcolor[HTML]{DCDCDC}\textbf{52.97}      & \cellcolor[HTML]{DCDCDC}\textbf{52.98} \\ \midrule
\multirow{11}{*}{STL-10}    & FA-PGD-AT   & 32.81      & 36.12 & 34.89 & 38.45      & 37.67 \\
                           & FA-TRADES   & 40.31      & 35.56 & 34.88 & 37.91      & 37.40 \\
                           & FA-Gen-AF   & 42.29      & 36.85 & 35.30 & 37.91      & 36.77 \\
                           & FP-PGD-AT   & 34.41      & 36.48 & 35.21 & 39.35      & 37.89 \\
                           & FP-TRADES   & 41.10      & 37.21 & 35.22 & 37.00      & 36.88 \\
                           & FP-Gen-AF   & 43.10      & 38.92 & 36.67 & 39.60      & 38.53 \\
                           & DBFAT       & 42.75      & 36.77 & 35.48 & 39.14      & 38.11 \\
                           & Per-Adv     & 32.92      & 36.83 & 35.01 & 38.25      & 38.01 \\
                           & Per-LoRA    & 43.06      & 38.90 & 36.82 & 39.25      & 38.44 \\
                           & Sylva       & 48.21      & 42.37 & 41.12 & 43.82      & 42.78 \\
                           & \cellcolor[HTML]{DCDCDC}Lorica(Ours) & \cellcolor[HTML]{DCDCDC}\textbf{49.90}      & \cellcolor[HTML]{DCDCDC}\textbf{43.95} & \cellcolor[HTML]{DCDCDC}\textbf{42.47} & \cellcolor[HTML]{DCDCDC}\textbf{44.81}      & \cellcolor[HTML]{DCDCDC}\textbf{44.02} \\ \midrule
\multirow{11}{*}{GTSRB}     & FA-PGD-AT   & 55.73      & 63.52 & 64.25 & 69.12      & 68.22 \\
                           & FA-TRADES   & 71.15      & 64.10 & 65.28 & 67.39      & 68.52 \\
                           & FA-Gen-AF   & 74.85      & 65.49 & 66.10 & 67.50      & 68.10 \\
                           & FP-PGD-AT   & 59.15      & 64.38 & 65.45 & 68.23      & 67.90 \\
                           & FP-TRADES   & 72.30      & 65.75 & 66.52 & 67.37      & 68.20 \\
                           & FP-Gen-AF   & 75.92      & 66.10 & 66.48 & 67.75      & 69.71 \\
                           & DBFAT       & 75.40      & 65.38 & 65.53 & 66.75      & 68.95 \\
                           & Per-Adv     & 60.73      & 63.06 & 64.25 & 67.21      & 67.53 \\
                           & Per-LoRA    & 74.33      & 65.80 & 65.71 & 67.05      & 68.32 \\
                           & Sylva       & 76.90      & 68.75 & \textbf{68.90} & 69.93      & 69.03 \\
                           & \cellcolor[HTML]{DCDCDC}Lorica(Ours) & \cellcolor[HTML]{DCDCDC}\textbf{79.21}     & \cellcolor[HTML]{DCDCDC}\textbf{70.13} & \cellcolor[HTML]{DCDCDC}68.84 & \cellcolor[HTML]{DCDCDC}\textbf{71.06}      & \cellcolor[HTML]{DCDCDC}\textbf{71.20} \\ \midrule
\multirow{11}{*}{CIFAR-100} & FA-PGD-AT   & 11.89      & 24.32 & 23.75 & 25.92      & 24.70 \\
                           & FA-TRADES   & 22.60      & 23.52 & 22.11 & 23.77      & 23.91 \\
                           & FA-Gen-AF   & 22.39      & 24.00 & 23.12 & 24.30      & 23.68 \\
                           & FP-PGD-AT   & 14.02      & 24.45 & 24.18 & 26.12      & 24.90 \\
                           & FP-TRADES   & 23.41      & 22.78 & 23.75 & 25.18      & 24.05 \\
                           & FP-Gen-AF   & 23.18      & 24.10 & 23.99 & 25.95      & 25.30 \\
                           & DBFAT       & 23.28      & 23.89 & 23.98 & 25.50      & 26.88 \\
                           & Per-Adv     & 14.65      & 23.87 & 22.64 & 23.85      & 24.04 \\
                           & Per-LoRA    & 23.27      & 24.20 & 23.75 & 24.89      & 25.06 \\
                           & Sylva       & 24.92      & 25.12 & 24.82 & 26.25      & 27.15 \\
                           & \cellcolor[HTML]{DCDCDC}Lorica(Ours) & \cellcolor[HTML]{DCDCDC}\textbf{25.86}     & \cellcolor[HTML]{DCDCDC}\textbf{25.57} & \cellcolor[HTML]{DCDCDC}\textbf{26.44}& \cellcolor[HTML]{DCDCDC}\textbf{26.71}     & \cellcolor[HTML]{DCDCDC}\textbf{27.88} \\ \bottomrule
\end{tabular}
}
\vspace{-3mm}
\end{table}

\begin{table}[!tb]
\centering
\caption{Performance comparison of \textit{Lorica} and baseline under different datasets and attack algorithms (ViT-L/16)}\label{tab:result_vit_l}
\scalebox{0.82}{
\renewcommand{\arraystretch}{0.8}  
\begin{tabular}{@{}c|c|ccccc@{}}
\toprule
Datasets                           &  Baselines           & Benign(BA) & FGSM  & PGD   & SparseFool & PAP   \\ \midrule
\multirow{11}{*}{CIFAR-10}  & FA-PGD-AT   & 40.50      & 47.21 & 43.89 & 49.45      & 45.58 \\
                           & FA-TRADES   & 52.37      & 46.18 & 44.47 & 47.00      & 44.64 \\
                           & FA-Gen-AF   & 54.75      & 48.17 & 46.87 & 47.27      & 46.78 \\
                           & FP-PGD-AT   & 42.14      & 48.62 & 45.09 & 50.02      & 46.87 \\
                           & FP-TRADES   & 54.17      & 48.02 & 46.68 & 48.78      & 46.56 \\
                           & FP-Gen-AF   & 56.07      & 49.94 & 47.83 & 50.10      & 48.48 \\
                           & DBFAT       & 54.92      & 47.01 & 46.13 & 50.20      & 49.72 \\
                           & Per-Adv     & 41.34      & 47.35 & 44.02 & 48.35      & 45.66 \\
                           & Per-LoRA    & 55.83      & 48.92 & 47.60 & 49.57      & 48.21 \\
                           & Sylva       & 61.51      & 54.01 & 56.39 & 54.12      & 53.26 \\
                           & \cellcolor[HTML]{DCDCDC}Lorica(Ours) & \cellcolor[HTML]{DCDCDC}\textbf{62.48}      & \cellcolor[HTML]{DCDCDC}\textbf{56.30} & \cellcolor[HTML]{DCDCDC}\textbf{56.89} & \cellcolor[HTML]{DCDCDC}\textbf{55.35}      & \cellcolor[HTML]{DCDCDC}\textbf{54.82} \\ \midrule
\multirow{11}{*}{STL-10}    & FA-PGD-AT   & 35.89      & 41.05 & 39.88 & 41.95      & 41.05 \\
                           & FA-TRADES   & 43.70      & 39.15 & 38.11 & 41.80      & 40.93 \\
                           & FA-Gen-AF   & 46.00      & 40.48 & 39.12 & 41.98      & 41.85 \\
                           & FP-PGD-AT   & 37.78      & 41.77 & 39.14 & 42.51      & 41.69 \\
                           & FP-TRADES   & 45.02      & 40.72 & 39.12 & 40.93      & 40.99 \\
                           & FP-Gen-AF   & 47.20      & 41.03 & 40.49 & 42.12      & 43.02 \\
                           & DBFAT       & 46.79      & 42.17 & 39.91 & 41.14      & 42.78 \\
                           & Per-Adv     & 37.95      & 41.23 & 39.48 & 41.52      & 41.12 \\
                           & Per-LoRA    & 46.95      & 41.78 & 41.20 & 42.05      & 42.11 \\
                           & Sylva       & 50.51      & 45.01 & 43.52 & 46.47      & 45.42 \\
                           & \cellcolor[HTML]{DCDCDC}Lorica(Ours) & \cellcolor[HTML]{DCDCDC}\textbf{52.14}      & \cellcolor[HTML]{DCDCDC}\textbf{47.22} & \cellcolor[HTML]{DCDCDC}\textbf{45.10} &\cellcolor[HTML]{DCDCDC}\textbf{47.22}      & \cellcolor[HTML]{DCDCDC}\textbf{46.39} \\ \midrule
\multirow{11}{*}{GTSRB}     & FA-PGD-AT   & 61.69      & 69.99 & 70.69 & 73.65      & 73.48 \\
                           & FA-TRADES   & 75.24      & 69.58 & 70.40 & 72.34      & 74.15 \\
                           & FA-Gen-AF   & 77.86      & 70.63 & 69.92 & 74.12      & 73.36 \\
                           & FP-PGD-AT   & 63.98      & 71.57 & 70.66 & 73.62      & 72.71 \\
                           & FP-TRADES   & 78.04      & 71.58 & 71.62 & 72.65      & 72.76 \\
                           & FP-Gen-AF   & 81.14      & 72.81 & 71.83 & 73.95      & 75.07 \\
                           & DBFAT       & 80.96      & 70.77 & 70.88 & 72.47      & 74.30 \\
                           & Per-Adv     & 65.52      & 69.29 & 69.37 & 72.55      & 73.18 \\
                           & Per-LoRA    & 78.83      & 71.52 & 70.25 & 72.36      & 74.01 \\
                           & Sylva       & 82.51      & 74.39 & 74.20 & 75.64      & 75.84 \\
                           & \cellcolor[HTML]{DCDCDC}Lorica(Ours) & \cellcolor[HTML]{DCDCDC}\textbf{83.02}      & \cellcolor[HTML]{DCDCDC}\textbf{75.20} & \cellcolor[HTML]{DCDCDC}\textbf{75.25} & \cellcolor[HTML]{DCDCDC}\textbf{76.44}      & \cellcolor[HTML]{DCDCDC}\textbf{76.80} \\ \midrule
\multirow{11}{*}{CIFAR-100} & FA-PGD-AT   & 14.35      & 26.59 & 26.04 & 28.29      & 27.33 \\
                           & FA-TRADES   & 24.73      & 25.50 & 24.93 & 25.95      & 26.14 \\
                           & FA-Gen-AF   & 24.23      & 26.05 & 25.81 & 26.81      & 26.09 \\
                           & FP-PGD-AT   & 15.82      & 26.47 & 26.04 & 28.86      & 27.46 \\
                           & FP-TRADES   & 25.32      & 24.42 & 26.05 & 27.32      & 26.33 \\
                           & FP-Gen-AF   & 25.70      & 26.11 & 27.89 & 28.15      & 27.30 \\
                           & DBFAT       & 25.66      & 25.78 & 26.91 & 27.79      & 29.04 \\
                           & Per-Adv     & 16.80      & 25.88 & 26.01 & 26.72      & 27.04 \\
                           & Per-LoRA    & 25.97      & 25.99 & 26.32 & 27.85      & 28.03 \\
                           & Sylva       & 29.62      & 28.33 & 28.87 & 29.77      & 29.44 \\
                           & \cellcolor[HTML]{DCDCDC}Lorica(Ours) & \cellcolor[HTML]{DCDCDC}\textbf{30.95}     & \cellcolor[HTML]{DCDCDC}\textbf{29.11}& \cellcolor[HTML]{DCDCDC}\textbf{29.35} & \cellcolor[HTML]{DCDCDC}\textbf{31.02}      & \cellcolor[HTML]{DCDCDC}\textbf{30.89} \\ \bottomrule
\end{tabular}
}
\vspace{-3mm}
\end{table}

\begin{figure}[!tb]
\centering
\subfloat[CIFAR-10]{\includegraphics[width=1.65in]{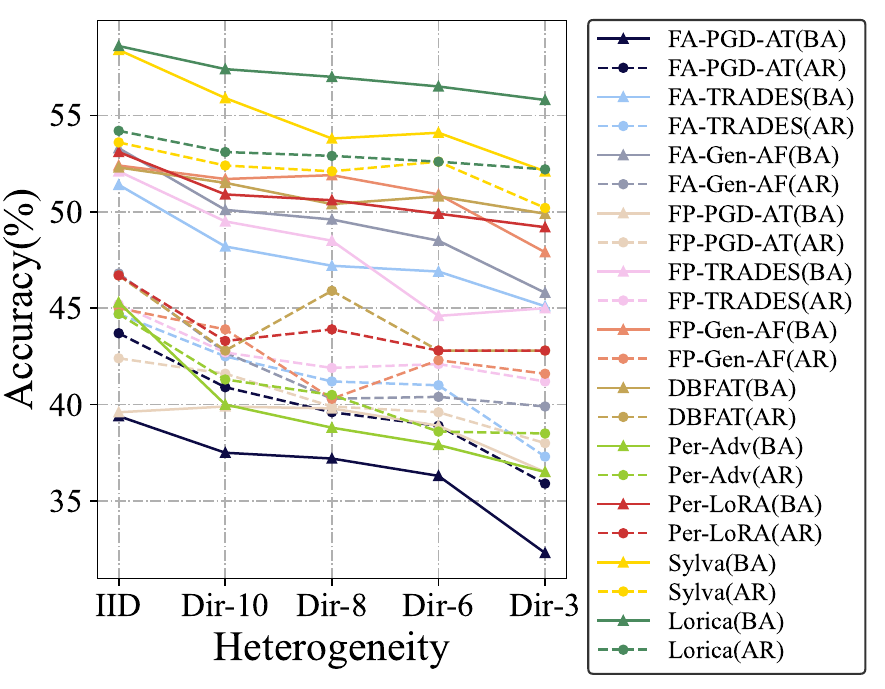}%
\label{fig:noniid_cifar-10_vit-t}}
\subfloat[STL-10]{\includegraphics[width=1.65in]{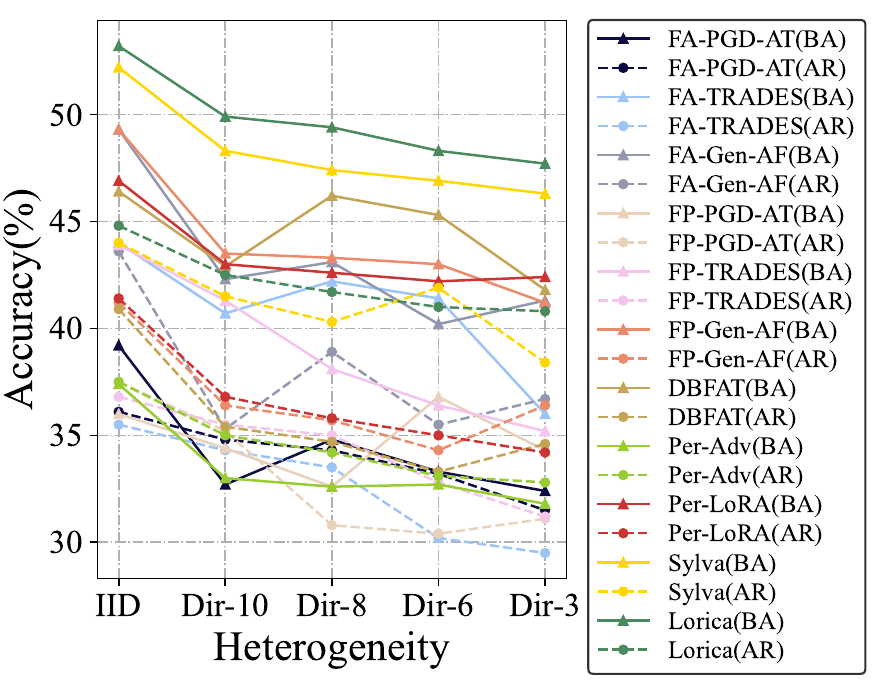}%
\label{fig:noniid_stl-10_vit-t}}

\vspace{-3mm} 
\subfloat[GTSRB]{\includegraphics[width=1.65in]{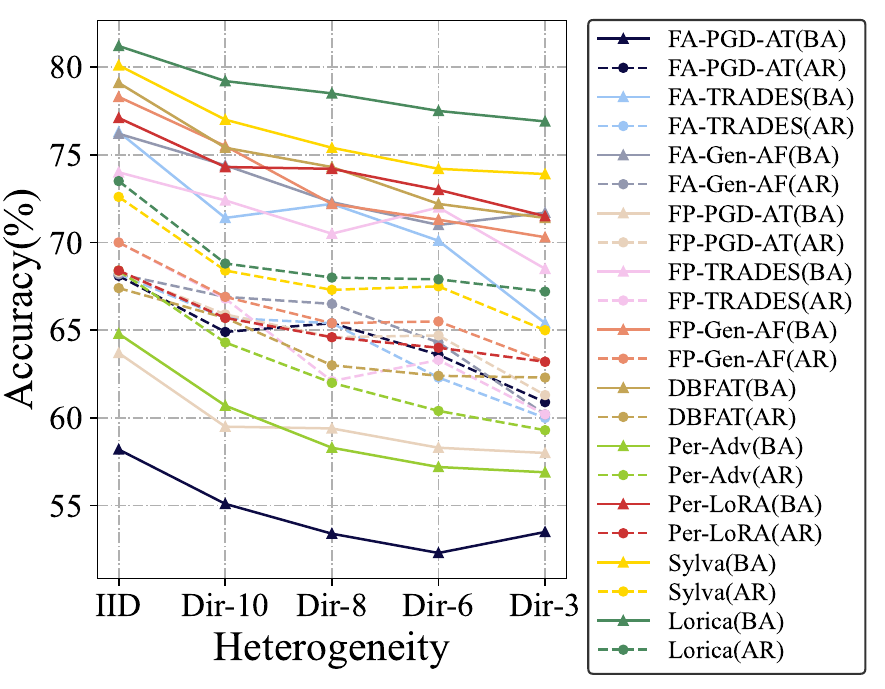}%
\label{fig:noniid_gtsrb_vit-t}}
\subfloat[CIFAR-100]{\includegraphics[width=1.65in]{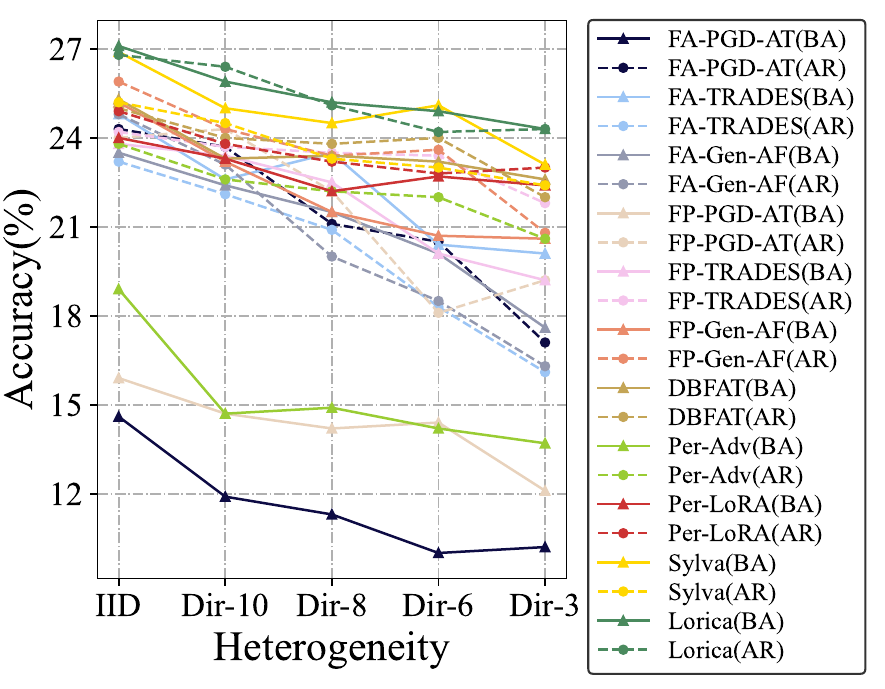}%
\label{fig:noniid_cifar-100_vit-t}}
\caption{Comparison of adversarial training performance under different heterogeneity levels (ViT-T/16)}
\label{fig:noniid_vit-t}
\vspace{-4mm}
\end{figure}

\begin{figure}[!tb]
\centering
\subfloat[CIFAR-10]{\includegraphics[width=1.65in]{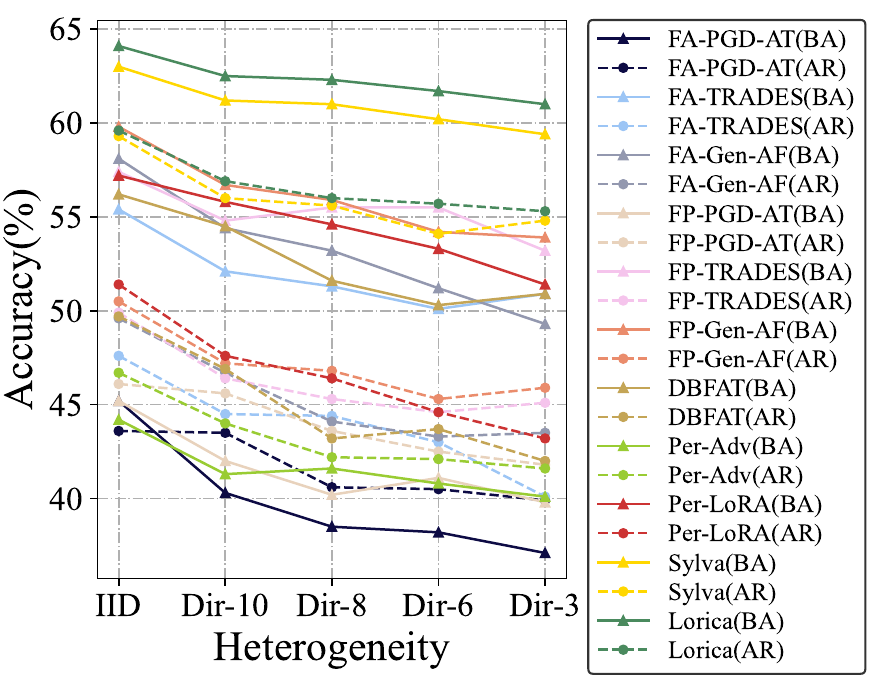}%
\label{fig:noniid_cifar-10_vit-l}}
\subfloat[STL-10]{\includegraphics[width=1.65in]{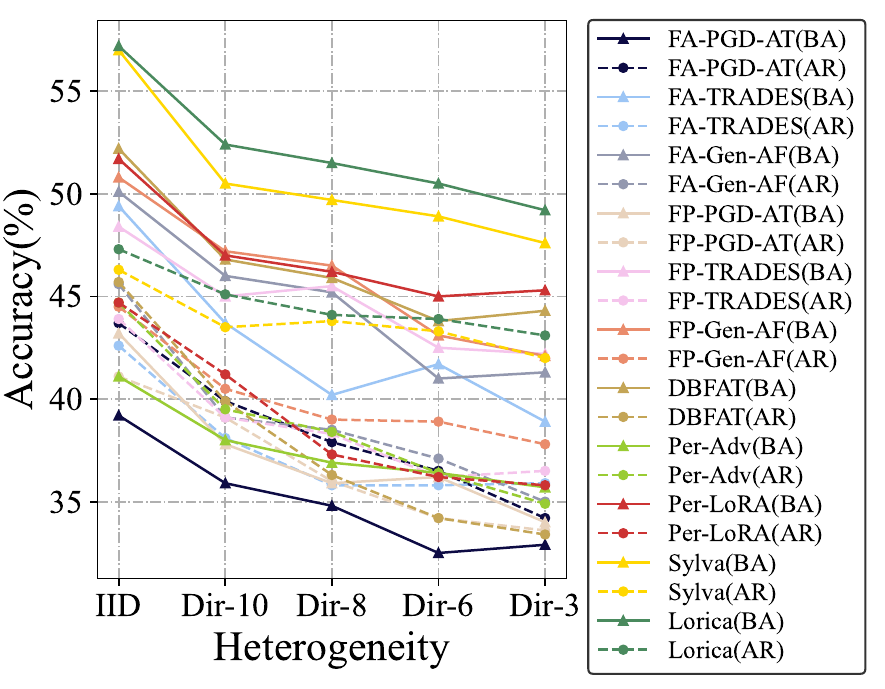}%
\label{fig:noniid_stl-10_vit-l}}

\vspace{-3mm} 
\subfloat[GTSRB]{\includegraphics[width=1.65in]{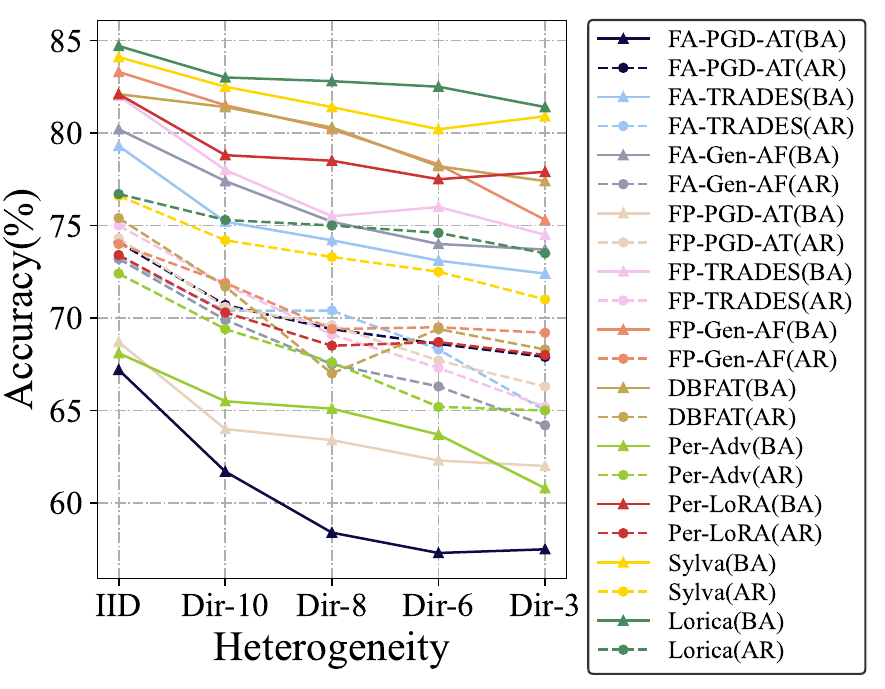}%
\label{fig:noniid_gtsrb_vit-l}}
\subfloat[CIFAR-100]{\includegraphics[width=1.65in]{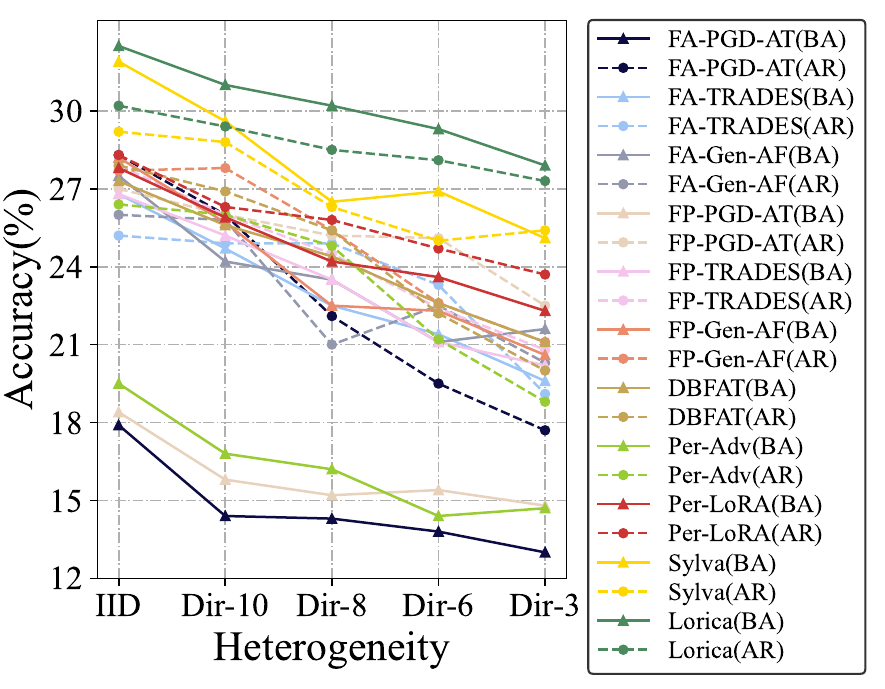}%
\label{fig:noniid_cifar-100_vit-l}}
\caption{Comparison of adversarial training performance under different heterogeneity levels (ViT-L/16)}
\label{fig:noniid_vit-l}
\vspace{-4mm}
\end{figure}

\subsection{Data Heterogeneity}\label{App data noniid}

We adopt the widely used data heterogeneity partitioning method based on the Dirichlet distribution~\cite{li2019convergence}, which effectively captures the non-IID nature of data distributions typically observed in real-world scenarios. For the CIFAR-10 dataset, we distribute the data among 15 clients with varying Dirichlet distribution parameters, ensuring that both the training and test sets for each client follow the same distribution. This method realistically simulates data heterogeneity, providing a robust setup for evaluating federated learning algorithms in non-IID environments.

The training set partitioning is visualized in \cref{fig:dir}, where the x-axis represents the clients and the y-axis shows the class sample counts, with different classes distinguished by colored bars. In the experiments, we use Dirichlet distribution parameters of 10, 6, and 3, along with an IID setup for comparison. Smaller Dirichlet parameters increase data heterogeneity, resulting in more imbalanced class distributions across clients. This approach effectively simulates realistic data heterogeneity by introducing varying class distributions, capturing natural class imbalances often encountered in real-world scenarios. It provides a more challenging evaluation setup, allowing for a comprehensive assessment of model robustness under diverse non-IID conditions, and tests the model's ability to generalize across clients with differing data distributions.

\begin{table}[!tb]
\centering
\caption{Detailed specifications of GPUs in real edge devices}\label{tab:result_gpu_status}
\renewcommand{\arraystretch}{0.9}
\scalebox{0.9}{
\begin{tabular}{@{}c|ccccc@{}}
\toprule
 &
  Type &
  \begin{tabular}[c]{@{}c@{}}Memory\\ (GB)\end{tabular} &
  \begin{tabular}[c]{@{}c@{}}Clock\\ (MHz)\end{tabular} &
  \begin{tabular}[c]{@{}c@{}}Bandwidth\\ (GB/s)\end{tabular} &
  \begin{tabular}[c]{@{}c@{}}FP16\\ (TFLOPS)\end{tabular} \\ \midrule
RTX 4090   & GDDR6X & 24 & 2235 & 1010  & 82.58 \\ 
RTX 3090   & GDDR6X & 24 & 1395 & 936   & 35.58 \\ 
RTX 2080-Ti & GDDR6  & 12 & 1350 & 768   & 30.14 \\ 
RTX 3060   & GDDR6  & 12 & 1320 & 360   & 12.74 \\ 
AGX Orin    & LPDDR5 & 32 & 930  & 204 & 10.65 \\ \bottomrule
\end{tabular}
}
\vspace{-3mm}
\end{table}

\begin{figure}[!tb]
\centering
\subfloat[GeForce RTX 3060]{\includegraphics[width=1.2in]{fig/real_devices-apollo.pdf}%
\label{fig:real_devices_apollo}}
\hspace{6mm} 
\subfloat[Jetson AGX Orin]{\includegraphics[width=1.2in]{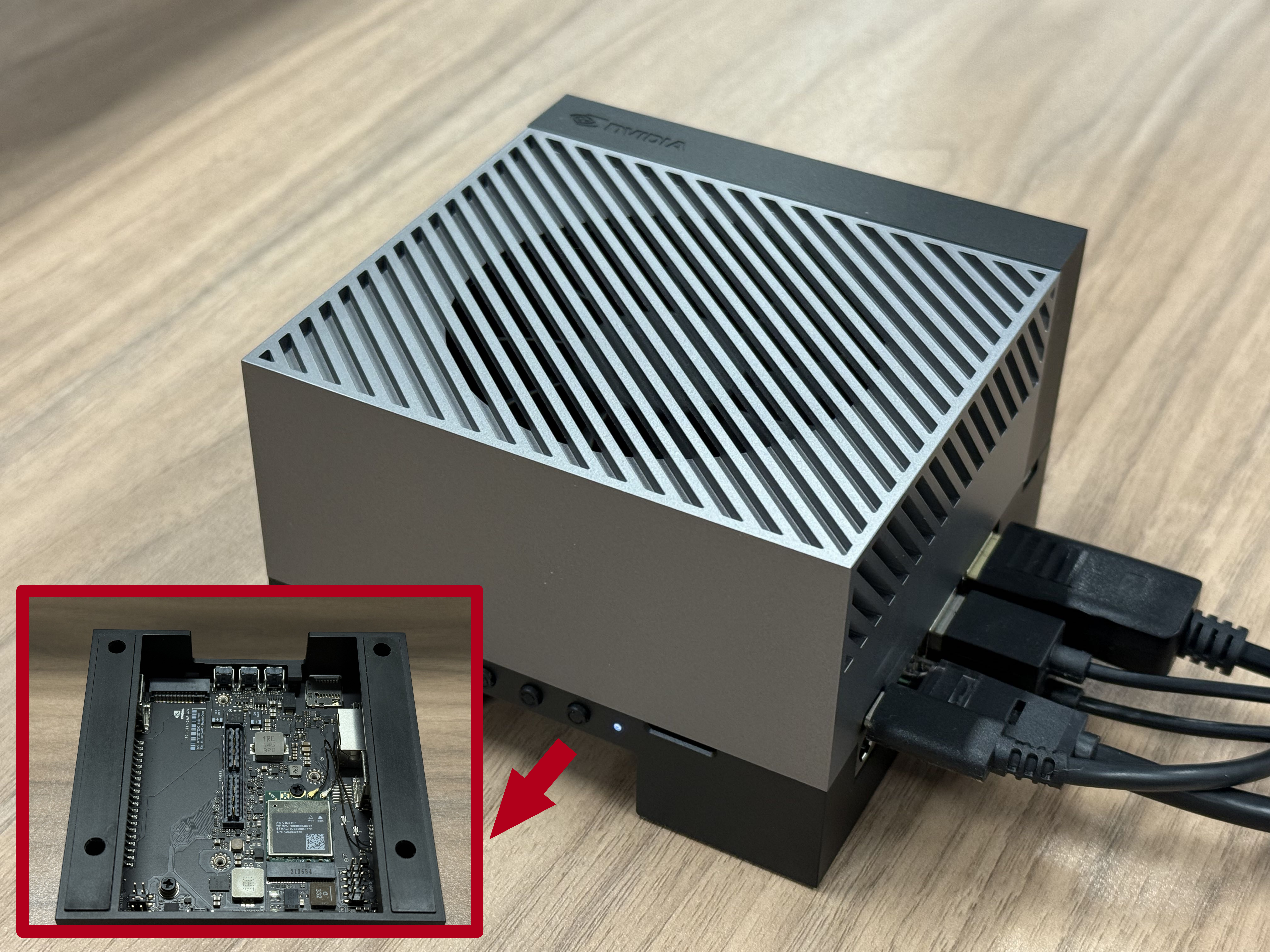}%
\label{fig:real_devices_orin}}
\caption{Commonly used GPUs in autonomous driving scenarios}
\label{fig:real_devices}
\vspace{-5mm}
\end{figure}

\begin{figure}[!tb]
\centering
\subfloat[ViT-T]{\includegraphics[width=1.1in]{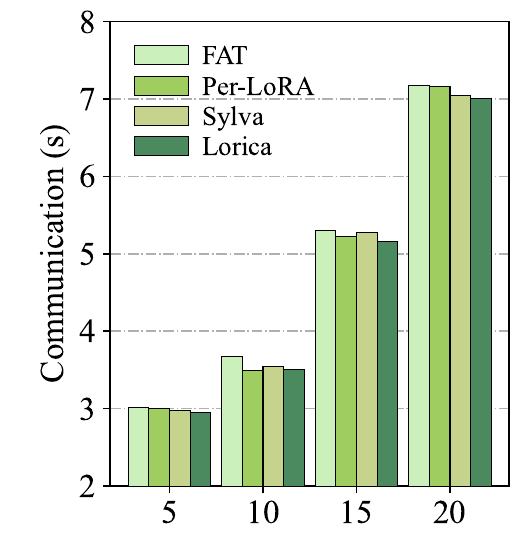}%
\label{fig:communication_vit_t}}
\subfloat[ViT-B]{\includegraphics[width=1.1in]{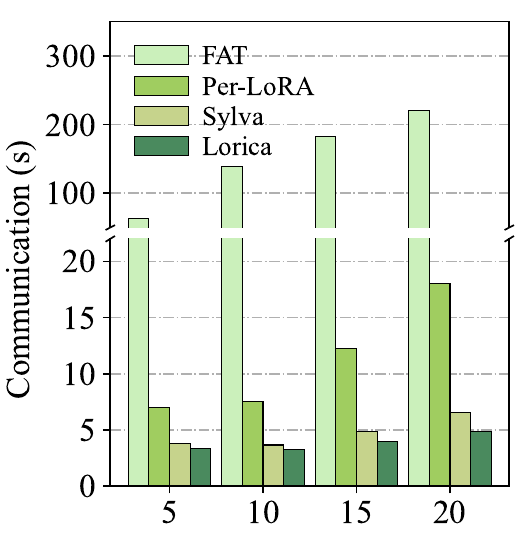}%
\label{fig:communication_vit_b}}
\subfloat[ViT-L]{\includegraphics[width=1.1in]{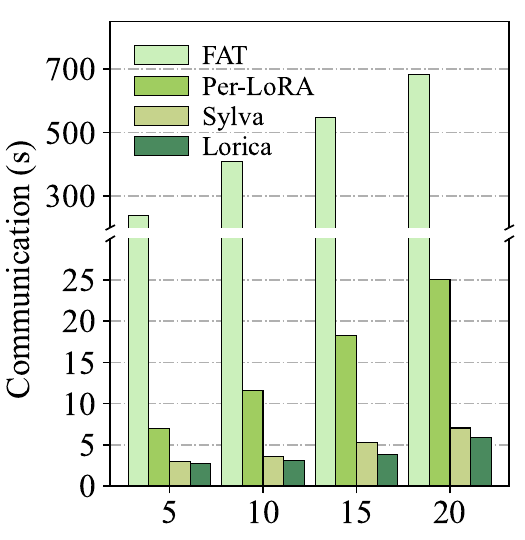}%
\label{fig:communication_vit_l}}
\caption{Relationship between the number of clients and communication time}
\label{fig:communication}
\vspace{-5mm}
\end{figure}

\section{Additional Results on the Performance of Adversarial Training (RQ1)}\label{App B}

\cref{tab:result_vit_b} provides a comprehensive comparison of \textit{Lorica} against baseline models utilizing the ViT-B architecture across multiple datasets. The results emphasize \textit{Lorica}'s superior performance in terms of both adversarial robustness and benign accuracy under a wide range of adversarial attack scenarios. Notably, \textit{Lorica} consistently surpasses the baseline methods, demonstrating its capability to effectively manage diverse attack conditions while maintaining exceptional levels of adversarial reliability.

To further evaluate the adaptability of \textit{Lorica}, we extend our analysis to models of varying sizes, specifically ViT-T and ViT-L, with detailed results presented in \cref{tab:result_vit_t} and \cref{tab:result_vit_l}. The findings highlight that \textit{Lorica} maintains strong defense capabilities across different model scales, delivering robust performance even as model complexity changes. \textit{Lorica} outperforms both traditional distributed algorithms and personalized algorithms, consistently demonstrating superior performance in terms of both robustness and adaptability. Notably, while the defense capabilities of baseline methods often degrade significantly with smaller models, such as ViT-T, \textit{Lorica} demonstrates exceptional adaptability. It consistently achieves an effective balance between adversarial robustness and benign accuracy, underscoring its versatility and resilience under varying conditions.
Moreover, \textit{Lorica} achieves overall better performance than \textit{Sylva}, both in terms of accuracy and robustness.
These findings highlight the remarkable versatility of \textit{Lorica}, demonstrating its suitability for a broad spectrum of model architectures and adversarial scenarios. Its consistent ability to deliver robust performance across diverse datasets and varying model scales underscores its potential as a dependable solution for adversarial training. This adaptability makes \textit{Lorica} an effective choice for applications ranging from simple to highly complex environments, further cementing its value in enhancing model resilience.

\begin{figure}[!tb]
\centering
\subfloat[Label-flipping]{\includegraphics[width=1.65in]{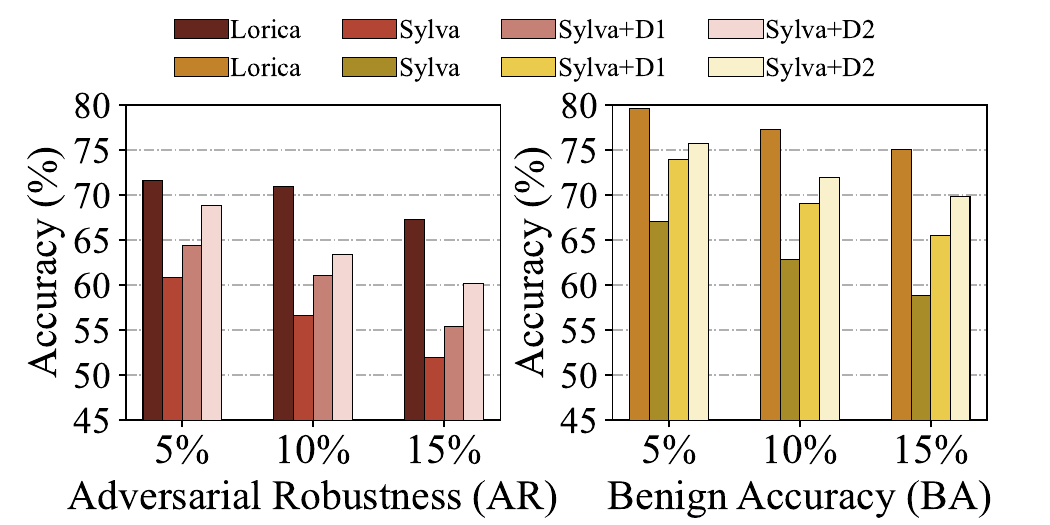}%
\label{fig:byzantine_1_gtsrb}}
\subfloat[MPAF]{\includegraphics[width=1.65in]{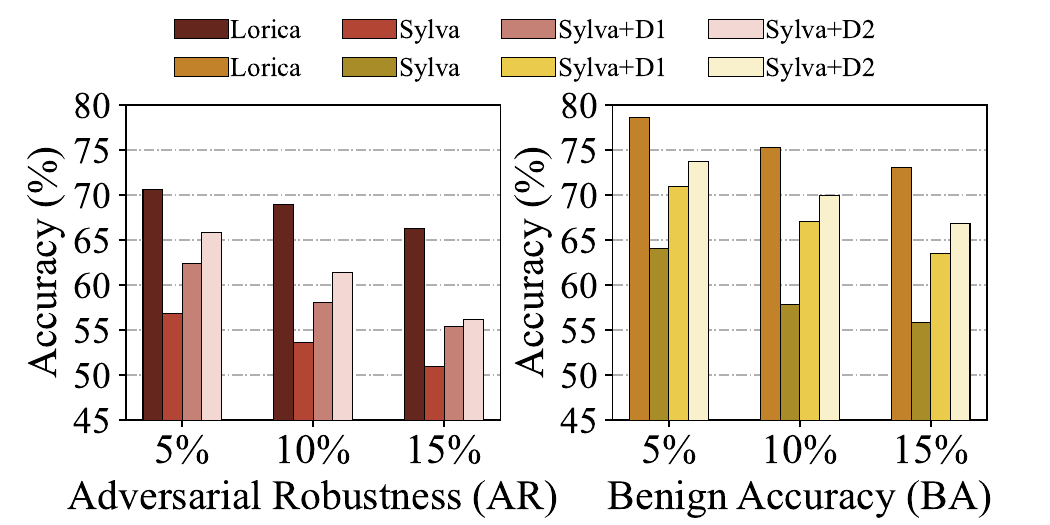}%
\label{fig:byzantine_2_gtsrb}}
\caption{Defense performance against Byzantine attacks on GTSRB}
\label{fig:byzantine_gtsrb}
\vspace{-5mm}
\end{figure}

\begin{figure}[!tb]
\centering
\subfloat[Label-flipping]{\includegraphics[width=1.65in]{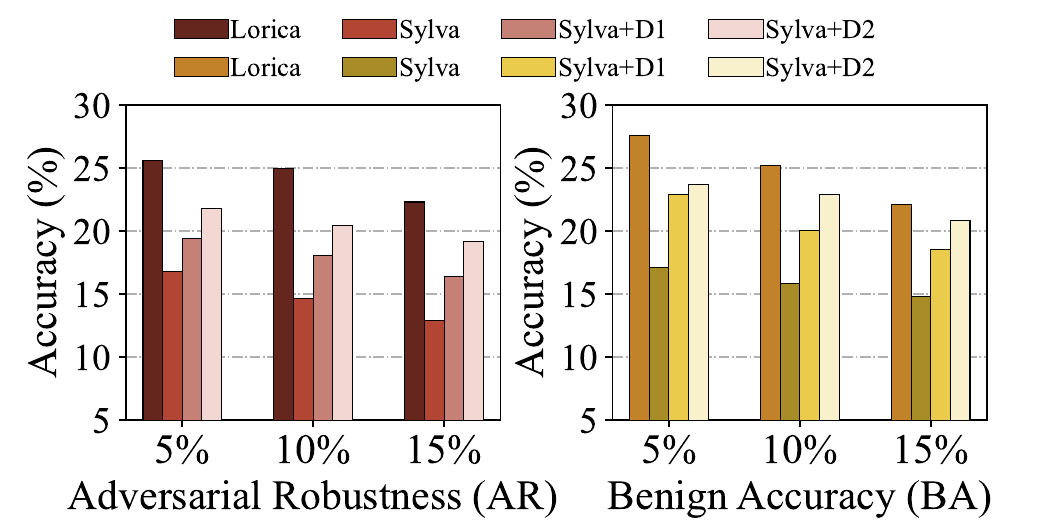}%
\label{fig:byzantine_1_cifar100}}
\subfloat[MPAF]{\includegraphics[width=1.65in]{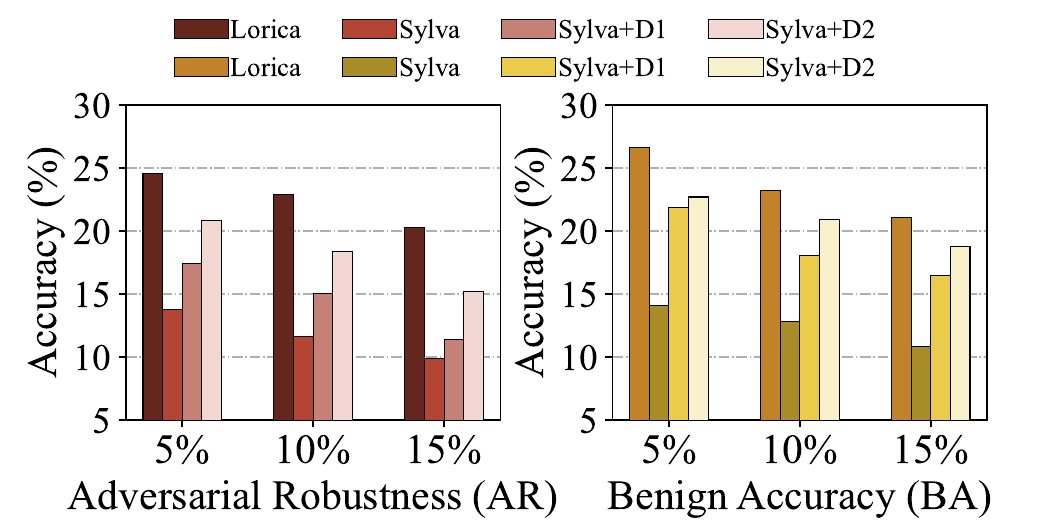}%
\label{fig:byzantine_2_cifar100}}
\caption{Defense performance against Byzantine attacks on CIFAR-100}
\label{fig:byzantine_cifar100}
\vspace{-5mm}
\end{figure}

\begin{figure}[!tb]
\centering
\subfloat[GTSRB]{\includegraphics[width=1.65in]{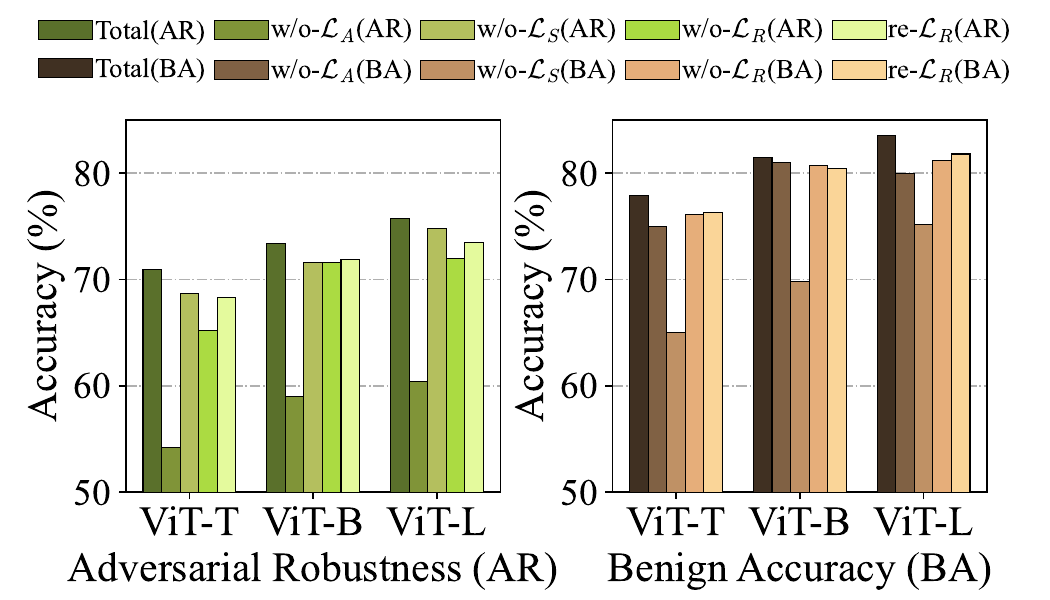}%
\label{fig:ablation_loss_gtsrb}}
\subfloat[CIFAR-100]{\includegraphics[width=1.65in]{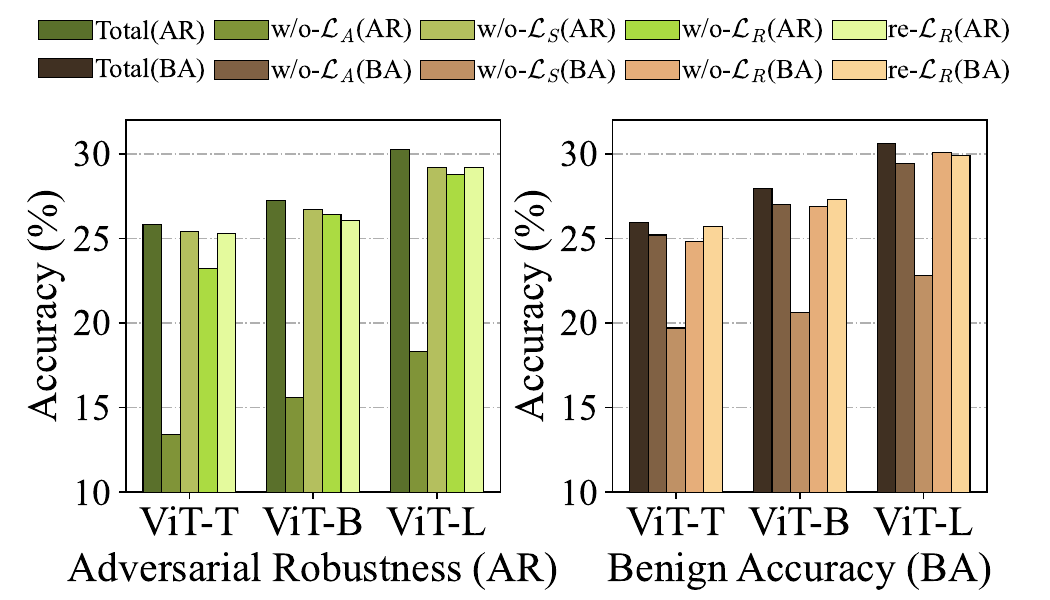}%
\label{fig:ablation_loss_cifar100}}
\caption{Impact of different loss modules on adversarial robustness and benign accuracy on GTSRB and CIFAR-100}
\label{fig:ablation_loss_app}
\vspace{-5mm}
\end{figure}

\section{Additional Results on the Performance Under Different Heterogeneity Levels (RQ2)}\label{App C}

In \cref{Sec4:RQ2}, we thoroughly analyze the adversarial defense performance of the ViT-B model under diverse heterogeneity conditions, with \textit{Lorica} demonstrating remarkable defense capabilities and robustness throughout. To gain deeper insights into how models of varying scales respond to heterogeneity, we extend the evaluation to the ViT-T and ViT-L models, systematically assessing their defense performance under different levels of heterogeneity, as depicted in \cref{fig:noniid_vit-t} and \cref{fig:noniid_vit-l}, respectively.

The results demonstrate that \textit{Lorica} consistently outperforms its counterparts across various model sizes, effectively maintaining both accuracy and robustness, even in highly challenging heterogeneous environments where data distribution and computational resources vary significantly. Notably, smaller models, such as ViT-T, exhibit more rapid convergence during training, enabling them to achieve relatively stable performance despite the increasing heterogeneity of the data.
This finding underscores \textit{Lorica}'s adaptability in managing the intricate balance between model complexity and environmental variability. Moreover, compared to the other two personalized baseline algorithms, the smaller performance fluctuations observed under heightened heterogeneity indicate that \textit{Lorica}'s personalized framework effectively mitigates the detrimental impacts of data heterogeneity by tailoring defense strategies to individual clients.
In addition, \textit{Lorica} demonstrates consistently superior performance to \textit{Sylva} across models of various scales, confirming the effectiveness of its personalization-driven optimization strategy.
These findings highlight the exceptional versatility of \textit{Lorica}, making it an ideal choice for adversarial training across systems characterized by diverse heterogeneity and a broad spectrum of model sizes.

\begin{figure}[!tb]
\centering
\subfloat[GTSRB]{\includegraphics[width=1.65in]{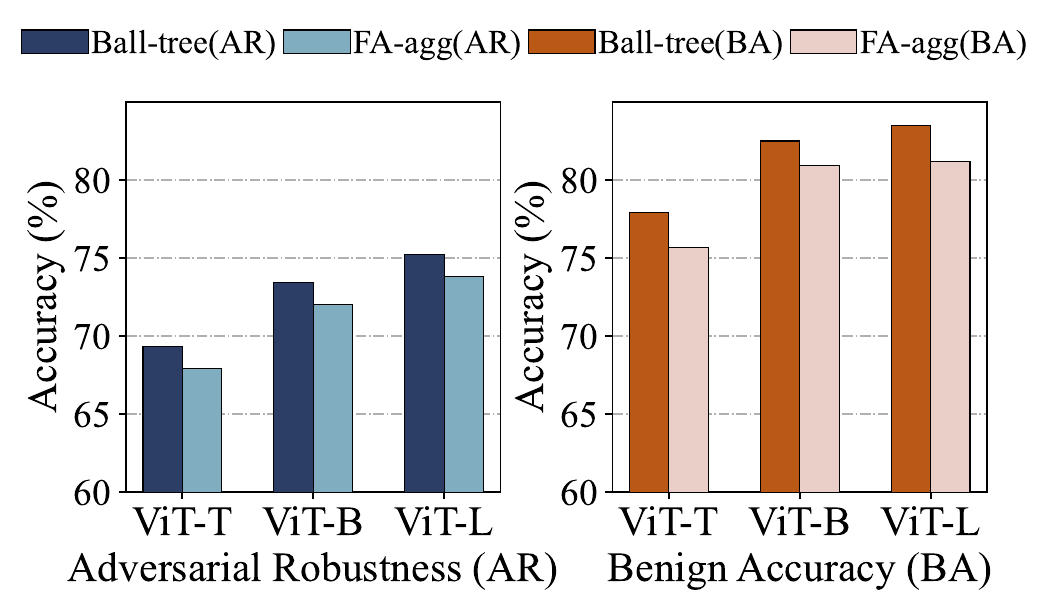}%
\label{fig:ablation_aggre_gtsrb}}
\subfloat[CIFAR-100]{\includegraphics[width=1.65in]{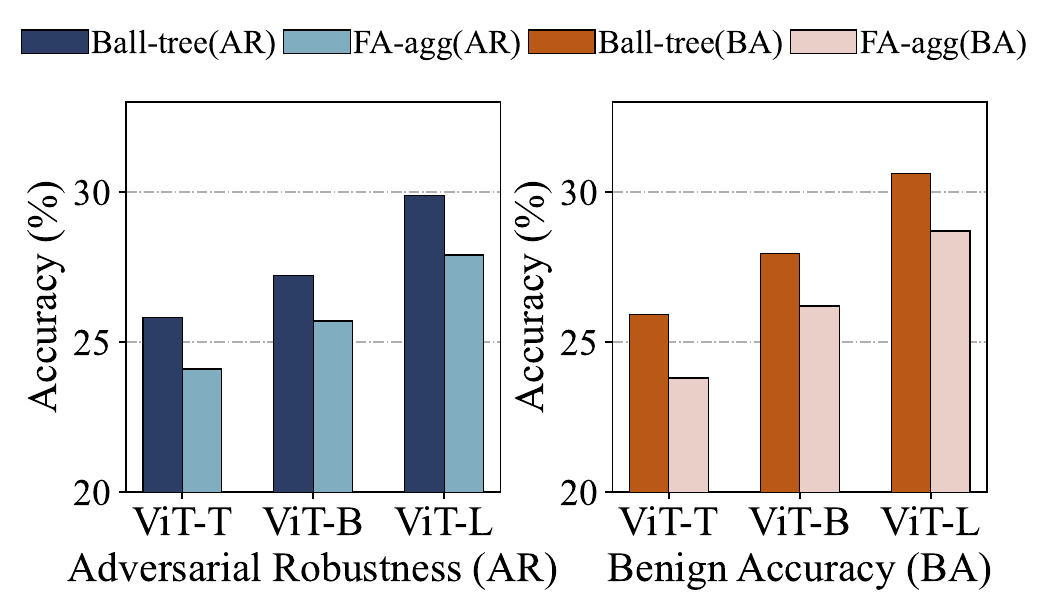}%
\label{fig:ablation_aggre_cifar100}}
\caption{Impact of ball-tree-based aggregation algorithm on GTSRB and CIFAR-100}
\label{fig:ablation_aggre_app}
\vspace{-5mm}
\end{figure}

\begin{figure}[!tb]
\centering
\subfloat[GTSRB]{\includegraphics[width=1.65in]{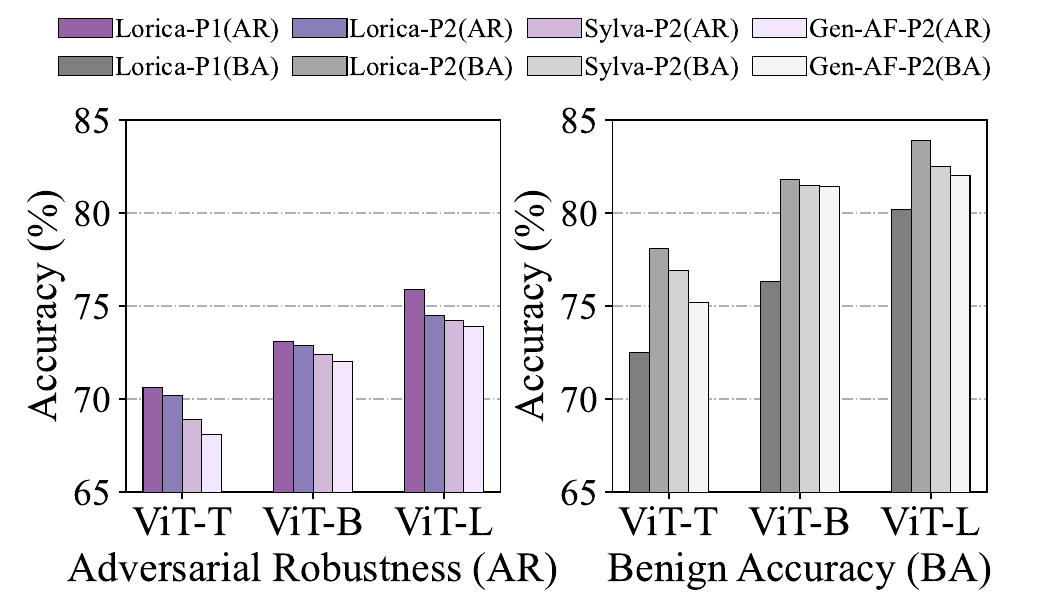}%
\label{fig:ablation_phase_gtsrb}}
\subfloat[CIFAR-100]{\includegraphics[width=1.65in]{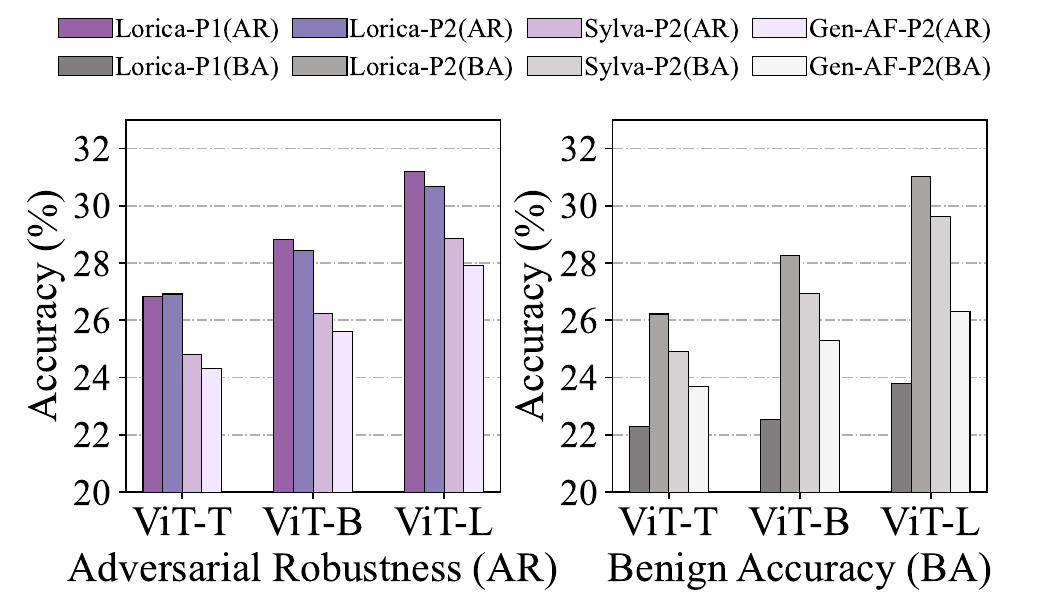}%
\label{fig:ablation_phase_cifar100}}
\caption{Impact of different phases on the trade-off between robustness and accuracy for GTSRB and CIFAR-100}
\label{fig:ablation_phase_app}
\vspace{-3mm}
\end{figure}

\begin{table}[!tb]
\centering
\caption{The Impact of Hyperparameters in STL-10}\label{tab:result_hyper_stl-10}
\scalebox{0.95}{
\begin{tabular}{@{}c|c|cc|c|c|cc@{}}
\toprule
Para                  & Value & AR    & BA    & Para                 & Value & AR    & BA    \\ \midrule
\multirow{4}{*}{$\gamma$}    & 0.9 & 44.07 & 52.13 & \multirow{4}{*}{$\beta$}                                                & 20  & 46.25 & 50.13 \\
                             & 0.7 & 44.30 & 51.67 &                                                                         & 10  & 46.91 & 50.37 \\
                             & 0.5 & 43.87 & 51.22 &                                                                         & 5   & 44.07 & 52.13 \\
                             & 0.3 & 43.15 & 50.48 &                                                                         & 1   & 44.75 & 51.66 \\ \midrule
\multirow{4}{*}{$\epsilon$}  & 0.9 & 44.07 & 52.13 & \multirow{4}{*}{$B$}                                                    & 5\%  & 42.76 & 51.50 \\
                             & 0.7 & 43.34 & 51.30 &                                                                         & 10\% & 44.07 & 52.13 \\
                             & 0.5 & 42.52 & 51.42 &                                                                         & 20\% & 44.79 & 52.26 \\
                             & 0.3 & 42.81 & 51.77 &                                                                         & 30\% & 44.36 & 53.15 \\ \midrule
\multirow{4}{*}{$r$}         & 8   & 44.89 & 52.53 & \multirow{4}{*}{\begin{tabular}[c]{@{}c@{}}PGD\\ Strength\end{tabular}} & 3   & 43.14 & 55.33 \\
                             & 4   & 44.07 & 52.13 &                                                                         & 5   & 44.53 & 53.80 \\
                             & 2   & 44.58 & 53.21 &                                                                         & 10  & 44.07 & 52.13 \\
                             & 1   & 43.77 & 52.86 &                                                                         & 20  & 45.60 & 50.51 \\ \bottomrule
\end{tabular}
}
\vspace{-3mm}
\end{table}

\section{Additional results on efficiency comparison across different system scales (RQ3)}\label{APP D}

In \cref{Sec4:RQ3}, we emphasize the efficiency advantages of \textit{Lorica} in real-world systems composed of diverse edge devices operating under adversarial training scenarios. To delve deeper into its communication efficiency, we perform additional experiments to assess the communication overhead in distributed systems of varying scales. By leveraging the edge devices described in \cref{Sec4:RQ3}, we simulate distributed systems with 5, 10, 15, and 20 clients, each performing adversarial training while concurrently communicating with the server. As the client count rises, network congestion emerges as a critical challenge, potentially undermining the timeliness and reliability of adversarial updates.

\cref{fig:communication} presents a comparative analysis of communication time for distributed training across three distinct models under different client counts. The results clearly demonstrate a proportional increase in communication time as the number of clients grows, primarily due to heightened network congestion. 
Crucially, \textit{Lorica} and Per-LoRA exhibit a significant advantage over traditional algorithms, particularly with larger models, by transmitting only the lightweight LoRA parameters. Since \textit{Lorica} only transmits the backbone's LoRA-FA parameters and keeps the classifier local to preserve personalization, it incurs lower communication overhead, making it more efficient for real-world deployments.
This approach drastically reduces communication overhead, underscoring \textit{Lorica}'s practicality and efficiency for deployment in real-world distributed systems. Its ability to mitigate communication challenges is especially valuable in scenarios involving large-scale models and high client densities, further cementing its suitability for such environments.

\begin{table}[!tb]
\centering
\caption{The Impact of Hyperparameters in GTSRB}\label{tab:result_hyper_gtsrb}
\scalebox{0.95}{
\begin{tabular}{@{}c|c|cc|c|c|cc@{}}
\toprule
Para                  & Value & AR    & BA    & Para                 & Value & AR    & BA    \\ \midrule
\multirow{4}{*}{$\gamma$}    & 0.9 & 73.92 & 81.54 & \multirow{4}{*}{$\beta$}                                                & 20  & 73.21 & 79.77 \\
                             & 0.7 & 73.33 & 80.57 &                                                                         & 10  & 73.58 & 81.62 \\
                             & 0.5 & 73.46 & 80.10 &                                                                         & 5   & 73.92 & 81.54 \\
                             & 0.3 & 72.92 & 79.59 &                                                                         & 1   & 72.39 & 81.42 \\ \midrule
\multirow{4}{*}{$\epsilon$}  & 0.9 & 73.92 & 81.54 & \multirow{4}{*}{$B$}                                                    & 5\%  & 71.80 & 80.04 \\
                             & 0.7 & 73.64 & 79.59 &                                                                         & 10\% & 73.92 & 81.54 \\
                             & 0.5 & 72.30 & 80.76 &                                                                         & 20\% & 73.36 & 81.38 \\
                             & 0.3 & 72.47 & 79.35 &                                                                         & 30\% & 72.69 & 80.52 \\ \midrule
\multirow{4}{*}{$r$}         & 8   & 73.88 & 80.74 & \multirow{4}{*}{\begin{tabular}[c]{@{}c@{}}PGD\\ Strength\end{tabular}} & 3   & 72.08 & 83.28 \\
                             & 4   & 73.92 & 81.54 &                                                                         & 5   & 72.31 & 82.52 \\
                             & 2   & 73.54 & 81.26 &                                                                         & 10  & 73.92 & 81.54 \\
                             & 1   & 72.68 & 79.67 &                                                                         & 20  & 74.71 & 79.27 \\ \bottomrule
\end{tabular}
}
\vspace{-3mm}
\end{table}

\begin{table}[!tb]
\centering
\caption{The Impact of Hyperparameters in CIFAR-100}\label{tab:result_hyper_cifar-100}
\scalebox{0.95}{
\begin{tabular}{@{}c|c|cc|c|c|cc@{}}
\toprule
Para                  & Value & AR    & BA    & Para                 & Value & AR    & BA    \\ \midrule
\multirow{4}{*}{$\gamma$}    & 0.9 & 27.77 & 27.93 & \multirow{4}{*}{$\beta$}                                                & 20  & 27.54 & 26.13 \\
                             & 0.7 & 27.06 & 27.35 &                                                                         & 10  & 27.48 & 26.10 \\
                             & 0.5 & 27.13 & 27.44 &                                                                         & 5   & 27.77 & 27.93 \\
                             & 0.3 & 26.15 & 26.57 &                                                                         & 1   & 25.44 & 29.47 \\ \midrule
\multirow{4}{*}{$\epsilon$}  & 0.9 & 27.77 & 27.93 & \multirow{4}{*}{$B$}                                                    & 5\%  & 26.67 & 26.58 \\
                             & 0.7 & 27.30 & 27.02 &                                                                         & 10\% & 27.77 & 27.93 \\
                             & 0.5 & 26.85 & 26.29 &                                                                         & 20\% & 27.33 & 28.01 \\
                             & 0.3 & 26.27 & 26.08 &                                                                         & 30\% & 27.32 & 28.12 \\ \midrule
\multirow{4}{*}{$r$}         & 8   & 27.55 & 28.13 & \multirow{4}{*}{\begin{tabular}[c]{@{}c@{}}PGD\\ Strength\end{tabular}} & 3   & 26.15 & 29.09 \\
                             & 4   & 27.77 & 27.93 &                                                                         & 5   & 26.71 & 28.68 \\
                             & 2   & 27.04 & 28.15 &                                                                         & 10  & 27.77 & 27.93 \\
                             & 1   & 27.01 & 27.49 &                                                                         & 20  & 28.19 & 26.02 \\ \bottomrule
\end{tabular}
}
\vspace{-3mm}
\end{table}

\section{Additional Results on the Performance of Byzantine Defence (RQ4)}\label{APP E}

In \cref{Sec4:RQ4}, we evaluate the defensive performance of \textit{Lorica} on CIFAR-10 and STL-10. In this subsection, we further extend our analysis to more challenging and complex datasets, namely GTSRB and CIFAR-100. The corresponding results are presented in \cref{fig:byzantine_gtsrb} and \cref{fig:byzantine_cifar100}.
The experimental findings demonstrate that, compared with Sylva and its enhanced variants, \textit{Lorica} consistently achieves superior robustness and accuracy across all evaluated datasets and under diverse attack settings. Notably, even on more fine-grained such as CIFAR-100 and domain-specific such as GTSRB classification tasks, where class diversity and distributional complexity are higher, \textit{Lorica} maintains stable defensive performance.
These results further indicate that \textit{Lorica} exhibits strong resilience against Byzantine behaviors, effectively mitigating the impact of malicious updates across diverse and complex data distributions.

\section{Additional Ablation Study Results (RQ5)}\label{APP F}

In \cref{Sec4:RQ5}, we evaluate the impact of \textit{Lorica}'s key modules on adversarial training models, focusing on CIFAR-10 and STL-10 datasets. This includes analyzing the loss function design, aggregation methods, and two-phase training framework, all critical to \textit{Lorica}'s performance. To validate these findings, we extend experiments to GTSRB and CIFAR-100 datasets.
The results, presented in \cref{fig:ablation_loss_app}, \cref{fig:ablation_aggre_app}, and \cref{fig:ablation_phase_app}, confirm the consistent contributions of each module. These trends align with the main text, demonstrating \textit{Lorica}'s robustness and strong generalization across datasets of varying complexity, highlighting its versatility in adversarial training scenarios.


We evaluate the impact of various hyperparameters, with detailed results shown in \cref{tab:result_hyper_stl-10}, \cref{tab:result_hyper_gtsrb}, and \cref{tab:result_hyper_cifar-100}. These findings align with those on CIFAR-10, revealing consistent trends across all datasets. Specifically, decreases in $\gamma$ and $\epsilon$ lead to declines in both adversarial robustness and benign accuracy, underscoring their importance in maintaining model performance.
In Phase 2, reducing $\beta$ slightly improves benign accuracy but significantly reduces adversarial robustness, highlighting the trade-off between these metrics. Different training budgets $B$ also exhibit a certain degree of sensitivity in the results.
These results stress the need for careful hyperparameter tuning to balance robustness and accuracy in adversarial training frameworks.

\begin{figure}[!tb]
\centering
\subfloat[CIFAR-10]{\includegraphics[width=1.65in]{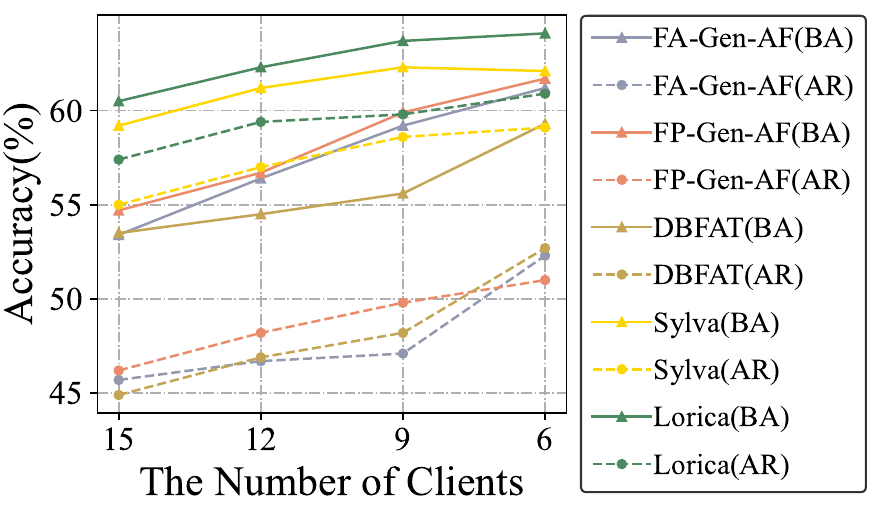}%
\label{fig:client_num_cifar-10_vit-b}}
\subfloat[STL-10]{\includegraphics[width=1.65in]{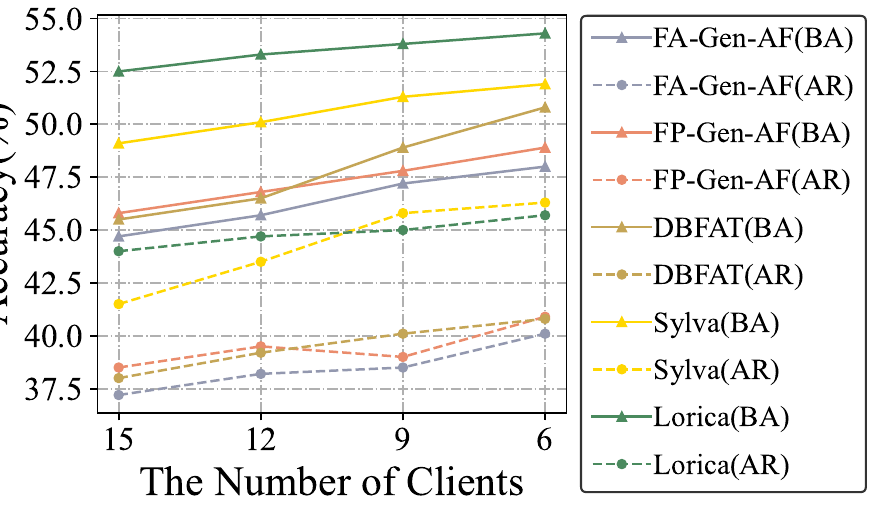}%
\label{fig:client_num_stl-10_vit-b}}

\vspace{-3mm} 
\subfloat[GTSRB]{\includegraphics[width=1.65in]{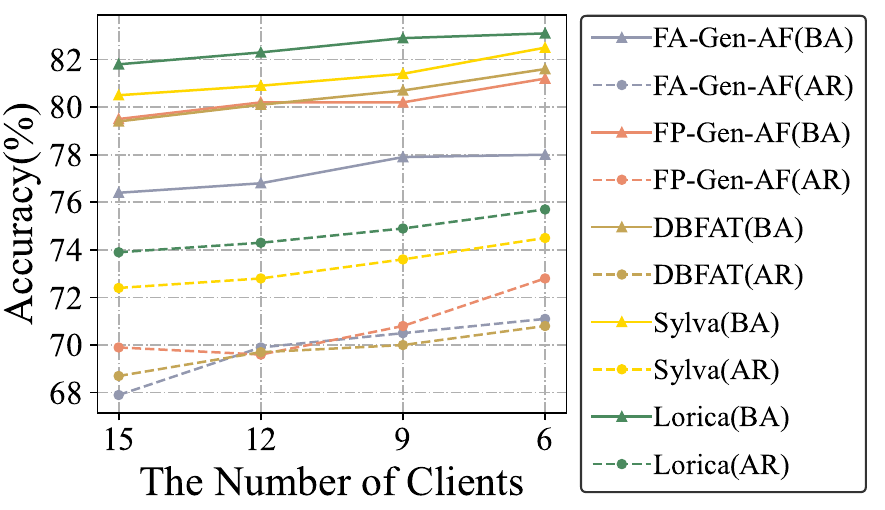}%
\label{fig:client_num_gtsrb_vit-b}}
\subfloat[CIFAR-100]{\includegraphics[width=1.65in]{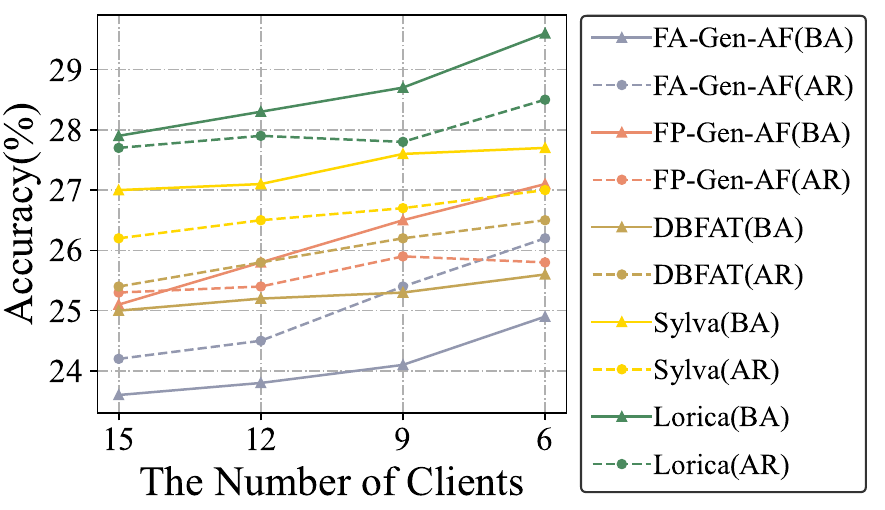}%
\label{fig:client_num_cifar-100_vit-b}}
\caption{Comparison of adversarial training performance under different number of clients (ViT-B/16)}
\label{fig:client_num}
\vspace{-3mm}
\end{figure}

To comprehensively evaluate the performance of the proposed framework, we conduct experiments under varying client numbers, as illustrated in \cref{fig:client_num}. For comparison, we select several high-performing baseline algorithms, including FA-Gen-AF, FP-Gen-AF, DBFAT and Sylva, to ensure a robust analysis. In the multi-client scenario, the number of clients is systematically varied among 15, 12, 9, and 6, with experiments conducted on four distinct datasets to assess both adversarial robustness and benign accuracy comprehensively.
The results reveal a clear trend: as the number of clients decreases, the overall performance improves, likely due to the reduced data heterogeneity associated with fewer clients. Notably, across all configurations, \textit{Lorica} consistently outperforms the baseline algorithms, demonstrating superior results in both adversarial robustness and benign accuracy. These findings underscore the framework's effectiveness and adaptability, highlighting its capacity to excel across a diverse range of multi-client settings, irrespective of the dataset or client configuration.

}

\end{document}